\documentclass[letterpaper]{article} % DO NOT CHANGE THIS
\usepackage{aaai2026}  % DO NOT CHANGE THIS
\usepackage{times}  % DO NOT CHANGE THIS
\usepackage{helvet}  % DO NOT CHANGE THIS
\usepackage{courier}  % DO NOT CHANGE THIS
\usepackage[hyphens]{url}  % DO NOT CHANGE THIS
\usepackage{graphicx} % DO NOT CHANGE THIS
\usepackage{natbib}  % DO NOT CHANGE THIS AND DO NOT ADD ANY OPTIONS TO IT
\usepackage{caption} % DO NOT CHANGE THIS AND DO NOT ADD ANY OPTIONS TO IT
\usepackage[T1]{fontenc}
\usepackage{amssymb,amsmath}
\usepackage{multirow}
\usepackage[utf8]{inputenc}
\usepackage{latexsym}
\usepackage{subcaption}
\usepackage{psfrag} % Note: 'cite' was removed as it conflicts with the mandatory 'natbib' package
\usepackage{algorithmic, algorithm}
\usepackage{booktabs}
\usepackage{acronym}
\usepackage{mathtools}
\usepackage{pgfplots}
\usepackage{tikz}
\usetikzlibrary{positioning, fit, backgrounds, arrows.meta, calc, shadows}
\mathtoolsset{showonlyrefs=true}

\newcommand{\oprocendsymbol}{\hbox{$\bullet$}}
\newcommand{\oprocend}{\relax\ifmmode\else\unskip\hfill\fi\oprocendsymbol}

\graphicspath{{epsfiles/}}

\title{Engagement-Aware Agentic Pursuit-Evasion}
\author{
    Ananya Acharya\textsuperscript{\rm 1},
    Trenton Goyette\textsuperscript{\rm 1},
    Masoud Ataei\textsuperscript{\rm 2},
    Adrian Stoica\textsuperscript{\rm 1},
    Vikas Dhiman\textsuperscript{\rm 2},
    \\
    Mohammad Javad Khojasteh\textsuperscript{\rm 1}
}
\affiliations{
    \textsuperscript{\rm 1}RIT, Rochester, NY\\
    \textsuperscript{\rm 2}University of Maine, Orono, ME\\
    
    \{aa2334, twg8622, mjkeme, adseee\}@rit.edu, 
    \{masoud.ataei, vikas.dhiman\}@maine.edu
}
\acrodef{mpc}[MPC]{model predictive control}
\acrodef{cbf}[CBF]{control barrier function}
\acrodef{roe}[RoE]{rules-of-engagement}
\acrodef{llm}[LLM]{large language model}

\begin{document}

\maketitle

%{\color{orange} should we change shield to defend? it seems we are not using shield anymore in the paper? or it is fine? we can just mention in the paper defend (shield) one time? maybe the second is better?}
\begin{abstract}
This paper presents a hierarchical multi-agent architecture in which independent large language model (LLM) planners perform strategic role assignment for attacking and defending robot teams, decoupled from low-level control execution. At each planning cycle, each team's LLM planner observes its own team in full but the opposing team only within its robots' combined field of view, then assigns each robot a tactical role -- e.g., hold a perimeter, neutralize an intruder on contact, or converge with teammates for capture -- together with a natural-language justification. Each robot independently executes its assigned role through a receding-horizon model predictive control (MPC) controller, followed by a discrete-time control barrier function (CBF) filter for safety and role-dependent engagement constraints. Differentiated capture and neutralization incentives require the defending planner to balance threat resolution against resource allocation under partial observability. We evaluate the framework across variable-sized adversary teams using both state-based tactical reasoning and image-based contact classification. Results show that collective team behavior can be adapted through high-level LLM role assignment while retaining the same underlying low-level control architecture. \footnote{Code available at: \url{https://github.com/ananya-ac/llm-pursuit-evasion.git}.}
\end{abstract}

%The source code is available\footnote{\url{https://github.com/ananya-ac/llm-pursuit-evasion.git} {\color{orange}since this is an ananymous submission we should not include this link for now. But it should be in arXiv version also it should be updated and working. thanks}}.%

\section{Introduction}
A growing body of work integrates foundation models -- large language and vision-language models pretrained on broad, task-agnostic data -- into robot planning and decision-making. Rather than training policies end-to-end for a single task, these approaches use foundation models to reason over the current task state in natural language and make high-level decisions in a zero-shot manner.~\cite{hu2023toward, firoozi2025foundation}. This shift is particularly relevant to planning specifically: rather than treating a foundation model as a low-level policy to be  fine-tuned~\cite{julg2025refined}, a language model can instead sit above a conventional control stack as a strategic layer, issuing structured, human-interpretable decisions that a separate, purpose-built controller then executes \cite{navaratnarajah2026zero}. This work situates that idea in an adversarial, multi-agent setting -- pursuit-evasion -- to test whether \acp{llm} can serve as  strategic planners under partial observability and competing objectives. %rather than in the fully cooperative  settings that has typically considered.

Pursuit-evasion problems provide a general framework for mathematically formalizing a wide range of applications, including surveillance, navigation, 
%nalysis of biological behaviors,
and conflict operations. The formulation considers multiple agents organized into two opposing teams: pursuers and evaders. The primary objective is to develop strategies that enable an autonomous agent to act effectively against its opponent~\cite{shishika2020review,weintraub2020introduction,chung2011search,weintraub2023range}.

We formulate pursuit-evasion as a finite-horizon dynamic game between a team of pursuers and a team of evaders, each team acting to optimize its own objective -- capture or interception for the pursuers, breach or escape for the evaders -- against the other's best response.
\begin{figure*}
    \centering
    \scalebox{0.7}{
        \begin{tikzpicture}[
    font=\sffamily\Large\bfseries,
    >={Stealth[length=3.5mm, width=2.5mm]},
    basebox/.style={
        rectangle, rounded corners=4pt, line width=1.5pt,
        minimum width=2.6cm, minimum height=1.1cm, align=center, text=black,
        draw=blue!70!black,
        drop shadow={opacity=0.15, shadow xshift=1mm, shadow yshift=-1mm}
    },
    roe/.style={basebox, minimum width=5.6cm, fill=blue!15},
    mpc/.style={basebox, fill=blue!10},
    cbf/.style={basebox, fill=blue!5},
    robot/.style={basebox, fill=gray!10, draw=gray!70!black},
    mainline/.style={->, line width=2pt, blue!70!black},
    envline/.style={->, line width=1.5pt, densely dotted, green!40!black},
    envbox/.style={rectangle, densely dotted, line width=1.5pt, draw=green!40!black, rounded corners=6pt, inner sep=0.45cm},
    feedbackline/.style={->, line width=1.5pt, black!65},
    freqlabel/.style={
        font=\sffamily\large, text=black, align=left,
        fill=white, fill opacity=0.85, text opacity=1,
        rounded corners=2pt, inner sep=2pt
    },
    ellipsis/.style={font=\sffamily\LARGE\bfseries, text=black}
]
% 1. Nodes -----------------------------------------------------------
\node[roe] (roe) {LLM RoE\\(Role + Target Assignment)};
\node[mpc, below=1.7cm of roe, xshift=-6.3cm] (mpc1) {MPC$_1$};
\node[mpc, below=1.7cm of roe] (mpc2) {MPC$_2$};
\node[mpc, below=1.7cm of roe, xshift=6.3cm]  (mpcN) {MPC$_N$};
\node[cbf, below=1.2cm of mpc1] (cbf1) {CBF$_1$};
\node[cbf, below=1.2cm of mpc2] (cbf2) {CBF$_2$};
\node[cbf, below=1.2cm of mpcN] (cbfN) {CBF$_N$};
\node[robot, below=1.2cm of cbf1] (r1) {Robot$_1$};
\node[robot, below=1.2cm of cbf2] (r2) {Robot$_2$};
\node[robot, below=1.2cm of cbfN] (rN) {Robot$_N$};
% Midpoint between the (now-centered) MPC2/CBF2 column and the MPCN/CBFN
% column -- used both for the elided-agent ellipses and the frequency labels
\coordinate (freqx) at ($(mpc2)!0.5!(mpcN)$);
% Ellipses marking the elided agents 3..N-1, sitting between column 2 and
% column N now that column 2 has moved to the figure's center
\node[ellipsis] at (freqx |- mpc1)  {$\cdots$};
\node[ellipsis] at (freqx |- cbf1)  {$\cdots$};
\node[ellipsis] at (freqx |- r1)    {$\cdots$};
% Dotted "Environment" box enclosing the whole robot row, label below it
\node[envbox, fit=(r1)(r2)(rN), label={[font=\sffamily\Large\bfseries, text=green!40!black]below:Environment}] (env) {};
% Sensing line's exit point: the box's west edge, low -- near the
% bottom-left corner, below Robot1's own height, so its vertical run sits
% clearly outside (further left of) column 1's feedback trunk rather than
% running alongside it
\coordinate (envLeftExit) at ($(env.south west)+(0,0.2)$);
\coordinate (envLeftOut) at ($(envLeftExit)+(-0.5,0)$);
% Entry point on RoE's west edge, offset below center -- the RoE->MPC1
% forward edge also exits via roe.west (at dead center), so entering there
% too would run the sensing line along that exact same horizontal strip
\coordinate (roeWestLow) at ($(roe.west)+(0,-0.4)$);
% Outward stubs for each robot's own feedback trunk -- computed straight
% off the robot node itself, so the line visibly emerges from the robot
% (and, for the outer columns, clears the green box's border along the way)
\coordinate (fb1x) at ($(r1.west)+(-0.6,0)$);
\coordinate (fb2x) at ($(r2.west)+(-0.6,0)$);
\coordinate (fbNx) at ($(rN.east)+(0.6,0)$);
% 2. Forward edges -----------------------------------------------------
\draw[mainline] (roe.west) -| (mpc1.north);
\draw[mainline] (roe.south) -- (mpc2.north);
\draw[mainline] (roe.east) -| (mpcN.north);
\node[freqlabel] at ($(mpc1.north -| freqx)+(0,1.3cm)$) {Low Frequency};
\draw[mainline] (mpc1.south) -- (cbf1.north);
\draw[mainline] (mpc2.south) -- (cbf2.north);
\draw[mainline] (mpcN.south) -- (cbfN.north);
\node[freqlabel] at ($(mpc1.south -| freqx)!0.5!(cbf1.north -| freqx)+(0,0.4cm)$) {Medium Frequency};
\draw[mainline] (cbf1.south) -- (r1.north);
\draw[mainline] (cbf2.south) -- (r2.north);
\draw[mainline] (cbfN.south) -- (rN.north);
\node[freqlabel] at ($(cbf1.south -| freqx)!0.5!(r1.north -| freqx)+(0,0.6cm)$) {High Frequency};
% 3. Environment edge: exits the box's west edge below Robot1's own
%    feedback stub, clear of the column-1 trunk, then enters RoE
%    horizontally from its west (left) side, below center so it doesn't
%    run along the same strip as the RoE->MPC1 forward edge
\draw[envline] (envLeftExit) -- (envLeftOut) |- (roeWestLow);
% 4. Feedback edges: each line emerges directly from its own robot (not
%    from the environment box's border), steps a visible amount outward,
%    then runs up its own column and branches inward twice -- once into
%    CBF, once continuing up into MPC.
\draw[feedbackline] (r1.west) -- (fb1x) |- (cbf1.west);
\draw[feedbackline] (fb1x |- cbf1.west) -- (fb1x |- mpc1.west) -- (mpc1.west);
\draw[feedbackline] (r2.west) -- (fb2x) |- (cbf2.west);
\draw[feedbackline] (fb2x |- cbf2.west) -- (fb2x |- mpc2.west) -- (mpc2.west);
\draw[feedbackline] (rN.east) -- (fbNx) |- (cbfN.east);
\draw[feedbackline] (fbNx |- cbfN.east) -- (fbNx |- mpcN.east) -- (mpcN.east);
\end{tikzpicture}
    }
    \caption{Engagement-Aware Agentic Pursuit-Evasion }
    \label{fig:roe}
\end{figure*}
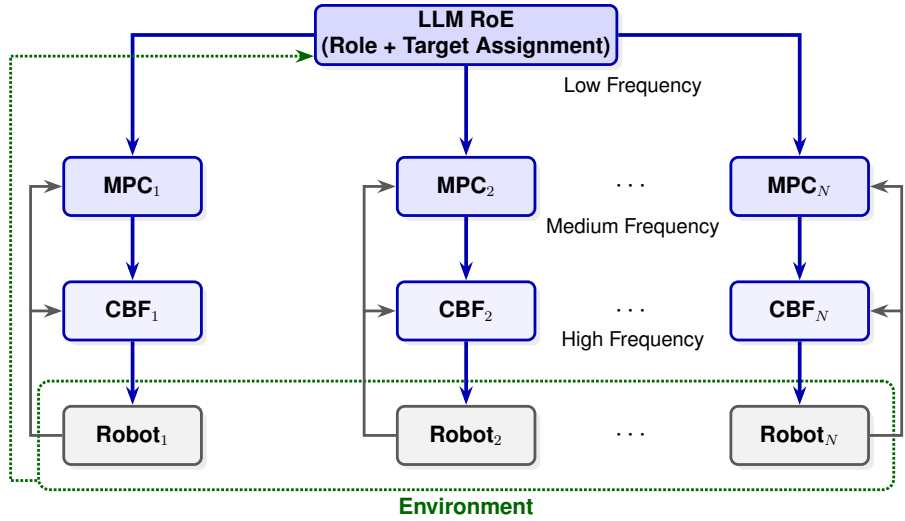
As team size grows, solving multi-agent pursuit-evasion as a single joint optimization becomes increasingly expensive. We therefore adopt a hierarchical architecture~\cite{matni2024towards} that separates strategic coordination from low-level control. Each robot independently solves a role-conditioned receding-horizon control problem and safety filter, avoiding a joint team-wide control solve. A slower strategic layer determines what objective each controller should pursue -- e.g., whether an agent should defend, engage a particular opponent, or reposition. In this first study, this strategic layer is centralized within each team: a single LLM planner assigns roles and targets to all active agents, while control execution remains per-agent.
Within this hierarchy, the two teams retain fixed opposing missions -- one defends a region and the other attempts to breach it -- while two independent LLM planners (one per team) adapt their agents' tactical roles to the evolving tactical picture.

 The two teams share the same underlying control architecture, but their tactical vocabularies are different: the defending team chooses among four roles (DEFEND, NEUTRALIZE, CAPTURE, RECON), reflecting its richer set of ways to intercept or contain a threat, while the attacking team chooses between only two (ATTACK, EVADE), reflecting its narrower, single-minded objective of breaching the region while avoiding capture. Within each team's mission, an agent's tactical assignment (e.g. hold position, pursue a specific threat, or scout) is treated as a control-design object rather than the direct output of a game-theoretic solve. This assignment is issued once per planning cycle for both teams, by an \ac{llm} planner explicitly over the tactical picture -- the \ac{llm}'s assignment accompanied by a natural-language justification for every agent. At the execution layer, each agent then independently solves its own receding-horizon  control problem conditioned on the assignment.

Beyond the choice of game formulation, the pursuer team's desired behavior is itself mission-dependent. Capturing and containing an evader requires different coordination than preventing it from reaching a protected region. Counter-UAV defense provides a motivating example: capture may be preferred for forensic analysis and attribution, whereas an imminent threat may require immediate neutralization.  We formalize this flexibility through mission-dependent \ac{roe}, which determine each pursuer’s tactical posture (DEFEND, NEUTRALIZE, CAPTURE, RECON). The strategic planner can therefore adapt how team resources are allocated as the tactical situation evolves, without changing the underlying control architecture.

In short, to enable different engagement modes without redesigning the underlying control logic, we propose an engagement-aware hierarchical control architecture that decouples strategic role assignment from low-level safety and engagement enforcement (see Fig.~\ref{fig:roe}): the LLM provides the strategic role assignment, each agent’s \ac{mpc} translates that role into a nominal control action, and a \ac{cbf}-based filter enforces safety and role-dependent engagement constraints
Critically, the assigned role parameterizes both the \ac{mpc} cost and the \ac{cbf} constraint set. For example, assigning NEUTRALIZE removes the pursuer --evader standoff constraint, permitting direct contact. Thus, switching from stand-off capture to direct neutralization requires only a change in the assigned role, which reparameterizes the MPC objective and active CBF constraints while preserving the same underlying control architecture.
Our contributions are as follows:
\begin{itemize}
\item An engagement-aware pursuit-evasion framework that treats rules of engagement as online tactical decision variables rather than fixed a priori;
\item An LLM-based strategic planning layer that assigns interpretable, adaptive tactical roles to each agent under field-of-view-limited observations, with an explicit natural-language justification;
\item A hierarchical execution architecture in which per-agent \ac{mpc} controllers and \ac{cbf}-based safety filters realize LLM-assigned roles without redesigning the underlying control architecture;
\item  A simulation-based benchmark for zero-shot LLM strategic planning in partially observed, adversarial multi-agent settings, including state-based tactical reasoning and image-based contact classification.
\end{itemize}

%------------------------------------------
\section{Related Work}
Large language models have increasingly been adopted as high-level planners for embodied agents, translating natural-language task descriptions and environment state into executable action sequences without task-specific training. Early work grounded a language model's proposed actions in a robot's actual affordances \cite{ahn2022can}, used a language model directly as a zero-shot sequential planner over admissible actions \cite{huang2022language}, and generated executable robot code or structured task programs from natural-language instructions \cite{liang2023code,singh2022progprompt}. More recent work extends this paradigm to multi-robot settings, where a language model coordinates several embodied agents through natural-language dialogue \cite{mandi2024roco} or is compared directly against a decentralized per-robot instantiation to characterize the centralization-communication tradeoff in LLM-driven multi-robot coordination \cite{chen2024scalable}. 

%{\color{blue}We extend this line of work to the adversarial, multi-agent pursuit-evasion setting: rather than a language model coordinating cooperative task allocation among embodied agents, it must reason about role assignment and target selection for a team under active opposition from an evading adversary.}

% This body of work, however, is almost exclusively cooperative: the language model plans on behalf of a single team pursuing a shared objective, with no adversary reasoning about or reacting to that team's plan. We instead study LLM-based planning in an adversarial, partially observed two-team setting, where the strategic layer must reason about an opponent's likely response  rather than only about task decomposition or inter-agent coordination.

Multi-agent pursuit-evasion has a broad literature spanning differential games, optimization-based control, reinforcement learning, and distributed multi-robot coordination. Prior work has studied heterogeneous pursuer--evader teams under uncertainty \cite{zhang2021pursuer}, cooperative multi-UAV target defense \cite{tong2023game}, reinforcement learning for decentralized pursuit \cite{zhang2022multi}, distributional soft actor-critic methods for underwater vehicle pursuit \cite{hou2023distributional}, limited-information model predictive control for pursuit-evasion games \cite{sani2021limited}, evader-agnostic pursuit in partially observable environments \cite{kalanther2025evader}, and online deep reinforcement learning for multi-UAV pursuit in unknown 3D environments \cite{chen2025online}. Distributed optimization frameworks provide another important direction, enabling multi-robot systems to make real-time local decisions that collectively achieve global objectives such as area coverage and target assignment \cite{jaleel2020distributed}.

Several works explicitly address prediction, uncertainty, and geometric guarantees. Shivam et al. \cite{shivam2019predictive} propose a predictive tracking framework based on Newton--Raphson flow, combined with an online-trained neural network to estimate evader strategies. GRAPE \cite{shah2019grape} models noisy evader localization using uncertainty ellipsoids and minimizes the evader's safe-reachable set, while Wang et al. \cite{wang2025distributed} formulate encirclement under unknown evader motion using distributed robust MPC.

Model-based and model-free distributed optimization methods provide another foundation for cooperative pursuit. Zhou et al. \cite{zhou2023distributed} formulate pursuit as an infinite-horizon differential game over graph-based formation and tracking errors, yielding Nash-equilibrium policies through Hamilton--Jacobi equations but requiring full observability and known dynamics. Dong et al. \cite{dong2024adaptive} reduce this dependence on explicit dynamics using a Q-learning and actor-critic formulation for linear-quadratic systems, though such approaches typically require sufficient exploration and introduce additional training complexity.
Safety-critical pursuit has also been studied through visibility and barrier-function constraints. Zhou et al. \cite{zhou2025control} maintain a moving evader within a pursuer's field of view under occlusions by combining a \ac{cbf} constraint with sampling-based planning. 

This work separates strategic planning from low-level control:
two independent LLM planners assign team-specific roles and targets
under partial observability, while each robot independently executes
its assignment through role-conditioned MPC and CBF-based safety
filtering. We evaluate this hierarchy using both state-based tactical
reasoning and image-based contact classification.

%------------------------------------------
\section{Approach}
\noindent Our architecture decomposes the multi-agent interaction into three hierarchical layers: a strategic planner, realized as an \ac{llm} agent, that assigns each agent a rule of engagement; a decentralized receding-horizon \ac{mpc} trajectory generator that independently realizes every agent's currently assigned role; and a discrete-time \ac{cbf}-QP safety and engagement filter that enforces inter-agent and adversarial constraints on the resulting commands.

% Let the dynamics of each agent be described by a discrete-time kinematic model. The state of agent $i$ is defined as $x_{i,k} \in \mathbb{R}^{n_x}$, comprising position $p_{i,k}$ and velocity $v_{i,k}$ vectors, and the control input is the acceleration $u_{i,k} \in \mathbb{R}^{n_u}$. The discrete-time update over time step $dt$ is given by:
% \begin{equation}
% x_{i,k+1} = f(x_{i,k}, u_{i,k})
% \end{equation}}
% {\color{red}vry confusing, to still have these dyanmics. As mentioned in appendix CBF and MPC intro can also be removed.}

\subsection{Rules of Engagement}
We consider pursuit-evasion scenarios where the pursuer team operates under mission-dependent \ac{roe} rather than pursuing a static objective.
While conventional formulations treat capture as the sole terminal goal, direct capture can be infeasible or disadvantageous when the evader has superior mobility or operates near sensitive regions where aggressive interception introduces collateral risk. In these settings, the pursuer team must dynamically transition across alternative engagement modes, including perimeter defense and neutralization.

Counter-UAV missions provide concrete motivation for this operational flexibility. When safe recovery is viable, capture is often preferred because it facilitates forensic analysis, attribution, and technical exploitation of the target's avionics, communication links, navigation payload, and hardware. Analysis of systems such as Shahed/Geran drones and the development of platforms like LUCAS underscore the intelligence and operational value of recovering adversarial UAVs~\cite{csis_shahed}. Conversely, when recovery is impractical or the target presents an imminent threat, the operational priority shifts to neutralization. This approach is exemplified in Ukraine by counter-UAV platforms such as the P1-SUN, Sting, and ZIRKA, which are designed to intercept hostile UAVs.

We implement \ac{roe} adaptation through discrete, per-agent roles assigned at each planning cycle by a high-level strategic planner -- realized in this work via an \ac{llm}. Each role (DEFEND, NEUTRALIZE, CAPTURE, RECON) jointly dictates the receding-horizon \ac{mpc} objective and, where relevant, the low-level \ac{cbf} safety constraints, realizing these operational modes end-to-end without modifying the underlying control architecture.
Under the \textit{defend} role, agents maintain designated stations along the protected perimeter, enforcing the deterrence posture. The \textit{recon} role instead diverts an agent to actively patrol toward a coverage region and sweep its sensor heading independently of its direction of travel, at no change to the \ac{cbf} constraints -- it trades perimeter presence or engagement capacity for expanded situational awareness. Both \textit{neutralize} and \textit{capture} instead drive agents toward the target using an identical interception \ac{mpc} cost, differing solely in their \ac{cbf} safety filtering. An agent assigned to \textit{neutralize} is exempted from the pursuer-evader separation constraint $D_{\text{safe, pe}}$, permitting direct kinetic contact that immediately terminates the target at the cost of permanently expending the intercepting agent. In \textit{capture} mode, $D_{\text{safe, pe}}$ remains strictly enforced; the target is captured via multi-agent containment, requiring at least $\kappa$ capturing agents to simultaneously enter the target's capture radius (we set $\kappa=2$ in our experiments [cf.~\cite{bopardikar2014k}])
%\footnote{We set $\kappa=2$ in our experiments, generalizing the single-pursuer formulation as in \cite{bopardikar2014k}.}
-- a coordinated maneuver that preserves all pursuer assets at the expense of time and agent allocation. This deliberate asymmetry forces the high-level strategic planner to weigh rapid target elimination against asset preservation, evaluated dynamically via the scoring formulation in Appendix~\ref{app:system-prompts} based on asset availability and threat urgency.

The simulation environment used here is intentionally low-fidelity: agents move under simplified double-integrator translational dynamics in a 2D arena, sensing is a noiseless conic field-of-view with perfect detection, and no communication delay or environmental disturbance is modeled. We make no claim that this setup is representative of real-world multi-robot deployment. The point of this work is narrower: to show that off-the-shelf large language models, used entirely zero-shot -- with no fine-tuning, task-specific training data, or reinforcement learning -- can serve as effective strategic planners for multi-agent tactical decision-making, producing interpretable, justified role assignments under a partially observed, adversarial state. Demonstrating this capability in a simplified but genuinely multi-agent, partially-observable, adversarial setting is a necessary first step before tackling the added complexity of higher-fidelity dynamics, sensing, or real hardware, which we leave to future work.

The interface between the strategic and execution layers is deliberately narrow. The strategic layer communicates with the execution layer through a single tuple per agent per cycle, a role and an optional target id, from which each agent's MPC cost and CBF constraint set are instantiated, accompanied by a one-sentence natural-language justification. This orchestrator-worker pattern (a planner assigning discrete role/target tuples to independently-executing workers) is not specific to pursuit-evasion; it mirrors the broader agentic-AI pattern of a coordinator delegating typed subtasks to specialized executors.

\subsection{Agent Dynamics and State Space}
\label{sec:agent-dynamics}
 
Every agent, pursuer or evader, evolves under the following discrete-time control-affine dynamics, discretized at a fixed timestep $dt$ via semi-implicit Euler integration:
\begin{equation}
x_{k+1} = f(x_k) + g\, u_k, \label{eq:dynamics}
\end{equation}
where the state $x$ and control input $u$ are
\begin{equation}
x = \begin{bmatrix} p^\top & v^\top & \theta \end{bmatrix}^\top \in \mathbb{R}^5, \qquad
u = \begin{bmatrix} a^\top & \omega \end{bmatrix}^\top \in \mathbb{R}^3,
\end{equation}
with $p, v \in \mathbb{R}^2$ its planar position and velocity, $\theta \in \mathbb{R}$ its heading, $a \in \mathbb{R}^2$ a commanded translational acceleration, and $\omega \in \mathbb{R}$ an independent commanded yaw rate. The drift and input matrices are
\begin{equation}
f(x_k) = \begin{bmatrix} p_k + dt\, v_k \\ v_k \\ \theta_k \end{bmatrix}, \qquad
g = \begin{bmatrix} dt^2 I_2 & \mathbf{0}_{2\times1} \\ dt\, I_2 & \mathbf{0}_{2\times1} \\ \mathbf{0}_{1\times2} & dt \end{bmatrix} \in \mathbb{R}^{5\times3},
\end{equation}
where $I_2$ is the $2\times2$ identity matrix. Heading evolves independently of position and velocity, such as a sensor turret mounted on a vehicle (e.g., rotating 360 camera). All controls and velocities are box-constrained per axis  -- $|a_x|,|a_y| \le a_{\max}$, $|v_x|,|v_y| \le v_{\max}$, and $|\omega| \le \omega_{\max}$.  
$p_{i,k}, v_{i,k} \in \mathbb{R}^2$ denote agent $i$'s current position and velocity at time step $k$. Moreover, $p_{ij} = p_i - p_j$, $v_{ij} = v_i - v_j$ denote the  relative position and velocity between agent $i$ and $j$.  

\subsection{Decentralized Receding-Horizon MPC}

We define two disjoint families of  receding horizon control problems for each team: the pursuer (defending) team's four roles -- \textsc{Defend}, \textsc{Recon}, \textsc{Neutralize}, \textsc{Capture} -- and the evader (attacking) team's two roles -- \textsc{Attack}, \textsc{Evade}. All six share the same receding-horizon template below, differing only in the role-specific cost term $J_{\text{role}}$; we define that template first, then $J_{\text{role}}$ for each pursuer role, then for each evader role.

Every pursuer $i$ (assigned role \textsc{Defend}, \textsc{Recon}, \textsc{Neutralize}, or \textsc{Capture}) and every evader $j$ (assigned role \textsc{Attack} or \textsc{Evade}) solves, from its current state $x_0$,
\begin{equation}
U^* = \arg\min_{U} \; w_u \sum_{k=0}^{N-1} \|u_k\|^2 + J_{\text{role}}(X)
\end{equation}
subject to its own dynamics \ref{eq:dynamics}, $x_0$ fixed to its current measured state, and actuation/velocity bounds $\|u_k\|_\infty \le a_{\max}$, $\|v_k\|_\infty \le v_{\max}$, enforced independently per axis, where $w_{\text{u}} > 0$ is a fixed weight penalizing control effort. $U=(u_0,\ldots,u_{N-1})$, $X=(x_1,\ldots,x_N)$ are the resulting control and state sequences over the horizon. 
%This $w_u$ is unrelated to the role-specific weights (e.g. $w_{\text{hold}}$, $w_{\text{sweep}}$) introduced within each $J_{\text{role}}$ below.
Only the role-specific term $J_{\text{role}}$ changes across roles. For roles whose cost is defined relative to the opposing team (\textsc{Defend}, \textsc{Neutralize}, \textsc{Capture}, \textsc{Evade}), this also determines which specific opposing agent it is solved against; \textsc{Recon} and \textsc{Attack} instead target a patrol waypoint, with no dependence on the opposing team's state at all.

\paragraph{Defend}
A pursuer assigned \textsc{Defend} holds a guard station on the perimeter of the protected region rather than closing on the evader. At each planning step it targets the closest point $p_{\text{bound}}$ on the (polygon-approximated) defended boundary to the nearest evader's \emph{current} position, and minimizes
\begin{equation}
J_{\text{defend}} = \sum_{k=1}^{N} w_{\text{defend}} \|p_{i,k} - p_{\text{bound}}\|^2 .
\end{equation}

\paragraph{Recon}
A pursuer assigned \textsc{Recon} is diverted from perimeter duty or engagement entirely to actively scout the arena, with no dependence on the evader's state. Its patrol target $p_{\text{target}}$ is a coverage-region center supplied by the strategic layer. Coverage regions are defined by discretizing the arena into a coarse fixed grid; each cell's staleness is the number of planning cycles elapsed since any pursuer's field of view last covered it, and the strategic layer assigns the region with the highest staleness (never-covered cells ranked above any finite value) as the target. Once reached, a directed target is held, while a random one is redrawn. The pursuer minimizes a hold-position term identical in form to \textsc{Defend}'s,

\begin{equation}
J_{\text{hold}} = \sum_{k=1}^{N} w_{\text{hold}} \|p_{i,k} - p_{\text{target}}\|^2,
\end{equation}
together with an independent bearing-sweep term that rotates its sensor heading $\theta_{i,k}$ sinusoidally about the direction of travel $\theta_c = \angle(p_{\text{target}} - p_{i,0})$, over a $2\pi/3$ arc and a fixed sweep period $T_s$:
\begin{equation}
J_{\text{sweep}} = \sum_{k=1}^{N} w_{\text{sweep}} \left(\theta_{i,k} - \theta_c - \tfrac{\pi}{3}\sin\!\big(\tfrac{2\pi (t_0 + k\,dt)}{T_s}\big)\right)^2,
\end{equation}
so that $J_{\text{recon}} = J_{\text{hold}} + J_{\text{sweep}}$. Translation and heading are governed by independent controls (Section~\ref{sec:agent-dynamics}), so the sweep is a genuine rotation of the sensor's field of view while the pursuer is in transit, not merely a byproduct of its direction of motion.

\paragraph{Neutralize and Capture}
\textsc{Neutralize} and \textsc{Capture} pursuers run the same intercept controller against their nearest evader; the roles differ only downstream, in whether the \ac{cbf} enforces or exempts the pursuer-evader standoff constraint for that agent. Since the evader is not solved jointly, its future trajectory is predicted open-loop under a constant-velocity assumption, $\bar p_{e,k} = p_{e,0} + k\, dt\, v_{e,0}$, and the pursuer minimizes
\begin{equation}
J_{\text{intercept}} = \sum_{k=1}^{N} w_e \|p_{i,k} - \bar p_{e,k}\|^2,
\end{equation}
tracking this predicted position directly rather than also matching a velocity-cutoff term against it; recomputing $\bar p_{e,k}$ from the evader's current state at every planning step still lets the controller correct for prediction drift each cycle, without committing to a closing velocity derived from the same stale linear extrapolation. An evader is captured once at least $\kappa$ \textsc{Capture}-role pursuers are simultaneously within the capture radius $r_c$ of it, or immediately upon direct contact with a single \textsc{Neutralize}-role pursuer.

\paragraph{Attack and Evade}
An evader assigned \textsc{Attack} advances directly on the defended region's center $p_{\text{def}}$, independent of the pursuers' plans:
\begin{equation}
J_{\text{attack}} = \sum_{k=1}^{N} w_{\text{prog}} \|p_{e,k} - p_{\text{def}}\|^2 .
\end{equation}
An evader assigned \textsc{Evade} instead reacts to the pursuers' just-computed nominal plans $\{\bar X_{p,i}\}_{i=1}^{N_p}$, maximizing its predicted separation from every pursuer's planned trajectory over the horizon:
\begin{equation}
J_{\text{evade}} = - w_{\text{avoid}} \sum_{k=1}^{N} \sum_{i=1}^{N_p} \|p_{e,k} - \bar p_{i,k}\|^2 .
\end{equation}
Because the avoidance term is a concave quadratic, this problem is solved with a general nonlinear solver, unlike the convex QPs used for the other three roles.

\subsection{Discrete-Time CBF Safety Filter}

%First history of  cbf, then cbf formulation and then solve as QP
Each agent's role-conditioned \ac{mpc} acts as a mid-level, task-oriented planner, while \acp{cbf}~\cite{8796030,9658123} serve as a low-level, one-step safety filter applied downstream of the MPC, independently of which role generated the nominal command. Under mild conditions, a CBF-based filter is necessary and sufficient to render a safe set forward-invariant, guaranteeing constraint satisfaction regardless of the nominal controller that generated the desired input \cite{8796030}; following this established practice, our discrete-time CBF filter minimally invasively corrects each agent's nominal MPC command to preserve safety.
To ensure collision avoidance between two agents $i$ and $j$, we define a velocity-aware kinematic barrier function $h(x_k)$ \cite{cheng2020safe,11443822,Borrmann2015},
\begin{equation}
h(x_k) = \frac{p_{ij}^T v_{ij}}{\|p_{ij}\|} + \sqrt{\alpha_{\max} (\|p_{ij}\| - D_{\text{safe}})} \ge 0
\end{equation}
 where $p_{ij}$  and $v_{ij}$ are defined in Section~\ref{sec:agent-dynamics}. 
Here $D_{\text{safe}}$ is the minimum separation required between agents $i$ and $j$ -- the pair-specific values used in this paper ($D_{\text{safe,p}}$ for pursuer-pursuer pairs, $D_{\text{safe,pe}}$ for pursuer-evader pairs, $D_{\text{safe,e}}$ for the evader's own filter).  Under appropriate assumptions on $\alpha_\text{max}$ stated in Lemma 2 of \citet{cheng2020safe}, the CBF defined above renders the safe set invariant.

Given the desired optimal control input $u_{\text{des}}$ generated by the agent's \ac{mpc} controller, the true control input applied to the system is the solution to a constrained Quadratic Program (QP) with decision variables $u$ and slack variables $\epsilon$:
\begin{equation}
u^*, \epsilon^* = \arg\min_{u, \epsilon} \frac{1}{2} \|u - u_{\text{des}}\|^2 + \lambda \epsilon^2
\end{equation}
Because $\epsilon$ is a free slack variable rather than hard-constrained to zero, the CBF condition above is enforced as a penalized soft constraint.
We implement these constraints for \textit{pursuer-pursuer} collision avoidance ($D_{\text{safe, p}}$) and \textit{pursuer-evader} standoff ($D_{\text{safe, pe}}$); the latter is omitted entirely for a pursuer assigned \textsc{Neutralize}, permitting the direct contact that role requires. A mirrored constraint ($D_{\text{safe, e}}$) is enforced on the evader's own control update.

\section{Experiments}
\label{sec:experiments}

To evaluate the performance of the proposed architecture, we conducted simulations within a 2D planar double integrator dynamics framework. Each agent's receding-horizon optimization problem is modeled using CasADi and solved independently: OSQP for the convex quadratic formulations used by the \textsc{Defend}, \textsc{Recon}, \textsc{Neutralize}/\textsc{Capture}, and \textsc{Attack} controllers, and IPOPT for the nonconvex formulation used by \textsc{Evade}. The lower level discrete \ac{cbf} safety filter bypasses this CasADi modeling layer entirely, assembling its sparse QP directly and solving it via OSQP's native interface, to maintain high control frequencies.

All experiments feature a team of four pursuers under decentralized control and central planning, defending against an evader presence whose scale we vary across trials -- from a single evader up to four in the state-only setting, or from one to four contact slots in the multi-modal setting (Section~\ref{sec:exp-multimodal}), where not every slot is guaranteed to resolve to a genuine evader. 

Two layers of fault tolerance keep a single unreliable model from invalidating a run: within a cycle, a role assignment that violates the structured-output schema is retried, and on repeated failure the affected team falls back to a fixed safe default (\textsc{Defend} for blue, \textsc{Attack} for red) for that cycle only, logged as a fallback event; at the episode level, an unhandled LLM API failure is recorded as that episode's outcome.

Within this dynamic framework, we evaluate the strategic planner under two observational modalities. In the state-information-only modality, the LLM planning agent observes only agent positions, velocities, and status. In the multi-modal modality, the agent additionally receives a static reference image of each detected contact -- a deliberate abstraction of perception rather than a physically-grounded rendering of the simulated scene, since the image carries no dependence on a contact's actual range, viewing angle, or motion, nor any sensor noise; Section~\ref{sec:exp-multimodal} discusses this abstraction and its implications for interpreting the resulting discrimination performance in full.
% a second modality additionally supplying image data of the attacking and defending drones is planned as a future extension and is not yet implemented.

\subsection{State-Only Strategic Planning}

\label{sec:exp-state-only}

\begin{figure}[t]
    \centering
    \begin{subfigure}[b]{0.48\linewidth}
        \centering
        \resizebox{\linewidth}{!}{% This file was created with tikzplotlib v0.10.1.
\begin{tikzpicture}

\definecolor{chocolate213940}{RGB}{213,94,0}
\definecolor{cornflowerblue86180233}{RGB}{86,180,233}
\definecolor{darkcyan0114178}{RGB}{0,114,178}
\definecolor{darkcyan0158115}{RGB}{0,158,115}
\definecolor{darkgrey176}{RGB}{176,176,176}
\definecolor{orange2301590}{RGB}{230,159,0}
\definecolor{palevioletred204121167}{RGB}{204,121,167}

\begin{axis}[
height=6cm,
tick align=outside,
tick pos=left,
title={Neutralizations (3s planning interval)},
width=9cm,
x grid style={darkgrey176},
xlabel={Number of evaders},
xmin=0.41, xmax=4.59,
xtick style={color=black},
xtick={1,2,3,4},
y grid style={darkgrey176},
ylabel={Mean neutralizations per game},
ymin=0, ymax=1.47,
ytick style={color=black}
]
\draw[draw=none,fill=chocolate213940] (axis cs:0.6,0) rectangle (axis cs:0.733333333333333,0.1);
\draw[draw=none,fill=chocolate213940] (axis cs:1.6,0) rectangle (axis cs:1.73333333333333,0);
\draw[draw=none,fill=chocolate213940] (axis cs:2.6,0) rectangle (axis cs:2.73333333333333,0.1);
\draw[draw=none,fill=chocolate213940] (axis cs:3.6,0) rectangle (axis cs:3.73333333333333,0.2);
\draw[draw=none,fill=orange2301590] (axis cs:0.733333333333333,0) rectangle (axis cs:0.866666666666667,0.2);
\draw[draw=none,fill=orange2301590] (axis cs:1.73333333333333,0) rectangle (axis cs:1.86666666666667,0.4);
\draw[draw=none,fill=orange2301590] (axis cs:2.73333333333333,0) rectangle (axis cs:2.86666666666667,0.8);
\draw[draw=none,fill=orange2301590] (axis cs:3.73333333333333,0) rectangle (axis cs:3.86666666666667,0.6);
\draw[draw=none,fill=cornflowerblue86180233] (axis cs:0.866666666666667,0) rectangle (axis cs:1,0.4);
\draw[draw=none,fill=cornflowerblue86180233] (axis cs:1.86666666666667,0) rectangle (axis cs:2,0.3);
\draw[draw=none,fill=cornflowerblue86180233] (axis cs:2.86666666666667,0) rectangle (axis cs:3,0.7);
\draw[draw=none,fill=cornflowerblue86180233] (axis cs:3.86666666666667,0) rectangle (axis cs:4,1);
\draw[draw=none,fill=palevioletred204121167] (axis cs:1,0) rectangle (axis cs:1.13333333333333,0);
\draw[draw=none,fill=palevioletred204121167] (axis cs:2,0) rectangle (axis cs:2.13333333333333,0.1);
\draw[draw=none,fill=palevioletred204121167] (axis cs:3,0) rectangle (axis cs:3.13333333333333,0.1);
\draw[draw=none,fill=palevioletred204121167] (axis cs:4,0) rectangle (axis cs:4.13333333333333,0);
\draw[draw=none,fill=darkcyan0114178] (axis cs:1.13333333333333,0) rectangle (axis cs:1.26666666666667,0.9);
\draw[draw=none,fill=darkcyan0114178] (axis cs:2.13333333333333,0) rectangle (axis cs:2.26666666666667,1.4);
\draw[draw=none,fill=darkcyan0114178] (axis cs:3.13333333333333,0) rectangle (axis cs:3.26666666666667,1.2);
\draw[draw=none,fill=darkcyan0114178] (axis cs:4.13333333333333,0) rectangle (axis cs:4.26666666666667,1);
\draw[draw=none,fill=darkcyan0158115] (axis cs:1.26666666666667,0) rectangle (axis cs:1.4,0);
\draw[draw=none,fill=darkcyan0158115] (axis cs:2.26666666666667,0) rectangle (axis cs:2.4,0.1);
\draw[draw=none,fill=darkcyan0158115] (axis cs:3.26666666666667,0) rectangle (axis cs:3.4,0);
\draw[draw=none,fill=darkcyan0158115] (axis cs:4.26666666666667,0) rectangle (axis cs:4.4,0);
\end{axis}

\end{tikzpicture}}
        \caption{Neutralizations, 3s}
        \label{fig:exp1-neutralizations-3s}
    \end{subfigure}
    \hfill
    \begin{subfigure}[b]{0.48\linewidth}
        \centering
        \resizebox{\linewidth}{!}{% This file was created with tikzplotlib v0.10.1.
\begin{tikzpicture}

\definecolor{chocolate213940}{RGB}{213,94,0}
\definecolor{cornflowerblue86180233}{RGB}{86,180,233}
\definecolor{darkcyan0114178}{RGB}{0,114,178}
\definecolor{darkcyan0158115}{RGB}{0,158,115}
\definecolor{darkgrey176}{RGB}{176,176,176}
\definecolor{orange2301590}{RGB}{230,159,0}
\definecolor{palevioletred204121167}{RGB}{204,121,167}

\begin{axis}[
height=6cm,
tick align=outside,
tick pos=left,
title={Captures (3s planning interval)},
width=9cm,
x grid style={darkgrey176},
xlabel={Number of evaders},
xmin=0.41, xmax=4.59,
xtick style={color=black},
xtick={1,2,3,4},
y grid style={darkgrey176},
ylabel={Mean captures per game},
ymin=0, ymax=1.365,
ytick style={color=black}
]
\draw[draw=none,fill=chocolate213940] (axis cs:0.6,0) rectangle (axis cs:0.733333333333333,0.6);
\draw[draw=none,fill=chocolate213940] (axis cs:1.6,0) rectangle (axis cs:1.73333333333333,1);
\draw[draw=none,fill=chocolate213940] (axis cs:2.6,0) rectangle (axis cs:2.73333333333333,1.2);
\draw[draw=none,fill=chocolate213940] (axis cs:3.6,0) rectangle (axis cs:3.73333333333333,1.3);
\draw[draw=none,fill=orange2301590] (axis cs:0.733333333333333,0) rectangle (axis cs:0.866666666666667,0.7);
\draw[draw=none,fill=orange2301590] (axis cs:1.73333333333333,0) rectangle (axis cs:1.86666666666667,0.2);
\draw[draw=none,fill=orange2301590] (axis cs:2.73333333333333,0) rectangle (axis cs:2.86666666666667,0.6);
\draw[draw=none,fill=orange2301590] (axis cs:3.73333333333333,0) rectangle (axis cs:3.86666666666667,1.1);
\draw[draw=none,fill=cornflowerblue86180233] (axis cs:0.866666666666667,0) rectangle (axis cs:1,0.1);
\draw[draw=none,fill=cornflowerblue86180233] (axis cs:1.86666666666667,0) rectangle (axis cs:2,0.3);
\draw[draw=none,fill=cornflowerblue86180233] (axis cs:2.86666666666667,0) rectangle (axis cs:3,0.4);
\draw[draw=none,fill=cornflowerblue86180233] (axis cs:3.86666666666667,0) rectangle (axis cs:4,0.6);
\draw[draw=none,fill=palevioletred204121167] (axis cs:1,0) rectangle (axis cs:1.13333333333333,0.1);
\draw[draw=none,fill=palevioletred204121167] (axis cs:2,0) rectangle (axis cs:2.13333333333333,0);
\draw[draw=none,fill=palevioletred204121167] (axis cs:3,0) rectangle (axis cs:3.13333333333333,0.1);
\draw[draw=none,fill=palevioletred204121167] (axis cs:4,0) rectangle (axis cs:4.13333333333333,0.2);
\draw[draw=none,fill=darkcyan0114178] (axis cs:1.13333333333333,0) rectangle (axis cs:1.26666666666667,0);
\draw[draw=none,fill=darkcyan0114178] (axis cs:2.13333333333333,0) rectangle (axis cs:2.26666666666667,0.3);
\draw[draw=none,fill=darkcyan0114178] (axis cs:3.13333333333333,0) rectangle (axis cs:3.26666666666667,0.4);
\draw[draw=none,fill=darkcyan0114178] (axis cs:4.13333333333333,0) rectangle (axis cs:4.26666666666667,0.9);
\draw[draw=none,fill=darkcyan0158115] (axis cs:1.26666666666667,0) rectangle (axis cs:1.4,1);
\draw[draw=none,fill=darkcyan0158115] (axis cs:2.26666666666667,0) rectangle (axis cs:2.4,1.1);
\draw[draw=none,fill=darkcyan0158115] (axis cs:3.26666666666667,0) rectangle (axis cs:3.4,0.9);
\draw[draw=none,fill=darkcyan0158115] (axis cs:4.26666666666667,0) rectangle (axis cs:4.4,1);
\end{axis}

\end{tikzpicture}}
        \caption{Captures, 3s}
        \label{fig:exp1-captures-3s}
    \end{subfigure}

    \vspace{0.6em}

    \begin{subfigure}[b]{0.48\linewidth}
        \centering
        \resizebox{\linewidth}{!}{% This file was created with tikzplotlib v0.10.1.
\begin{tikzpicture}

\definecolor{chocolate213940}{RGB}{213,94,0}
\definecolor{cornflowerblue86180233}{RGB}{86,180,233}
\definecolor{darkcyan0114178}{RGB}{0,114,178}
\definecolor{darkcyan0158115}{RGB}{0,158,115}
\definecolor{darkgrey176}{RGB}{176,176,176}
\definecolor{orange2301590}{RGB}{230,159,0}
\definecolor{palevioletred204121167}{RGB}{204,121,167}

\begin{axis}[
height=6cm,
tick align=outside,
tick pos=left,
title={Neutralizations (5s planning interval)},
width=9cm,
x grid style={darkgrey176},
xlabel={Number of evaders},
xmin=0.41, xmax=4.59,
xtick style={color=black},
xtick={1,2,3,4},
y grid style={darkgrey176},
ylabel={Mean neutralizations per game},
ymin=0, ymax=1.05,
ytick style={color=black}
]
\draw[draw=none,fill=chocolate213940] (axis cs:0.6,0) rectangle (axis cs:0.733333333333333,0);
\draw[draw=none,fill=chocolate213940] (axis cs:1.6,0) rectangle (axis cs:1.73333333333333,0);
\draw[draw=none,fill=chocolate213940] (axis cs:2.6,0) rectangle (axis cs:2.73333333333333,0.1);
\draw[draw=none,fill=chocolate213940] (axis cs:3.6,0) rectangle (axis cs:3.73333333333333,0.3);
\draw[draw=none,fill=orange2301590] (axis cs:0.733333333333333,0) rectangle (axis cs:0.866666666666667,0);
\draw[draw=none,fill=orange2301590] (axis cs:1.73333333333333,0) rectangle (axis cs:1.86666666666667,0.3);
\draw[draw=none,fill=orange2301590] (axis cs:2.73333333333333,0) rectangle (axis cs:2.86666666666667,0.4);
\draw[draw=none,fill=orange2301590] (axis cs:3.73333333333333,0) rectangle (axis cs:3.86666666666667,0.3);
\draw[draw=none,fill=cornflowerblue86180233] (axis cs:0.866666666666667,0) rectangle (axis cs:1,0.3);
\draw[draw=none,fill=cornflowerblue86180233] (axis cs:1.86666666666667,0) rectangle (axis cs:2,0.3);
\draw[draw=none,fill=cornflowerblue86180233] (axis cs:2.86666666666667,0) rectangle (axis cs:3,0.5);
\draw[draw=none,fill=cornflowerblue86180233] (axis cs:3.86666666666667,0) rectangle (axis cs:4,0.7);
\draw[draw=none,fill=palevioletred204121167] (axis cs:1,0) rectangle (axis cs:1.13333333333333,0);
\draw[draw=none,fill=palevioletred204121167] (axis cs:2,0) rectangle (axis cs:2.13333333333333,0.1);
\draw[draw=none,fill=palevioletred204121167] (axis cs:3,0) rectangle (axis cs:3.13333333333333,0);
\draw[draw=none,fill=palevioletred204121167] (axis cs:4,0) rectangle (axis cs:4.13333333333333,0.2);
\draw[draw=none,fill=darkcyan0114178] (axis cs:1.13333333333333,0) rectangle (axis cs:1.26666666666667,0.7);
\draw[draw=none,fill=darkcyan0114178] (axis cs:2.13333333333333,0) rectangle (axis cs:2.26666666666667,1);
\draw[draw=none,fill=darkcyan0114178] (axis cs:3.13333333333333,0) rectangle (axis cs:3.26666666666667,1);
\draw[draw=none,fill=darkcyan0114178] (axis cs:4.13333333333333,0) rectangle (axis cs:4.26666666666667,1);
\draw[draw=none,fill=darkcyan0158115] (axis cs:1.26666666666667,0) rectangle (axis cs:1.4,0);
\draw[draw=none,fill=darkcyan0158115] (axis cs:2.26666666666667,0) rectangle (axis cs:2.4,0);
\draw[draw=none,fill=darkcyan0158115] (axis cs:3.26666666666667,0) rectangle (axis cs:3.4,0);
\draw[draw=none,fill=darkcyan0158115] (axis cs:4.26666666666667,0) rectangle (axis cs:4.4,0.1);
\end{axis}

\end{tikzpicture}}
        \caption{Neutralizations, 5s}
        \label{fig:exp1-neutralizations-5s}
    \end{subfigure}
    \hfill
    \begin{subfigure}[b]{0.48\linewidth}
        \centering
        \resizebox{\linewidth}{!}{% This file was created with tikzplotlib v0.10.1.
\begin{tikzpicture}

\definecolor{chocolate213940}{RGB}{213,94,0}
\definecolor{cornflowerblue86180233}{RGB}{86,180,233}
\definecolor{darkcyan0114178}{RGB}{0,114,178}
\definecolor{darkcyan0158115}{RGB}{0,158,115}
\definecolor{darkgrey176}{RGB}{176,176,176}
\definecolor{orange2301590}{RGB}{230,159,0}
\definecolor{palevioletred204121167}{RGB}{204,121,167}

\begin{axis}[
height=6cm,
tick align=outside,
tick pos=left,
title={Captures (5s planning interval)},
width=9cm,
x grid style={darkgrey176},
xlabel={Number of evaders},
xmin=0.41, xmax=4.59,
xtick style={color=black},
xtick={1,2,3,4},
y grid style={darkgrey176},
ylabel={Mean captures per game},
ymin=0, ymax=1.785,
ytick style={color=black}
]
\draw[draw=none,fill=chocolate213940] (axis cs:0.6,0) rectangle (axis cs:0.733333333333333,0.8);
\draw[draw=none,fill=chocolate213940] (axis cs:1.6,0) rectangle (axis cs:1.73333333333333,1.2);
\draw[draw=none,fill=chocolate213940] (axis cs:2.6,0) rectangle (axis cs:2.73333333333333,1.4);
\draw[draw=none,fill=chocolate213940] (axis cs:3.6,0) rectangle (axis cs:3.73333333333333,1.4);
\draw[draw=none,fill=orange2301590] (axis cs:0.733333333333333,0) rectangle (axis cs:0.866666666666667,0.7);
\draw[draw=none,fill=orange2301590] (axis cs:1.73333333333333,0) rectangle (axis cs:1.86666666666667,1.1);
\draw[draw=none,fill=orange2301590] (axis cs:2.73333333333333,0) rectangle (axis cs:2.86666666666667,1);
\draw[draw=none,fill=orange2301590] (axis cs:3.73333333333333,0) rectangle (axis cs:3.86666666666667,1.7);
\draw[draw=none,fill=cornflowerblue86180233] (axis cs:0.866666666666667,0) rectangle (axis cs:1,0.3);
\draw[draw=none,fill=cornflowerblue86180233] (axis cs:1.86666666666667,0) rectangle (axis cs:2,0.3);
\draw[draw=none,fill=cornflowerblue86180233] (axis cs:2.86666666666667,0) rectangle (axis cs:3,0.6);
\draw[draw=none,fill=cornflowerblue86180233] (axis cs:3.86666666666667,0) rectangle (axis cs:4,0.6);
\draw[draw=none,fill=palevioletred204121167] (axis cs:1,0) rectangle (axis cs:1.13333333333333,0.2);
\draw[draw=none,fill=palevioletred204121167] (axis cs:2,0) rectangle (axis cs:2.13333333333333,0.3);
\draw[draw=none,fill=palevioletred204121167] (axis cs:3,0) rectangle (axis cs:3.13333333333333,0.7);
\draw[draw=none,fill=palevioletred204121167] (axis cs:4,0) rectangle (axis cs:4.13333333333333,0.4);
\draw[draw=none,fill=darkcyan0114178] (axis cs:1.13333333333333,0) rectangle (axis cs:1.26666666666667,0.2);
\draw[draw=none,fill=darkcyan0114178] (axis cs:2.13333333333333,0) rectangle (axis cs:2.26666666666667,0.5);
\draw[draw=none,fill=darkcyan0114178] (axis cs:3.13333333333333,0) rectangle (axis cs:3.26666666666667,0.8);
\draw[draw=none,fill=darkcyan0114178] (axis cs:4.13333333333333,0) rectangle (axis cs:4.26666666666667,1.4);
\draw[draw=none,fill=darkcyan0158115] (axis cs:1.26666666666667,0) rectangle (axis cs:1.4,0.9);
\draw[draw=none,fill=darkcyan0158115] (axis cs:2.26666666666667,0) rectangle (axis cs:2.4,1.3);
\draw[draw=none,fill=darkcyan0158115] (axis cs:3.26666666666667,0) rectangle (axis cs:3.4,1.2);
\draw[draw=none,fill=darkcyan0158115] (axis cs:4.26666666666667,0) rectangle (axis cs:4.4,1.2);
\end{axis}

\end{tikzpicture}}
        \caption{Captures, 5s}
        \label{fig:exp1-captures-5s}
    \end{subfigure}

    \resizebox{0.95\linewidth}{!}{% This file was created with tikzplotlib v0.10.1.
\begin{tikzpicture}

\definecolor{chocolate213940}{RGB}{213,94,0}
\definecolor{cornflowerblue86180233}{RGB}{86,180,233}
\definecolor{darkcyan0114178}{RGB}{0,114,178}
\definecolor{darkcyan0158115}{RGB}{0,158,115}
\definecolor{darkgrey176}{RGB}{176,176,176}
\definecolor{orange2301590}{RGB}{230,159,0}
\definecolor{palevioletred204121167}{RGB}{204,121,167}

\begin{axis}[
height=1.6cm,
hide x axis,
hide y axis,
legend cell align={left},
legend columns=3,
legend style={
  fill opacity=0.8,
  draw opacity=1,
  text opacity=1,
  at={(0.5,0.5)},
  anchor=center,
  draw=none
},
tick align=outside,
tick pos=left,
width=14cm,
x grid style={darkgrey176},
xmin=0, xmax=1,
xtick style={color=black},
y grid style={darkgrey176},
ymin=0, ymax=1,
ytick style={color=black}
]
\draw[draw=none,fill=chocolate213940] (axis cs:-1.4,0) rectangle (axis cs:-0.6,1);
\addlegendimage{ybar,ybar legend,draw=none,fill=chocolate213940}
\addlegendentry{Claude Haiku 4.5}

\draw[draw=none,fill=orange2301590] (axis cs:-1.4,0) rectangle (axis cs:-0.6,1);
\addlegendimage{ybar,ybar legend,draw=none,fill=orange2301590}
\addlegendentry{DeepSeek Chat}

\draw[draw=none,fill=cornflowerblue86180233] (axis cs:-1.4,0) rectangle (axis cs:-0.6,1);
\addlegendimage{ybar,ybar legend,draw=none,fill=cornflowerblue86180233}
\addlegendentry{GPT-5 Nano}

\draw[draw=none,fill=palevioletred204121167] (axis cs:-1.4,0) rectangle (axis cs:-0.6,1);
\addlegendimage{ybar,ybar legend,draw=none,fill=palevioletred204121167}
\addlegendentry{Gemini 2.5 Flash}

\draw[draw=none,fill=darkcyan0114178] (axis cs:-1.4,0) rectangle (axis cs:-0.6,1);
\addlegendimage{ybar,ybar legend,draw=none,fill=darkcyan0114178}
\addlegendentry{Llama 3.3 70B}

\draw[draw=none,fill=darkcyan0158115] (axis cs:-1.4,0) rectangle (axis cs:-0.6,1);
\addlegendimage{ybar,ybar legend,draw=none,fill=darkcyan0158115}
\addlegendentry{Rule-Based}

\end{axis}

\end{tikzpicture}}
    \caption{Experiment I: mean number of pursuer-initiated eliminations per game as a function of the number of evaders, by model, split by planning interval (top row: 3s; bottom row: 5s). \textsc{Neutralize} (\subref{fig:exp1-neutralizations-3s}, \subref{fig:exp1-neutralizations-5s}) resolves fast with a single agent at the cost of permanently disabling it, while \textsc{Capture} (\subref{fig:exp1-captures-3s}, \subref{fig:exp1-captures-5s}) requires two or more agents to converge but preserves the whole force; the two roles are directly comparable at every evader count since both share the same underlying scoring, and the two rows isolate how much of that behavior is sensitive to how often the RoE layer gets to replan.}
    \label{fig:exp1-eliminations}
\end{figure}

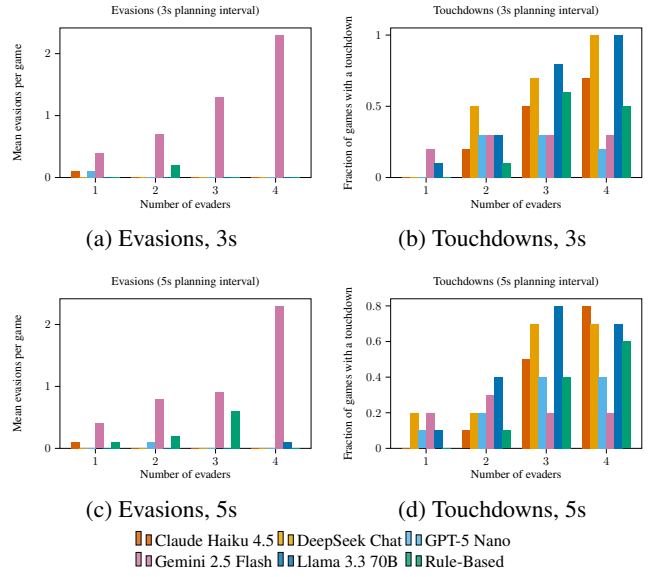
\begin{figure}[t]
    \centering
    \begin{subfigure}[b]{0.48\linewidth}
        \centering
        \resizebox{\linewidth}{!}{% This file was created with tikzplotlib v0.10.1.
\begin{tikzpicture}

\definecolor{chocolate213940}{RGB}{213,94,0}
\definecolor{cornflowerblue86180233}{RGB}{86,180,233}
\definecolor{darkcyan0114178}{RGB}{0,114,178}
\definecolor{darkcyan0158115}{RGB}{0,158,115}
\definecolor{darkgrey176}{RGB}{176,176,176}
\definecolor{orange2301590}{RGB}{230,159,0}
\definecolor{palevioletred204121167}{RGB}{204,121,167}

\begin{axis}[
height=6cm,
tick align=outside,
tick pos=left,
title={Evasions (3s planning interval)},
width=9cm,
x grid style={darkgrey176},
xlabel={Number of evaders},
xmin=0.41, xmax=4.59,
xtick style={color=black},
xtick={1,2,3,4},
y grid style={darkgrey176},
ylabel={Mean evasions per game},
ymin=0, ymax=2.415,
ytick style={color=black}
]
\draw[draw=none,fill=chocolate213940] (axis cs:0.6,0) rectangle (axis cs:0.733333333333333,0.1);
\draw[draw=none,fill=chocolate213940] (axis cs:1.6,0) rectangle (axis cs:1.73333333333333,0);
\draw[draw=none,fill=chocolate213940] (axis cs:2.6,0) rectangle (axis cs:2.73333333333333,0);
\draw[draw=none,fill=chocolate213940] (axis cs:3.6,0) rectangle (axis cs:3.73333333333333,0);
\draw[draw=none,fill=orange2301590] (axis cs:0.733333333333333,0) rectangle (axis cs:0.866666666666667,0);
\draw[draw=none,fill=orange2301590] (axis cs:1.73333333333333,0) rectangle (axis cs:1.86666666666667,0);
\draw[draw=none,fill=orange2301590] (axis cs:2.73333333333333,0) rectangle (axis cs:2.86666666666667,0);
\draw[draw=none,fill=orange2301590] (axis cs:3.73333333333333,0) rectangle (axis cs:3.86666666666667,0);
\draw[draw=none,fill=cornflowerblue86180233] (axis cs:0.866666666666667,0) rectangle (axis cs:1,0.1);
\draw[draw=none,fill=cornflowerblue86180233] (axis cs:1.86666666666667,0) rectangle (axis cs:2,0);
\draw[draw=none,fill=cornflowerblue86180233] (axis cs:2.86666666666667,0) rectangle (axis cs:3,0);
\draw[draw=none,fill=cornflowerblue86180233] (axis cs:3.86666666666667,0) rectangle (axis cs:4,0);
\draw[draw=none,fill=palevioletred204121167] (axis cs:1,0) rectangle (axis cs:1.13333333333333,0.4);
\draw[draw=none,fill=palevioletred204121167] (axis cs:2,0) rectangle (axis cs:2.13333333333333,0.7);
\draw[draw=none,fill=palevioletred204121167] (axis cs:3,0) rectangle (axis cs:3.13333333333333,1.3);
\draw[draw=none,fill=palevioletred204121167] (axis cs:4,0) rectangle (axis cs:4.13333333333333,2.3);
\draw[draw=none,fill=darkcyan0114178] (axis cs:1.13333333333333,0) rectangle (axis cs:1.26666666666667,0);
\draw[draw=none,fill=darkcyan0114178] (axis cs:2.13333333333333,0) rectangle (axis cs:2.26666666666667,0);
\draw[draw=none,fill=darkcyan0114178] (axis cs:3.13333333333333,0) rectangle (axis cs:3.26666666666667,0);
\draw[draw=none,fill=darkcyan0114178] (axis cs:4.13333333333333,0) rectangle (axis cs:4.26666666666667,0);
\draw[draw=none,fill=darkcyan0158115] (axis cs:1.26666666666667,0) rectangle (axis cs:1.4,0);
\draw[draw=none,fill=darkcyan0158115] (axis cs:2.26666666666667,0) rectangle (axis cs:2.4,0.2);
\draw[draw=none,fill=darkcyan0158115] (axis cs:3.26666666666667,0) rectangle (axis cs:3.4,0);
\draw[draw=none,fill=darkcyan0158115] (axis cs:4.26666666666667,0) rectangle (axis cs:4.4,0);
\end{axis}

\end{tikzpicture}}
        \caption{Evasions, 3s}
        \label{fig:exp1-evasions-3s}
    \end{subfigure}
    \hfill
    \begin{subfigure}[b]{0.48\linewidth}
        \centering
        \resizebox{\linewidth}{!}{% This file was created with tikzplotlib v0.10.1.
\begin{tikzpicture}

\definecolor{chocolate213940}{RGB}{213,94,0}
\definecolor{cornflowerblue86180233}{RGB}{86,180,233}
\definecolor{darkcyan0114178}{RGB}{0,114,178}
\definecolor{darkcyan0158115}{RGB}{0,158,115}
\definecolor{darkgrey176}{RGB}{176,176,176}
\definecolor{orange2301590}{RGB}{230,159,0}
\definecolor{palevioletred204121167}{RGB}{204,121,167}

\begin{axis}[
height=6cm,
tick align=outside,
tick pos=left,
title={Touchdowns (3s planning interval)},
width=9cm,
x grid style={darkgrey176},
xlabel={Number of evaders},
xmin=0.41, xmax=4.59,
xtick style={color=black},
xtick={1,2,3,4},
y grid style={darkgrey176},
ylabel={Fraction of games with a touchdown},
ymin=0, ymax=1.05,
ytick style={color=black}
]
\draw[draw=none,fill=chocolate213940] (axis cs:0.6,0) rectangle (axis cs:0.733333333333333,0);
\draw[draw=none,fill=chocolate213940] (axis cs:1.6,0) rectangle (axis cs:1.73333333333333,0.2);
\draw[draw=none,fill=chocolate213940] (axis cs:2.6,0) rectangle (axis cs:2.73333333333333,0.5);
\draw[draw=none,fill=chocolate213940] (axis cs:3.6,0) rectangle (axis cs:3.73333333333333,0.7);
\draw[draw=none,fill=orange2301590] (axis cs:0.733333333333333,0) rectangle (axis cs:0.866666666666667,0);
\draw[draw=none,fill=orange2301590] (axis cs:1.73333333333333,0) rectangle (axis cs:1.86666666666667,0.5);
\draw[draw=none,fill=orange2301590] (axis cs:2.73333333333333,0) rectangle (axis cs:2.86666666666667,0.7);
\draw[draw=none,fill=orange2301590] (axis cs:3.73333333333333,0) rectangle (axis cs:3.86666666666667,1);
\draw[draw=none,fill=cornflowerblue86180233] (axis cs:0.866666666666667,0) rectangle (axis cs:1,0);
\draw[draw=none,fill=cornflowerblue86180233] (axis cs:1.86666666666667,0) rectangle (axis cs:2,0.3);
\draw[draw=none,fill=cornflowerblue86180233] (axis cs:2.86666666666667,0) rectangle (axis cs:3,0.3);
\draw[draw=none,fill=cornflowerblue86180233] (axis cs:3.86666666666667,0) rectangle (axis cs:4,0.2);
\draw[draw=none,fill=palevioletred204121167] (axis cs:1,0) rectangle (axis cs:1.13333333333333,0.2);
\draw[draw=none,fill=palevioletred204121167] (axis cs:2,0) rectangle (axis cs:2.13333333333333,0.3);
\draw[draw=none,fill=palevioletred204121167] (axis cs:3,0) rectangle (axis cs:3.13333333333333,0.3);
\draw[draw=none,fill=palevioletred204121167] (axis cs:4,0) rectangle (axis cs:4.13333333333333,0.3);
\draw[draw=none,fill=darkcyan0114178] (axis cs:1.13333333333333,0) rectangle (axis cs:1.26666666666667,0.1);
\draw[draw=none,fill=darkcyan0114178] (axis cs:2.13333333333333,0) rectangle (axis cs:2.26666666666667,0.3);
\draw[draw=none,fill=darkcyan0114178] (axis cs:3.13333333333333,0) rectangle (axis cs:3.26666666666667,0.8);
\draw[draw=none,fill=darkcyan0114178] (axis cs:4.13333333333333,0) rectangle (axis cs:4.26666666666667,1);
\draw[draw=none,fill=darkcyan0158115] (axis cs:1.26666666666667,0) rectangle (axis cs:1.4,0);
\draw[draw=none,fill=darkcyan0158115] (axis cs:2.26666666666667,0) rectangle (axis cs:2.4,0.1);
\draw[draw=none,fill=darkcyan0158115] (axis cs:3.26666666666667,0) rectangle (axis cs:3.4,0.6);
\draw[draw=none,fill=darkcyan0158115] (axis cs:4.26666666666667,0) rectangle (axis cs:4.4,0.5);
\end{axis}

\end{tikzpicture}}
        \caption{Touchdowns, 3s}
        \label{fig:exp1-touchdowns-3s}
    \end{subfigure}

    \vspace{0.6em}

    \begin{subfigure}[b]{0.48\linewidth}
        \centering
        \resizebox{\linewidth}{!}{% This file was created with tikzplotlib v0.10.1.
\begin{tikzpicture}

\definecolor{chocolate213940}{RGB}{213,94,0}
\definecolor{cornflowerblue86180233}{RGB}{86,180,233}
\definecolor{darkcyan0114178}{RGB}{0,114,178}
\definecolor{darkcyan0158115}{RGB}{0,158,115}
\definecolor{darkgrey176}{RGB}{176,176,176}
\definecolor{orange2301590}{RGB}{230,159,0}
\definecolor{palevioletred204121167}{RGB}{204,121,167}

\begin{axis}[
height=6cm,
tick align=outside,
tick pos=left,
title={Evasions (5s planning interval)},
width=9cm,
x grid style={darkgrey176},
xlabel={Number of evaders},
xmin=0.41, xmax=4.59,
xtick style={color=black},
xtick={1,2,3,4},
y grid style={darkgrey176},
ylabel={Mean evasions per game},
ymin=0, ymax=2.415,
ytick style={color=black}
]
\draw[draw=none,fill=chocolate213940] (axis cs:0.6,0) rectangle (axis cs:0.733333333333333,0.1);
\draw[draw=none,fill=chocolate213940] (axis cs:1.6,0) rectangle (axis cs:1.73333333333333,0);
\draw[draw=none,fill=chocolate213940] (axis cs:2.6,0) rectangle (axis cs:2.73333333333333,0);
\draw[draw=none,fill=chocolate213940] (axis cs:3.6,0) rectangle (axis cs:3.73333333333333,0);
\draw[draw=none,fill=orange2301590] (axis cs:0.733333333333333,0) rectangle (axis cs:0.866666666666667,0);
\draw[draw=none,fill=orange2301590] (axis cs:1.73333333333333,0) rectangle (axis cs:1.86666666666667,0);
\draw[draw=none,fill=orange2301590] (axis cs:2.73333333333333,0) rectangle (axis cs:2.86666666666667,0);
\draw[draw=none,fill=orange2301590] (axis cs:3.73333333333333,0) rectangle (axis cs:3.86666666666667,0);
\draw[draw=none,fill=cornflowerblue86180233] (axis cs:0.866666666666667,0) rectangle (axis cs:1,0);
\draw[draw=none,fill=cornflowerblue86180233] (axis cs:1.86666666666667,0) rectangle (axis cs:2,0.1);
\draw[draw=none,fill=cornflowerblue86180233] (axis cs:2.86666666666667,0) rectangle (axis cs:3,0);
\draw[draw=none,fill=cornflowerblue86180233] (axis cs:3.86666666666667,0) rectangle (axis cs:4,0);
\draw[draw=none,fill=palevioletred204121167] (axis cs:1,0) rectangle (axis cs:1.13333333333333,0.4);
\draw[draw=none,fill=palevioletred204121167] (axis cs:2,0) rectangle (axis cs:2.13333333333333,0.8);
\draw[draw=none,fill=palevioletred204121167] (axis cs:3,0) rectangle (axis cs:3.13333333333333,0.9);
\draw[draw=none,fill=palevioletred204121167] (axis cs:4,0) rectangle (axis cs:4.13333333333333,2.3);
\draw[draw=none,fill=darkcyan0114178] (axis cs:1.13333333333333,0) rectangle (axis cs:1.26666666666667,0);
\draw[draw=none,fill=darkcyan0114178] (axis cs:2.13333333333333,0) rectangle (axis cs:2.26666666666667,0);
\draw[draw=none,fill=darkcyan0114178] (axis cs:3.13333333333333,0) rectangle (axis cs:3.26666666666667,0);
\draw[draw=none,fill=darkcyan0114178] (axis cs:4.13333333333333,0) rectangle (axis cs:4.26666666666667,0.1);
\draw[draw=none,fill=darkcyan0158115] (axis cs:1.26666666666667,0) rectangle (axis cs:1.4,0.1);
\draw[draw=none,fill=darkcyan0158115] (axis cs:2.26666666666667,0) rectangle (axis cs:2.4,0.2);
\draw[draw=none,fill=darkcyan0158115] (axis cs:3.26666666666667,0) rectangle (axis cs:3.4,0.6);
\draw[draw=none,fill=darkcyan0158115] (axis cs:4.26666666666667,0) rectangle (axis cs:4.4,0);
\end{axis}

\end{tikzpicture}}
        \caption{Evasions, 5s}
        \label{fig:exp1-evasions-5s}
    \end{subfigure}
    \hfill
    \begin{subfigure}[b]{0.48\linewidth}
        \centering
        \resizebox{\linewidth}{!}{% This file was created with tikzplotlib v0.10.1.
\begin{tikzpicture}

\definecolor{chocolate213940}{RGB}{213,94,0}
\definecolor{cornflowerblue86180233}{RGB}{86,180,233}
\definecolor{darkcyan0114178}{RGB}{0,114,178}
\definecolor{darkcyan0158115}{RGB}{0,158,115}
\definecolor{darkgrey176}{RGB}{176,176,176}
\definecolor{orange2301590}{RGB}{230,159,0}
\definecolor{palevioletred204121167}{RGB}{204,121,167}

\begin{axis}[
height=6cm,
tick align=outside,
tick pos=left,
title={Touchdowns (5s planning interval)},
width=9cm,
x grid style={darkgrey176},
xlabel={Number of evaders},
xmin=0.41, xmax=4.59,
xtick style={color=black},
xtick={1,2,3,4},
y grid style={darkgrey176},
ylabel={Fraction of games with a touchdown},
ymin=0, ymax=0.84,
ytick style={color=black}
]
\draw[draw=none,fill=chocolate213940] (axis cs:0.6,0) rectangle (axis cs:0.733333333333333,0);
\draw[draw=none,fill=chocolate213940] (axis cs:1.6,0) rectangle (axis cs:1.73333333333333,0.1);
\draw[draw=none,fill=chocolate213940] (axis cs:2.6,0) rectangle (axis cs:2.73333333333333,0.5);
\draw[draw=none,fill=chocolate213940] (axis cs:3.6,0) rectangle (axis cs:3.73333333333333,0.8);
\draw[draw=none,fill=orange2301590] (axis cs:0.733333333333333,0) rectangle (axis cs:0.866666666666667,0.2);
\draw[draw=none,fill=orange2301590] (axis cs:1.73333333333333,0) rectangle (axis cs:1.86666666666667,0.2);
\draw[draw=none,fill=orange2301590] (axis cs:2.73333333333333,0) rectangle (axis cs:2.86666666666667,0.7);
\draw[draw=none,fill=orange2301590] (axis cs:3.73333333333333,0) rectangle (axis cs:3.86666666666667,0.7);
\draw[draw=none,fill=cornflowerblue86180233] (axis cs:0.866666666666667,0) rectangle (axis cs:1,0.1);
\draw[draw=none,fill=cornflowerblue86180233] (axis cs:1.86666666666667,0) rectangle (axis cs:2,0.2);
\draw[draw=none,fill=cornflowerblue86180233] (axis cs:2.86666666666667,0) rectangle (axis cs:3,0.4);
\draw[draw=none,fill=cornflowerblue86180233] (axis cs:3.86666666666667,0) rectangle (axis cs:4,0.4);
\draw[draw=none,fill=palevioletred204121167] (axis cs:1,0) rectangle (axis cs:1.13333333333333,0.2);
\draw[draw=none,fill=palevioletred204121167] (axis cs:2,0) rectangle (axis cs:2.13333333333333,0.3);
\draw[draw=none,fill=palevioletred204121167] (axis cs:3,0) rectangle (axis cs:3.13333333333333,0.2);
\draw[draw=none,fill=palevioletred204121167] (axis cs:4,0) rectangle (axis cs:4.13333333333333,0.2);
\draw[draw=none,fill=darkcyan0114178] (axis cs:1.13333333333333,0) rectangle (axis cs:1.26666666666667,0.1);
\draw[draw=none,fill=darkcyan0114178] (axis cs:2.13333333333333,0) rectangle (axis cs:2.26666666666667,0.4);
\draw[draw=none,fill=darkcyan0114178] (axis cs:3.13333333333333,0) rectangle (axis cs:3.26666666666667,0.8);
\draw[draw=none,fill=darkcyan0114178] (axis cs:4.13333333333333,0) rectangle (axis cs:4.26666666666667,0.7);
\draw[draw=none,fill=darkcyan0158115] (axis cs:1.26666666666667,0) rectangle (axis cs:1.4,0);
\draw[draw=none,fill=darkcyan0158115] (axis cs:2.26666666666667,0) rectangle (axis cs:2.4,0.1);
\draw[draw=none,fill=darkcyan0158115] (axis cs:3.26666666666667,0) rectangle (axis cs:3.4,0.4);
\draw[draw=none,fill=darkcyan0158115] (axis cs:4.26666666666667,0) rectangle (axis cs:4.4,0.6);
\end{axis}

\end{tikzpicture}}
        \caption{Touchdowns, 5s}
        \label{fig:exp1-touchdowns-5s}
    \end{subfigure}

    \resizebox{0.95\linewidth}{!}{% This file was created with tikzplotlib v0.10.1.
\begin{tikzpicture}

\definecolor{chocolate213940}{RGB}{213,94,0}
\definecolor{cornflowerblue86180233}{RGB}{86,180,233}
\definecolor{darkcyan0114178}{RGB}{0,114,178}
\definecolor{darkcyan0158115}{RGB}{0,158,115}
\definecolor{darkgrey176}{RGB}{176,176,176}
\definecolor{orange2301590}{RGB}{230,159,0}
\definecolor{palevioletred204121167}{RGB}{204,121,167}

\begin{axis}[
height=1.6cm,
hide x axis,
hide y axis,
legend cell align={left},
legend columns=3,
legend style={
  fill opacity=0.8,
  draw opacity=1,
  text opacity=1,
  at={(0.5,0.5)},
  anchor=center,
  draw=none
},
tick align=outside,
tick pos=left,
width=14cm,
x grid style={darkgrey176},
xmin=0, xmax=1,
xtick style={color=black},
y grid style={darkgrey176},
ymin=0, ymax=1,
ytick style={color=black}
]
\draw[draw=none,fill=chocolate213940] (axis cs:-1.4,0) rectangle (axis cs:-0.6,1);
\addlegendimage{ybar,ybar legend,draw=none,fill=chocolate213940}
\addlegendentry{Claude Haiku 4.5}

\draw[draw=none,fill=orange2301590] (axis cs:-1.4,0) rectangle (axis cs:-0.6,1);
\addlegendimage{ybar,ybar legend,draw=none,fill=orange2301590}
\addlegendentry{DeepSeek Chat}

\draw[draw=none,fill=cornflowerblue86180233] (axis cs:-1.4,0) rectangle (axis cs:-0.6,1);
\addlegendimage{ybar,ybar legend,draw=none,fill=cornflowerblue86180233}
\addlegendentry{GPT-5 Nano}

\draw[draw=none,fill=palevioletred204121167] (axis cs:-1.4,0) rectangle (axis cs:-0.6,1);
\addlegendimage{ybar,ybar legend,draw=none,fill=palevioletred204121167}
\addlegendentry{Gemini 2.5 Flash}

\draw[draw=none,fill=darkcyan0114178] (axis cs:-1.4,0) rectangle (axis cs:-0.6,1);
\addlegendimage{ybar,ybar legend,draw=none,fill=darkcyan0114178}
\addlegendentry{Llama 3.3 70B}

\draw[draw=none,fill=darkcyan0158115] (axis cs:-1.4,0) rectangle (axis cs:-0.6,1);
\addlegendimage{ybar,ybar legend,draw=none,fill=darkcyan0158115}
\addlegendentry{Rule-Based}

\end{axis}

\end{tikzpicture}}
    \caption{Experiment I: defensive failure modes as a function of the number of evaders, by model, split by planning interval (top row: 3s; bottom row: 5s). Evasions (\subref{fig:exp1-evasions-3s}, \subref{fig:exp1-evasions-5s}) count evaders that leave the arena unresolved, reported as a mean per game; touchdowns (\subref{fig:exp1-touchdowns-3s}, \subref{fig:exp1-touchdowns-5s}) report the fraction of games in which the defended region was breached, an episode-ending outcome capped at one per game.}
    \label{fig:exp1-failures}
\end{figure}

Before introducing visual input, we first establish how the agentic RoE Planner performs when it reasons purely over structured state -- positions, velocities, and role/status flags for every agent. We sweep three axes against this state-only planner. The blue team is fixed at four pursuers; we vary (i) the size of the red team, $N_e \in {1,2,3,4}$, (ii) the planner's replanning latency, ${3, 5}$ simulated seconds, and (iii) the model instantiating both the blue and red \ac{roe} planners. All other simulation parameters are held fixed across the sweep: a $dt=0.1,\text{s}$ integration step, an $N=20$-step \ac{mpc} horizon, a $600$-step ($60,\text{s}$) episode budget, and both the pursuer-pursuer and pursuer-evader \ac{cbf} filters enabled.
For each of the $4$ red-team sizes we run $10$ independently seeded trials, randomizing every agent's initial position and velocity (blues around the defended perimeter, reds near the corner of the arena diametrically opposite it); this is repeated for every combination of the $2$ planning latencies and $6$ candidate models -- DeepSeek V3, Qwen3-30B, Claude Haiku 4.5, GPT-5 Nano, Llama 3.3 70B, and Gemini 2.5 Flash.

Each episode terminates on one of: a perimeter breach, a blue-blue collision, a solver failure at either the per-agent \ac{mpc} or \ac{cbf}-QP stage, or resolution of every red agent via capture, neutralization, or escape; an episode that exhausts its step budget without any of these is logged as incomplete. For every episode we record the terminal outcome, steps completed, number of reds captured versus escaped, the resulting blue score (Appendix~\ref{app:system-prompt-blue}), and the number of blues permanently disabled via \textsc{Neutralize}. Independently of outcome, every planning cycle in every episode is logged at the level of individual agents: for each active blue and red, the role it was assigned and the model's one-sentence natural-language justification for that assignment, giving a complete, per-decision reasoning trace across the full sweep rather than only aggregate statistics.

Histograms of all four end conditions per agent are shown in Figure~\ref{fig:exp1-eliminations} and Figure~\ref{fig:exp1-failures}. Qwen3-30B-A3B is excluded from these results because it failed to reliably produce schema-compliant structured output, forcing the safe-default fallback on a large fraction of planning cycles for both teams. The rule-based heuristic planner for the attacking team assigns \textsc{Evade} once a detected blue agent is within a fixed minimum distance of that specific red, and \textsc{Attack} otherwise \ref{app:rule-based}. This baseline is competitive with several LLMs: it achieves the second-highest capture rate of any model (behind only ClaudeHaiku4.5), a touchdown rate lower than three of the five LLMs, and few neutralizations, indicating that its threats are resolved overwhelmingly through multi-agent \textsc{Capture} rather than single-agent contact. Among the five LLMs, tactical style diverges sharply despite the identical role vocabulary: Llama3.370B relies overwhelmingly on \textsc{Neutralize}, trading a permanently disabled agent for a fast single-agent kill, yet still posts the highest touchdown rate of any model, suggesting this aggression comes at the expense of perimeter coverage. ClaudeHaiku4.5 and DeepSeekChat instead favor \textsc{Capture}, preserving their force at the cost of committing two agents per engagement -- though the two diverge sharply in outcome, with Claude's touchdown rate middling and DeepSeek's nearly as high as Llama's, so favoring \textsc{Capture} over \textsc{Neutralize} does not by itself predict defensive success. GPT-5Nano achieves the lowest touchdown rate overall despite only moderate engagement volume, indicating its assignments are more efficiently timed or targeted rather than simply more numerous. Gemini2.5Flash shows the most evasions among the LLMs -- more reds are left to exit the arena unaddressed -- yet still holds a comparably low touchdown rate, consistent with a containment-oriented strategy that tolerates some threats escaping rather than overcommitting the defense.
\subsection{Multi-Modal Strategic Planning}
\label{sec:exp-multimodal}

\begin{figure}[t]
    \centering
    \begin{subfigure}[b]{0.48\linewidth}
        \centering
        \resizebox{\linewidth}{!}{% This file was created with tikzplotlib v0.10.1.
\begin{tikzpicture}

\definecolor{chocolate213940}{RGB}{213,94,0}
\definecolor{cornflowerblue86180233}{RGB}{86,180,233}
\definecolor{darkgrey176}{RGB}{176,176,176}
\definecolor{palevioletred204121167}{RGB}{204,121,167}

\begin{axis}[
height=6cm,
tick align=outside,
tick pos=left,
title={Neutralizations (3s planning interval)},
width=9cm,
x grid style={darkgrey176},
xlabel={Number of contact slots},
xmin=0.41, xmax=4.59,
xtick style={color=black},
xtick={1,2,3,4},
y grid style={darkgrey176},
ylabel={Mean neutralizations per game},
ymin=0, ymax=0.315,
ytick style={color=black}
]
\draw[draw=none,fill=chocolate213940] (axis cs:0.6,0) rectangle (axis cs:0.866666666666667,0.2);
\draw[draw=none,fill=chocolate213940] (axis cs:1.6,0) rectangle (axis cs:1.86666666666667,0.2);
\draw[draw=none,fill=chocolate213940] (axis cs:2.6,0) rectangle (axis cs:2.86666666666667,0.2);
\draw[draw=none,fill=chocolate213940] (axis cs:3.6,0) rectangle (axis cs:3.86666666666667,0);
\draw[draw=none,fill=cornflowerblue86180233] (axis cs:0.866666666666667,0) rectangle (axis cs:1.13333333333333,0.2);
\draw[draw=none,fill=cornflowerblue86180233] (axis cs:1.86666666666667,0) rectangle (axis cs:2.13333333333333,0.2);
\draw[draw=none,fill=cornflowerblue86180233] (axis cs:2.86666666666667,0) rectangle (axis cs:3.13333333333333,0.3);
\draw[draw=none,fill=cornflowerblue86180233] (axis cs:3.86666666666667,0) rectangle (axis cs:4.13333333333333,0.3);
\draw[draw=none,fill=palevioletred204121167] (axis cs:1.13333333333333,0) rectangle (axis cs:1.4,0);
\draw[draw=none,fill=palevioletred204121167] (axis cs:2.13333333333333,0) rectangle (axis cs:2.4,0);
\draw[draw=none,fill=palevioletred204121167] (axis cs:3.13333333333333,0) rectangle (axis cs:3.4,0);
\draw[draw=none,fill=palevioletred204121167] (axis cs:4.13333333333333,0) rectangle (axis cs:4.4,0);
\end{axis}

\end{tikzpicture}}
        \caption{Neutralizations, 3s}
        \label{fig:exp2-neutralizations-3s}
    \end{subfigure}
    \hfill
    \begin{subfigure}[b]{0.48\linewidth}
        \centering
        \resizebox{\linewidth}{!}{% This file was created with tikzplotlib v0.10.1.
\begin{tikzpicture}

\definecolor{chocolate213940}{RGB}{213,94,0}
\definecolor{cornflowerblue86180233}{RGB}{86,180,233}
\definecolor{darkgrey176}{RGB}{176,176,176}
\definecolor{palevioletred204121167}{RGB}{204,121,167}

\begin{axis}[
height=6cm,
tick align=outside,
tick pos=left,
title={Captures (3s planning interval)},
width=9cm,
x grid style={darkgrey176},
xlabel={Number of contact slots},
xmin=0.41, xmax=4.59,
xtick style={color=black},
xtick={1,2,3,4},
y grid style={darkgrey176},
ylabel={Mean captures per game},
ymin=0, ymax=1.89,
ytick style={color=black}
]
\draw[draw=none,fill=chocolate213940] (axis cs:0.6,0) rectangle (axis cs:0.866666666666667,0.6);
\draw[draw=none,fill=chocolate213940] (axis cs:1.6,0) rectangle (axis cs:1.86666666666667,0.4);
\draw[draw=none,fill=chocolate213940] (axis cs:2.6,0) rectangle (axis cs:2.86666666666667,0.7);
\draw[draw=none,fill=chocolate213940] (axis cs:3.6,0) rectangle (axis cs:3.86666666666667,1.8);
\draw[draw=none,fill=cornflowerblue86180233] (axis cs:0.866666666666667,0) rectangle (axis cs:1.13333333333333,0.4);
\draw[draw=none,fill=cornflowerblue86180233] (axis cs:1.86666666666667,0) rectangle (axis cs:2.13333333333333,0.3);
\draw[draw=none,fill=cornflowerblue86180233] (axis cs:2.86666666666667,0) rectangle (axis cs:3.13333333333333,0.6);
\draw[draw=none,fill=cornflowerblue86180233] (axis cs:3.86666666666667,0) rectangle (axis cs:4.13333333333333,0.4);
\draw[draw=none,fill=palevioletred204121167] (axis cs:1.13333333333333,0) rectangle (axis cs:1.4,0);
\draw[draw=none,fill=palevioletred204121167] (axis cs:2.13333333333333,0) rectangle (axis cs:2.4,0.1);
\draw[draw=none,fill=palevioletred204121167] (axis cs:3.13333333333333,0) rectangle (axis cs:3.4,0.1);
\draw[draw=none,fill=palevioletred204121167] (axis cs:4.13333333333333,0) rectangle (axis cs:4.4,0.1);
\end{axis}

\end{tikzpicture}}
        \caption{Captures, 3s}
        \label{fig:exp2-captures-3s}
    \end{subfigure}

    \vspace{0.6em}

    \begin{subfigure}[b]{0.48\linewidth}
        \centering
        \resizebox{\linewidth}{!}{% This file was created with tikzplotlib v0.10.1.
\begin{tikzpicture}

\definecolor{chocolate213940}{RGB}{213,94,0}
\definecolor{cornflowerblue86180233}{RGB}{86,180,233}
\definecolor{darkgrey176}{RGB}{176,176,176}
\definecolor{palevioletred204121167}{RGB}{204,121,167}

\begin{axis}[
height=6cm,
tick align=outside,
tick pos=left,
title={Neutralizations (5s planning interval)},
width=9cm,
x grid style={darkgrey176},
xlabel={Number of contact slots},
xmin=0.41, xmax=4.59,
xtick style={color=black},
xtick={1,2,3,4},
y grid style={darkgrey176},
ylabel={Mean neutralizations per game},
ymin=0, ymax=0.315,
ytick style={color=black}
]
\draw[draw=none,fill=chocolate213940] (axis cs:0.6,0) rectangle (axis cs:0.866666666666667,0.1);
\draw[draw=none,fill=chocolate213940] (axis cs:1.6,0) rectangle (axis cs:1.86666666666667,0.2);
\draw[draw=none,fill=chocolate213940] (axis cs:2.6,0) rectangle (axis cs:2.86666666666667,0.1);
\draw[draw=none,fill=chocolate213940] (axis cs:3.6,0) rectangle (axis cs:3.86666666666667,0.3);
\draw[draw=none,fill=cornflowerblue86180233] (axis cs:0.866666666666667,0) rectangle (axis cs:1.13333333333333,0);
\draw[draw=none,fill=cornflowerblue86180233] (axis cs:1.86666666666667,0) rectangle (axis cs:2.13333333333333,0.1);
\draw[draw=none,fill=cornflowerblue86180233] (axis cs:2.86666666666667,0) rectangle (axis cs:3.13333333333333,0);
\draw[draw=none,fill=cornflowerblue86180233] (axis cs:3.86666666666667,0) rectangle (axis cs:4.13333333333333,0.1);
\draw[draw=none,fill=palevioletred204121167] (axis cs:1.13333333333333,0) rectangle (axis cs:1.4,0);
\draw[draw=none,fill=palevioletred204121167] (axis cs:2.13333333333333,0) rectangle (axis cs:2.4,0.1);
\draw[draw=none,fill=palevioletred204121167] (axis cs:3.13333333333333,0) rectangle (axis cs:3.4,0);
\draw[draw=none,fill=palevioletred204121167] (axis cs:4.13333333333333,0) rectangle (axis cs:4.4,0.2);
\end{axis}

\end{tikzpicture}}
        \caption{Neutralizations, 5s}
        \label{fig:exp2-neutralizations-5s}
    \end{subfigure}
    \hfill
    \begin{subfigure}[b]{0.48\linewidth}
        \centering
        \resizebox{\linewidth}{!}{% This file was created with tikzplotlib v0.10.1.
\begin{tikzpicture}

\definecolor{chocolate213940}{RGB}{213,94,0}
\definecolor{cornflowerblue86180233}{RGB}{86,180,233}
\definecolor{darkgrey176}{RGB}{176,176,176}
\definecolor{palevioletred204121167}{RGB}{204,121,167}

\begin{axis}[
height=6cm,
tick align=outside,
tick pos=left,
title={Captures (5s planning interval)},
width=9cm,
x grid style={darkgrey176},
xlabel={Number of contact slots},
xmin=0.41, xmax=4.59,
xtick style={color=black},
xtick={1,2,3,4},
y grid style={darkgrey176},
ylabel={Mean captures per game},
ymin=0, ymax=2.1,
ytick style={color=black}
]
\draw[draw=none,fill=chocolate213940] (axis cs:0.6,0) rectangle (axis cs:0.866666666666667,0.6);
\draw[draw=none,fill=chocolate213940] (axis cs:1.6,0) rectangle (axis cs:1.86666666666667,0.8);
\draw[draw=none,fill=chocolate213940] (axis cs:2.6,0) rectangle (axis cs:2.86666666666667,1.8);
\draw[draw=none,fill=chocolate213940] (axis cs:3.6,0) rectangle (axis cs:3.86666666666667,2);
\draw[draw=none,fill=cornflowerblue86180233] (axis cs:0.866666666666667,0) rectangle (axis cs:1.13333333333333,0.4);
\draw[draw=none,fill=cornflowerblue86180233] (axis cs:1.86666666666667,0) rectangle (axis cs:2.13333333333333,0.1);
\draw[draw=none,fill=cornflowerblue86180233] (axis cs:2.86666666666667,0) rectangle (axis cs:3.13333333333333,0.6);
\draw[draw=none,fill=cornflowerblue86180233] (axis cs:3.86666666666667,0) rectangle (axis cs:4.13333333333333,0.6);
\draw[draw=none,fill=palevioletred204121167] (axis cs:1.13333333333333,0) rectangle (axis cs:1.4,0.1);
\draw[draw=none,fill=palevioletred204121167] (axis cs:2.13333333333333,0) rectangle (axis cs:2.4,0.2);
\draw[draw=none,fill=palevioletred204121167] (axis cs:3.13333333333333,0) rectangle (axis cs:3.4,0.3);
\draw[draw=none,fill=palevioletred204121167] (axis cs:4.13333333333333,0) rectangle (axis cs:4.4,0.3);
\end{axis}

\end{tikzpicture}}
        \caption{Captures, 5s}
        \label{fig:exp2-captures-5s}
    \end{subfigure}

    \resizebox{0.95\linewidth}{!}{% This file was created with tikzplotlib v0.10.1.
\begin{tikzpicture}

\definecolor{chocolate213940}{RGB}{213,94,0}
\definecolor{cornflowerblue86180233}{RGB}{86,180,233}
\definecolor{darkgrey176}{RGB}{176,176,176}
\definecolor{palevioletred204121167}{RGB}{204,121,167}

\begin{axis}[
height=1.6cm,
hide x axis,
hide y axis,
legend cell align={left},
legend columns=3,
legend style={
  fill opacity=0.8,
  draw opacity=1,
  text opacity=1,
  at={(0.5,0.5)},
  anchor=center,
  draw=none
},
tick align=outside,
tick pos=left,
width=14cm,
x grid style={darkgrey176},
xmin=0, xmax=1,
xtick style={color=black},
y grid style={darkgrey176},
ymin=0, ymax=1,
ytick style={color=black}
]
\draw[draw=none,fill=chocolate213940] (axis cs:-1.4,0) rectangle (axis cs:-0.6,1);
\addlegendimage{ybar,ybar legend,draw=none,fill=chocolate213940}
\addlegendentry{Claude Haiku 4.5}

\draw[draw=none,fill=cornflowerblue86180233] (axis cs:-1.4,0) rectangle (axis cs:-0.6,1);
\addlegendimage{ybar,ybar legend,draw=none,fill=cornflowerblue86180233}
\addlegendentry{GPT-5 Nano}

\draw[draw=none,fill=palevioletred204121167] (axis cs:-1.4,0) rectangle (axis cs:-0.6,1);
\addlegendimage{ybar,ybar legend,draw=none,fill=palevioletred204121167}
\addlegendentry{Gemini 2.5 Flash}

\end{axis}

\end{tikzpicture}}
    \caption{Experiment II: mean number of pursuer-initiated eliminations per game as a function of the number of contact slots offered, by vision-capable model, split by planning interval (top row: 3s; bottom row: 5s).}
    \label{fig:exp2-eliminations}
\end{figure}
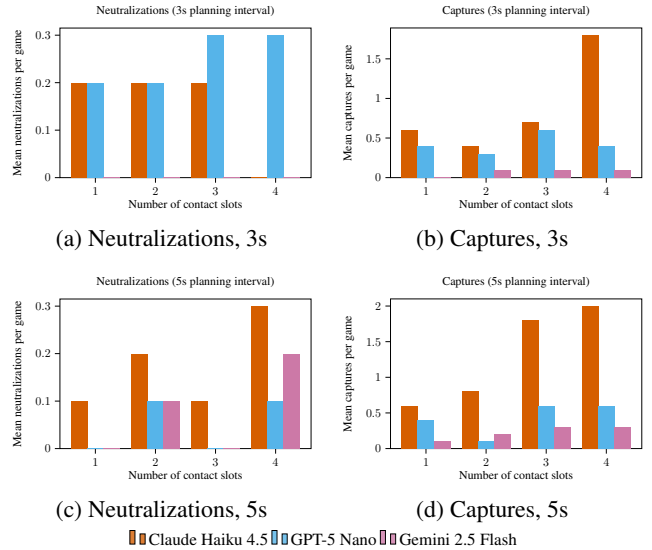

\begin{figure}[t]
    \centering
    \begin{subfigure}[b]{0.48\linewidth}
        \centering
        \resizebox{\linewidth}{!}{% This file was created with tikzplotlib v0.10.1.
\begin{tikzpicture}

\definecolor{chocolate213940}{RGB}{213,94,0}
\definecolor{cornflowerblue86180233}{RGB}{86,180,233}
\definecolor{darkgrey176}{RGB}{176,176,176}
\definecolor{palevioletred204121167}{RGB}{204,121,167}

\begin{axis}[
height=6cm,
tick align=outside,
tick pos=left,
title={Evasions (3s planning interval)},
width=9cm,
x grid style={darkgrey176},
xlabel={Number of contact slots},
xmin=0.41, xmax=4.59,
xtick style={color=black},
xtick={1,2,3,4},
y grid style={darkgrey176},
ylabel={Mean evasions per game},
ymin=0, ymax=2.205,
ytick style={color=black}
]
\draw[draw=none,fill=chocolate213940] (axis cs:0.6,0) rectangle (axis cs:0.866666666666667,0.1);
\draw[draw=none,fill=chocolate213940] (axis cs:1.6,0) rectangle (axis cs:1.86666666666667,0.2);
\draw[draw=none,fill=chocolate213940] (axis cs:2.6,0) rectangle (axis cs:2.86666666666667,0.4);
\draw[draw=none,fill=chocolate213940] (axis cs:3.6,0) rectangle (axis cs:3.86666666666667,0);
\draw[draw=none,fill=cornflowerblue86180233] (axis cs:0.866666666666667,0) rectangle (axis cs:1.13333333333333,0);
\draw[draw=none,fill=cornflowerblue86180233] (axis cs:1.86666666666667,0) rectangle (axis cs:2.13333333333333,0);
\draw[draw=none,fill=cornflowerblue86180233] (axis cs:2.86666666666667,0) rectangle (axis cs:3.13333333333333,0);
\draw[draw=none,fill=cornflowerblue86180233] (axis cs:3.86666666666667,0) rectangle (axis cs:4.13333333333333,0);
\draw[draw=none,fill=palevioletred204121167] (axis cs:1.13333333333333,0) rectangle (axis cs:1.4,0.6);
\draw[draw=none,fill=palevioletred204121167] (axis cs:2.13333333333333,0) rectangle (axis cs:2.4,0.4);
\draw[draw=none,fill=palevioletred204121167] (axis cs:3.13333333333333,0) rectangle (axis cs:3.4,1.3);
\draw[draw=none,fill=palevioletred204121167] (axis cs:4.13333333333333,0) rectangle (axis cs:4.4,2.1);
\end{axis}

\end{tikzpicture}}
        \caption{Evasions, 3s}
        \label{fig:exp2-evasions-3s}
    \end{subfigure}
    \hfill
    \begin{subfigure}[b]{0.48\linewidth}
        \centering
        \resizebox{\linewidth}{!}{% This file was created with tikzplotlib v0.10.1.
\begin{tikzpicture}

\definecolor{chocolate213940}{RGB}{213,94,0}
\definecolor{cornflowerblue86180233}{RGB}{86,180,233}
\definecolor{darkgrey176}{RGB}{176,176,176}
\definecolor{palevioletred204121167}{RGB}{204,121,167}

\begin{axis}[
height=6cm,
tick align=outside,
tick pos=left,
title={Touchdowns (3s planning interval)},
width=9cm,
x grid style={darkgrey176},
xlabel={Number of contact slots},
xmin=0.41, xmax=4.59,
xtick style={color=black},
xtick={1,2,3,4},
y grid style={darkgrey176},
ylabel={Fraction of games with a touchdown},
ymin=0, ymax=0.525,
ytick style={color=black}
]
\draw[draw=none,fill=chocolate213940] (axis cs:0.6,0) rectangle (axis cs:0.866666666666667,0);
\draw[draw=none,fill=chocolate213940] (axis cs:1.6,0) rectangle (axis cs:1.86666666666667,0.2);
\draw[draw=none,fill=chocolate213940] (axis cs:2.6,0) rectangle (axis cs:2.86666666666667,0.3);
\draw[draw=none,fill=chocolate213940] (axis cs:3.6,0) rectangle (axis cs:3.86666666666667,0.4);
\draw[draw=none,fill=cornflowerblue86180233] (axis cs:0.866666666666667,0) rectangle (axis cs:1.13333333333333,0);
\draw[draw=none,fill=cornflowerblue86180233] (axis cs:1.86666666666667,0) rectangle (axis cs:2.13333333333333,0.2);
\draw[draw=none,fill=cornflowerblue86180233] (axis cs:2.86666666666667,0) rectangle (axis cs:3.13333333333333,0.3);
\draw[draw=none,fill=cornflowerblue86180233] (axis cs:3.86666666666667,0) rectangle (axis cs:4.13333333333333,0.5);
\draw[draw=none,fill=palevioletred204121167] (axis cs:1.13333333333333,0) rectangle (axis cs:1.4,0.2);
\draw[draw=none,fill=palevioletred204121167] (axis cs:2.13333333333333,0) rectangle (axis cs:2.4,0.2);
\draw[draw=none,fill=palevioletred204121167] (axis cs:3.13333333333333,0) rectangle (axis cs:3.4,0.2);
\draw[draw=none,fill=palevioletred204121167] (axis cs:4.13333333333333,0) rectangle (axis cs:4.4,0.3);
\end{axis}

\end{tikzpicture}}
        \caption{Touchdowns, 3s}
        \label{fig:exp2-touchdowns-3s}
    \end{subfigure}

    \vspace{0.6em}

    \begin{subfigure}[b]{0.48\linewidth}
        \centering
        \resizebox{\linewidth}{!}{% This file was created with tikzplotlib v0.10.1.
\begin{tikzpicture}

\definecolor{chocolate213940}{RGB}{213,94,0}
\definecolor{cornflowerblue86180233}{RGB}{86,180,233}
\definecolor{darkgrey176}{RGB}{176,176,176}
\definecolor{palevioletred204121167}{RGB}{204,121,167}

\begin{axis}[
height=6cm,
tick align=outside,
tick pos=left,
title={Evasions (5s planning interval)},
width=9cm,
x grid style={darkgrey176},
xlabel={Number of contact slots},
xmin=0.41, xmax=4.59,
xtick style={color=black},
xtick={1,2,3,4},
y grid style={darkgrey176},
ylabel={Mean evasions per game},
ymin=0, ymax=1.155,
ytick style={color=black}
]
\draw[draw=none,fill=chocolate213940] (axis cs:0.6,0) rectangle (axis cs:0.866666666666667,0.1);
\draw[draw=none,fill=chocolate213940] (axis cs:1.6,0) rectangle (axis cs:1.86666666666667,0);
\draw[draw=none,fill=chocolate213940] (axis cs:2.6,0) rectangle (axis cs:2.86666666666667,0);
\draw[draw=none,fill=chocolate213940] (axis cs:3.6,0) rectangle (axis cs:3.86666666666667,0.1);
\draw[draw=none,fill=cornflowerblue86180233] (axis cs:0.866666666666667,0) rectangle (axis cs:1.13333333333333,0.2);
\draw[draw=none,fill=cornflowerblue86180233] (axis cs:1.86666666666667,0) rectangle (axis cs:2.13333333333333,0.1);
\draw[draw=none,fill=cornflowerblue86180233] (axis cs:2.86666666666667,0) rectangle (axis cs:3.13333333333333,0);
\draw[draw=none,fill=cornflowerblue86180233] (axis cs:3.86666666666667,0) rectangle (axis cs:4.13333333333333,0.1);
\draw[draw=none,fill=palevioletred204121167] (axis cs:1.13333333333333,0) rectangle (axis cs:1.4,0.5);
\draw[draw=none,fill=palevioletred204121167] (axis cs:2.13333333333333,0) rectangle (axis cs:2.4,0.2);
\draw[draw=none,fill=palevioletred204121167] (axis cs:3.13333333333333,0) rectangle (axis cs:3.4,0.4);
\draw[draw=none,fill=palevioletred204121167] (axis cs:4.13333333333333,0) rectangle (axis cs:4.4,1.1);
\end{axis}

\end{tikzpicture}}
        \caption{Evasions, 5s}
        \label{fig:exp2-evasions-5s}
    \end{subfigure}
    \hfill
    \begin{subfigure}[b]{0.48\linewidth}
        \centering
        \resizebox{\linewidth}{!}{% This file was created with tikzplotlib v0.10.1.
\begin{tikzpicture}

\definecolor{chocolate213940}{RGB}{213,94,0}
\definecolor{cornflowerblue86180233}{RGB}{86,180,233}
\definecolor{darkgrey176}{RGB}{176,176,176}
\definecolor{palevioletred204121167}{RGB}{204,121,167}

\begin{axis}[
height=6cm,
tick align=outside,
tick pos=left,
title={Touchdowns (5s planning interval)},
width=9cm,
x grid style={darkgrey176},
xlabel={Number of contact slots},
xmin=0.41, xmax=4.59,
xtick style={color=black},
xtick={1,2,3,4},
y grid style={darkgrey176},
ylabel={Fraction of games with a touchdown},
ymin=0, ymax=0.63,
ytick style={color=black}
]
\draw[draw=none,fill=chocolate213940] (axis cs:0.6,0) rectangle (axis cs:0.866666666666667,0);
\draw[draw=none,fill=chocolate213940] (axis cs:1.6,0) rectangle (axis cs:1.86666666666667,0.2);
\draw[draw=none,fill=chocolate213940] (axis cs:2.6,0) rectangle (axis cs:2.86666666666667,0.2);
\draw[draw=none,fill=chocolate213940] (axis cs:3.6,0) rectangle (axis cs:3.86666666666667,0.4);
\draw[draw=none,fill=cornflowerblue86180233] (axis cs:0.866666666666667,0) rectangle (axis cs:1.13333333333333,0.1);
\draw[draw=none,fill=cornflowerblue86180233] (axis cs:1.86666666666667,0) rectangle (axis cs:2.13333333333333,0.6);
\draw[draw=none,fill=cornflowerblue86180233] (axis cs:2.86666666666667,0) rectangle (axis cs:3.13333333333333,0.5);
\draw[draw=none,fill=cornflowerblue86180233] (axis cs:3.86666666666667,0) rectangle (axis cs:4.13333333333333,0.5);
\draw[draw=none,fill=palevioletred204121167] (axis cs:1.13333333333333,0) rectangle (axis cs:1.4,0.1);
\draw[draw=none,fill=palevioletred204121167] (axis cs:2.13333333333333,0) rectangle (axis cs:2.4,0.4);
\draw[draw=none,fill=palevioletred204121167] (axis cs:3.13333333333333,0) rectangle (axis cs:3.4,0.3);
\draw[draw=none,fill=palevioletred204121167] (axis cs:4.13333333333333,0) rectangle (axis cs:4.4,0.4);
\end{axis}

\end{tikzpicture}}
        \caption{Touchdowns, 5s}
        \label{fig:exp2-touchdowns-5s}
    \end{subfigure}

    \resizebox{0.95\linewidth}{!}{% This file was created with tikzplotlib v0.10.1.
\begin{tikzpicture}

\definecolor{chocolate213940}{RGB}{213,94,0}
\definecolor{cornflowerblue86180233}{RGB}{86,180,233}
\definecolor{darkgrey176}{RGB}{176,176,176}
\definecolor{palevioletred204121167}{RGB}{204,121,167}

\begin{axis}[
height=1.6cm,
hide x axis,
hide y axis,
legend cell align={left},
legend columns=3,
legend style={
  fill opacity=0.8,
  draw opacity=1,
  text opacity=1,
  at={(0.5,0.5)},
  anchor=center,
  draw=none
},
tick align=outside,
tick pos=left,
width=14cm,
x grid style={darkgrey176},
xmin=0, xmax=1,
xtick style={color=black},
y grid style={darkgrey176},
ymin=0, ymax=1,
ytick style={color=black}
]
\draw[draw=none,fill=chocolate213940] (axis cs:-1.4,0) rectangle (axis cs:-0.6,1);
\addlegendimage{ybar,ybar legend,draw=none,fill=chocolate213940}
\addlegendentry{Claude Haiku 4.5}

\draw[draw=none,fill=cornflowerblue86180233] (axis cs:-1.4,0) rectangle (axis cs:-0.6,1);
\addlegendimage{ybar,ybar legend,draw=none,fill=cornflowerblue86180233}
\addlegendentry{GPT-5 Nano}

\draw[draw=none,fill=palevioletred204121167] (axis cs:-1.4,0) rectangle (axis cs:-0.6,1);
\addlegendimage{ybar,ybar legend,draw=none,fill=palevioletred204121167}
\addlegendentry{Gemini 2.5 Flash}

\end{axis}

\end{tikzpicture}}
    \caption{Experiment II: defensive failure modes as a function of the number of contact slots offered, by vision-capable model, split by planning interval (top row: 3s; bottom row: 5s).}
    \label{fig:exp2-failures}
\end{figure}
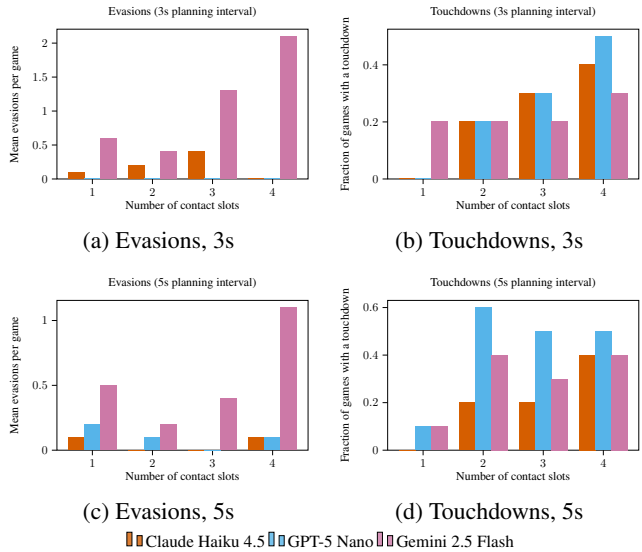

The task this perception is meant to support is genuine visual discrimination, not a trivial one. Since this requires interpreting an attached image, we restrict this experiment to the three of the six aforementioned models capable of multi-modal (image) input -- Claude Haiku 4.5, GPT-5 Nano, and Gemini 2.5 Flash; the remaining three, which accept text only, are not applicable here. Rather than a fixed red team, each of $N_{\text{slot}}$ contact slots independently resolves, once per episode, to one of three kinds: a real antagonistic drone (probability $0.7$, a genuine threat and the only kind counted toward the win condition), a harmless bird (probability $p_b=0.15$), or a harmless civilian drone (probability $p_c=0.15$). Reds and decoy contacts are shuffled into a single unlabeled contacts roster, so the model never sees which slot is which -- it must judge each contact's attached image on its own. Unlike the bird, which is silhouette-distinguishable at a glance, the civilian drone is deliberately the hard case: it is a quadcopter too, and can look very similar to the antagonistic drone, so the planner must rely on cues an image genuinely carries (e.g. color, markings, build) rather than shape alone. To guarantee every episode has a resolvable objective, the per-slot draw is resampled until at least one slot resolves to a real drone; the realized composition (how many of each kind) is logged per episode rather than assumed from $N_{\text{slot}}$ alone, since it is a random outcome of this draw and not a fixed count.

Histograms of all four end conditions per agent, aggregated over contact-slot counts, are shown in Figure~\ref{fig:exp2-eliminations} and Figure~\ref{fig:exp2-failures}. Unlike experiment I, no model required exclusion here -- schema-compliance failures were negligible across all three vision-capable models. ClaudeHaiku4.5 and GPT-5Nano account for the large majority of neutralizations and captures between them, with Claude in particular achieving both the highest capture rate of any model and the lowest touchdown rate, indicating an effective, well-targeted engagement policy despite the added visual-discrimination burden. Gemini2.5Flash, by contrast, both engages far less often and evades far more -- 0.825 mean evasions per game against at most 0.113 for the other two -- and the reasoning logs help explain why: across its 80 games, Gemini commits an agent to a civilian drone 134 times, more than either other model, despite completing the fewest real captures and neutralizations overall, so a disproportionate share of its already-sparse engagement attempts are misdirected at the harder decoy category rather than the genuine threat. This asymmetry holds across all three models: bird engagements are essentially absent (at most one, across 240 games total), while civilian engagements number in the dozens to low hundreds per model, confirming that the bird decoy is reliably ruled out by its winged silhouette as intended, while the visually similar civilian drone remains a genuine source of confusion even for the strongest models.

\begin{figure}[t]
    \centering
    \begin{subfigure}[t]{0.48\linewidth}
        \centering
        \includegraphics[width=\linewidth]{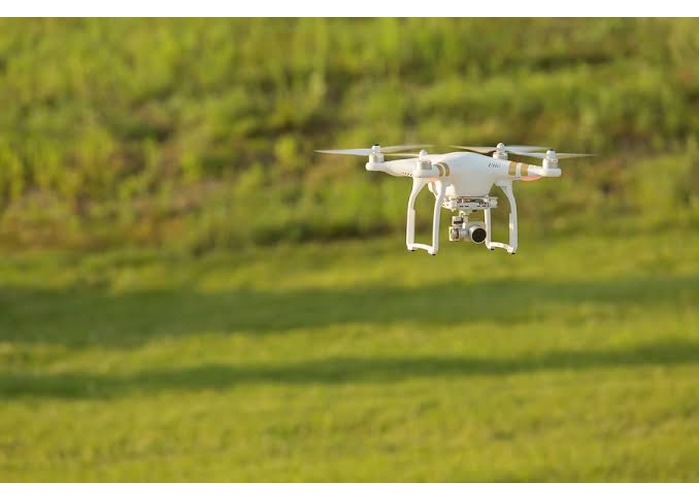}

        \parbox[t][2.5cm][t]{\linewidth}{
        \small\itshape\centering
        ``Contact 1 appears to be a harmless civilian drone, and with two agents committed to a high-priority threat, agent 2 should hold a defensive position near the perimeter.''
        }
        \caption{Correct identification}
        \label{fig:civilian-correct-id}
    \end{subfigure}
    \hfill
    \begin{subfigure}[t]{0.48\linewidth}
        \centering
        \includegraphics[width=\linewidth]{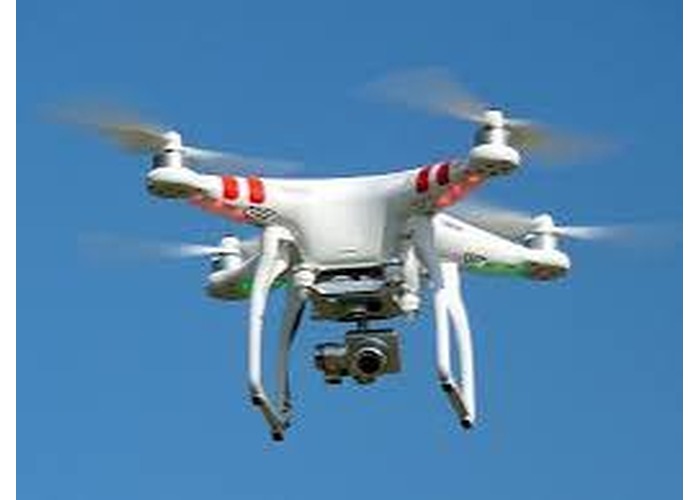}

        \parbox[t][2.5cm][t]{\linewidth}{
        \small\itshape\centering
        ``Agent 0 will CAPTURE contact 3 because it appears to be a hostile drone, and capturing allows us to preserve our asset while still neutralizing the threat.''
        }
        \caption{Failed identification}
        \label{fig:civilian-failed-id}
    \end{subfigure}
    \caption{Contrasting outcomes on the harder decoy case (\texttt{google/gemini-2.5-flash}, 3\,s planning interval). Both contacts are confirmed civilian drones by ground truth. In (\subref{fig:civilian-correct-id}), the planner correctly reads the image as a harmless civilian drone and assigns \textsc{Defend} instead of engaging. In (\subref{fig:civilian-failed-id}), the planner misreads the visually similar civilian drone as a hostile target and commits an agent to \textsc{Capture} against it -- a wasted engagement, since two agents are needed to converge on a target that poses no actual threat.}
    \label{fig:civilian-id-comparison}
\end{figure}

\section{Conclusion and Future Work}

This work demonstrates a control architecture for multi-agent adversarial pursuit-evasion that combines decentralized control with centralized, per-team strategic planning. Each agent's motion is governed by its own independent model-predictive controller, filtered through a control barrier function safety layer, while a single centralized planner per team assigns every active agent on that team a role, re-evaluated on a slower cadence than the underlying decentralized control loop. We instantiate this centralized planning layer a LLM operating over structured, natural-language tactical observations, and evaluate it where each team's planner only perceives opposing agents within its combined field of view, rather than assuming full state knowledge. We further extend this to a multi-modal setting in which the planner is given an image of each detected contact and must visually distinguish genuine threats from harmless decoys before committing to an engagement. Across both the state-only and vision-augmented settings, the framework demonstrates that an LLM can serve as an effective centralized tactical orchestrator.
The underlying simulation is intentionally low-fidelity:  the framework is best understood as a benchmark for LLM-driven agentic planning in cyber-physical systems, rather than a deployable pursuit-evasion controller in itself. Moving to a higher-fidelity environment is left to future work.

% \section*{ACKNOWLEDGMENTS}
% The authors acknowledge Research Computing at the Rochester Institute of Technology for providing computational resources and support that have contributed to the research results reported in this publication~\cite{https://doi.org/10.34788/0s3g-qd15}.

\bibliography{mybib}

\clearpage

\appendix

\begin{table*}[t]
\centering
\caption{Structured-output schema fields accompanying each prompt. Every field is enforced via a strict JSON schema (\texttt{response\_format: json\_schema}, \texttt{strict: true}) -- a response violating it is retried, and on repeated failure the affected team falls back to a safe default role for that cycle (Section~\ref{sec:experiments}).}
\label{tab:schema-fields}
\begin{tabular}{lll}
\toprule
Field & Team & Description \\
\midrule
\texttt{agent\_id} & Both & Integer id of the agent being assigned. \\
\texttt{role} & Both & One of the enumerated roles for that team. \\
\texttt{target\_red\_id} & BLUE & Detected red to engage; required ($\ge 0$) for \textsc{Neutralize}/\textsc{Capture}, $-1$ otherwise. \\
\texttt{target\_region\_id} & BLUE & Coverage region to patrol; required ($\ge 0$) for \textsc{Recon}, $-1$ otherwise. \\
\texttt{reasoning} & Both & One-sentence natural-language justification for the assignment. \\
\bottomrule
\end{tabular}
\end{table*}

\section{System Prompts}
\label{app:system-prompts}

Both teams' strategic planning layers are realized as an LLM called once per planning cycle with a fixed system prompt (below) and a per-cycle user message containing the current tactical observation (Section~\ref{sec:experiments}). The model's response is constrained to a strict JSON schema (Table~\ref{tab:schema-fields}); the prompt text and the schema jointly determine what the model can express, so both are reproduced here for completeness. Placeholder values in the BLUE prompt (\texttt{capture\_points\_neutralize}, \texttt{capture\_points\_capture}, \texttt{capture\_threshold}) are filled in at construction time from the same constants that govern the simulation's actual scoring, so the model is never told an incentive structure that diverges from what it is actually evaluated against; the values shown below are the defaults used in all reported experiments.

\subsection{Experiment I}
\subsubsection{Defending Team (BLUE)}
\label{app:system-prompt-blue}

\begin{quote}
\ttfamily\small
You are the tactical commander for the BLUE team defending a region against RED intruders in a 2D pursuit-evasion simulation. Each planning cycle, you assign every active blue agent exactly one role:

- DEFEND: hold a guard station on the perimeter of the defended region.

\medskip

- NEUTRALIZE: a single blue closes to physical contact with a red agent. No collision-avoidance safety margin is applied against red for this role -- contact is allowed and intended. Resolves fast, with just one agent. Consequence: that blue is permanently disabled afterward (it takes no further part in the engagement) and the kill is worth only 1.0 points.

\medskip

- CAPTURE: 2 or more blues converge on the same red agent simultaneously and hold within capture\_radius while maintaining a collision-avoidance safety margin -- red is captured by proximity without any contact. Slower and costlier up front (2 committed agents instead of 1), but every participating blue remains fully operational afterward, and the kill is worth 2.0 points -- more than NEUTRALIZE.

\medskip

- RECON: actively patrols toward a coverage region you choose (see the coverage list below) instead of holding a perimeter station or engaging -- it walks there while sweeping its sensor back and forth, then holds and keeps sweeping once it arrives. Use it to re-establish eyes on ground your team hasn't swept recently, not as a way to park an agent idle.

\medskip

This is a real tradeoff, not a preference: NEUTRALIZE trades a permanently lost asset and lower points for speed and certainty with a single agent. CAPTURE preserves your whole force and scores higher, but ties up 2 agents at once and only resolves once they arrive together, so a fast or already-close red may breach or escape before CAPTURE converges. RECON commits an agent away from engagement or perimeter duty in exchange for reducing the chance an undetected red is sitting in a region you haven't looked at recently. Weigh this against how many active (non-disabled) blues you have left, how many red agents are still active, and how urgent this particular red is (close to breaching vs. far away). Disabled blues are listed in the observation and are unavailable -- do not assign them a role.

Your team does not have full battlefield awareness: the red\_agents list below only includes reds detected by your team's combined field of view (each blue senses a conic region ahead of its current heading) -- there may be additional reds beyond what's listed that your team simply cannot currently see. A short or empty red\_agents list does not mean the defended region is safe, only that nothing is currently detected. The coverage list shows, for a fixed grid of regions spanning the arena, how many steps it's been since any blue's sensor last swept each one (-1 means never swept) -- higher numbers mean a red could have been sitting there undetected for longer, and are the main signal for where to send RECON.

For every agent you assign NEUTRALIZE or CAPTURE, you must also set target\_red\_id to the id of the specific red agent (from the red\_agents list above) it should engage -- you decide which threat each agent goes after, not just what role it plays. target\_red\_id must be one of the ids currently listed under red\_agents (you can only target a red your team has actually detected); for DEFEND and RECON, set target\_red\_id to -1. When assigning CAPTURE, coordinate: multiple agents assigned CAPTURE against the same target\_red\_id converge on that one red together, since capture requires several blues on the same red at once.

For every agent you assign RECON, you must also set target\_region\_id to the id of the specific coverage region (from the coverage list above) it should patrol toward -- prefer regions with a high (or -1, i.e. never-swept) steps\_since\_seen value. target\_region\_id must be one of the ids currently listed under coverage; for DEFEND, NEUTRALIZE, and CAPTURE, set target\_region\_id to -1. If multiple agents are assigned RECON, prefer spreading them across different regions rather than sending them all to the same one.

Assign roles based on the full tactical picture: how many red agents you can currently see, how close each is to blues and to the defended region, and how many blues you can commit to intercepting without leaving the perimeter undefended or over-spending your force. You do not need to use every role every cycle. Every active blue agent id provided must receive exactly one role, and for each one you must give a one-sentence reasoning explaining specifically why that agent got that role and target (referencing the actual tactical picture -- distances, force levels, urgency, coverage staleness -- not a generic restatement of the role's definition).
\end{quote}

\subsubsection{Attacking Team (RED)}
\label{app:system-prompt-red}

\begin{quote}
\ttfamily\small
You are the tactical commander for the RED team attempting to breach a defended region in a 2D pursuit-evasion simulation, while avoiding capture by BLUE agents. Each planning cycle, you assign every red agent exactly one role:

- ATTACK: advance toward the defended region, attempting to breach it.

\medskip

- EVADE: flee from nearby threatening blue agents instead of advancing.

\medskip

EVADE only ever makes sense once you have actually detected a blue agent -- you cannot react to a threat you haven't seen. If the blue\_agents list below is empty, no threat is currently detected and every red agent should ATTACK. Blue agents marked status=disabled in the observation are permanently out of action (they were expended neutralizing a red earlier) and pose no threat -- ignore them entirely when judging danger.

Your team does not have full battlefield awareness: the blue\_agents list below only includes blues detected by your team's combined field of view (each red senses a conic region ahead of its current heading) -- there may be additional blues beyond what's listed that your team simply cannot currently see.

Every red agent id provided must receive exactly one role, and for each one you must give a one-sentence reasoning explaining specifically why that agent got that role (referencing the actual tactical picture -- distances, threats, opportunity -- not a generic restatement of the role's definition).
\end{quote}

\begin{table*}[t]
    \centering
    \caption{Experiment I (state-only): mean end-condition counts per game, touchdown rate, incomplete games, and total non-parseable (bad) planner responses, broken out by planning interval (40 games per model per interval unless noted).}
    \label{tab:exp1-summary}
    \resizebox{\textwidth}{!}{%
    \begin{tabular}{llcccccc}
        \toprule
        Model & Interval & Neutralizations & Captures & Evasions & Touchdown Rate & Incomplete & Bad Responses \\
        \midrule
        \multirow{2}{*}{Claude Haiku 4.5}   & 3s & 0.100 & 1.025 & 0.025 & 0.350 & 16 & 0 \\
                                             & 5s & 0.100 & 1.200 & 0.025 & 0.350 & 12 & 0 \\
        \midrule
        \multirow{2}{*}{DeepSeek Chat}      & 3s & 0.500 & 0.650 & 0.000 & 0.550 & 7  & 7 \\
                                             & 5s & 0.250 & 1.125 & 0.000 & 0.450 & 5  & 7 \\
        \midrule
        \multirow{2}{*}{GPT-5 Nano}         & 3s & 0.600 & 0.350 & 0.025 & 0.200 & 25 & 0 \\
                                             & 5s & 0.450 & 0.450 & 0.025 & 0.275 & 20 & 1 \\
        \midrule
        \multirow{2}{*}{Gemini 2.5 Flash}   & 3s & 0.050 & 0.100 & 1.175 & 0.275 & 13 & 1 \\
                                             & 5s & 0.075 & 0.400 & 1.100 & 0.225 & 14 & 0 \\
        \midrule
        \multirow{2}{*}{Llama 3.3 70B}      & 3s & 1.125 & 0.400 & 0.000 & 0.550 & 0  & 0 \\
                                             & 5s & 0.925 & 0.725 & 0.025 & 0.500 & 3  & 0 \\
        \midrule
        \multirow{2}{*}{Qwen3 30B A3B}      & 3s & 0.000 & 1.825 & 0.000 & 0.450 & 3  & 518 \\
                                             & 5s & 0.000 & 1.750 & 0.000 & 0.525 & 0  & 345 \\
        \midrule
        \multirow{2}{*}{Rule-Based}         & 3s & 0.025 & 1.000 & 0.050 & 0.300 & 12 & 0 \\
                                             & 5s & 0.025 & 1.150 & 0.225 & 0.275 & 10 & 0 \\
        \bottomrule
    \end{tabular}%
    }
\end{table*}

\begin{table*}[t]
    \centering
    \caption{Experiment II (multi-modal): mean end-condition counts per game, touchdown rate, incomplete games, and total non-parseable (bad) planner responses, broken out by planning interval (40 games per model per interval).}
    \label{tab:exp2-summary}
    \resizebox{\textwidth}{!}{%
    \begin{tabular}{llcccccc}
        \toprule
        Model & Interval & Neutralizations & Captures & Evasions & Touchdown Rate & Incomplete & Bad Responses \\
        \midrule
        \multirow{2}{*}{Claude Haiku 4.5}   & 3s & 0.150 & 0.875 & 0.175 & 0.225 & 6  & 0 \\
                                             & 5s & 0.175 & 1.300 & 0.050 & 0.200 & 4  & 0 \\
        \midrule
        \multirow{2}{*}{GPT-5 Nano}         & 3s & 0.250 & 0.425 & 0.000 & 0.250 & 16 & 7 \\
                                             & 5s & 0.050 & 0.425 & 0.100 & 0.425 & 16 & 1 \\
        \midrule
        \multirow{2}{*}{Gemini 2.5 Flash}   & 3s & 0.000 & 0.075 & 1.100 & 0.225 & 10 & 0 \\
                                             & 5s & 0.075 & 0.225 & 0.550 & 0.300 & 11 & 0 \\
        \bottomrule
    \end{tabular}%
    }
\end{table*}

\subsubsection{Rule-Based Baseline}
\label{app:rule-based}

We also implement a deterministic, non-LLM heuristic planner for both teams, behind the same \texttt{RolePlanner}/\texttt{RedRolePlanner} interface, used as a free (zero-API-cost) baseline throughout our sweeps. Unlike the LLM planners, it provides no natural-language justification.

\paragraph{Defending team.} Let $\mathcal{B}$ be the set of active (non-disabled) blues and $\mathcal{R}_d \subseteq \mathcal{R}$ the reds currently detected by the team's combined field of view. If $\mathcal{R}_d = \varnothing$, every blue in $\mathcal{B}$ is assigned \textsc{Defend} (modulo reconnaissance, below). Otherwise, each blue $i \in \mathcal{B}$ is matched to its nearest detected red, $j^*(i) = \arg\min_{j \in \mathcal{R}_d} \|p_i - p_j\|$, and blues are ranked by this distance, closest first.

This ranking is used to split $\mathcal{B}$ into three groups. First, a fixed number of the closest blues are set aside to engage at all; everyone ranked below this cutoff is assigned \textsc{Defend}. This cutoff is
\begin{equation}
n_{\text{intercept}} = \max\big(k_{\min},\, \lceil |\mathcal{B}| \cdot f_{\text{neut}} \rceil\big).
\end{equation}
Second, within that engaging group, the closest blues are further split into two roles: the nearest
\begin{equation}
n_{\text{capture}} = \mathrm{round}(n_{\text{intercept}} \cdot f_{\text{cap}})
\end{equation}
are assigned \textsc{Capture} (targeting their own $j^*(i)$), and the remaining $n_{\text{intercept}} - n_{\text{capture}}$ -- closer than the \textsc{Defend} group, but further than the \textsc{Capture} group -- are assigned \textsc{Neutralize} (also targeting their own $j^*(i)$). This particular split by proximity rank is one arbitrary way to give the baseline non-trivial engagement behavior; we make no claim that it is an effective or well-tuned strategy in its own right. $f_{\text{neut}}, k_{\min}, f_{\text{cap}}$ are fixed fractions/minimums (defaults: $f_{\text{neut}}=0.5$, $k_{\min}=1$, $f_{\text{cap}}=0.5$).

Independently of detection, up to a fraction $f_{\text{recon}}$ of the blues otherwise assigned \textsc{Defend} ($\mathcal{D}$) are redirected to \textsc{Recon} (default $f_{\text{recon}}=0$, i.e. never offered): coverage regions are ranked by steps since last swept, never-swept regions ranked first, and the
\begin{equation}
n_{\text{recon}} = \min\big(\max(k_{\min}^{\text{recon}},\, \mathrm{round}(|\mathcal{D}| \cdot f_{\text{recon}})),\; |\mathcal{D}|,\; |\text{regions}|\big)
\end{equation}
stalest regions are each assigned to a distinct blue in $\mathcal{D}$, one blue per region.

\paragraph{Attacking team.} Each red is independently assigned \textsc{Evade} if a detected blue agent is within a fixed minimum distance of that red, and \textsc{Attack} (advance directly on the defended region) otherwise. Reconnaissance is not offered to the attacking team.

\subsubsection{Scoring System}
\label{app:scoring-system}

The defending team accumulates a score over the course of an episode based on how each attacking agent is ultimately resolved. Three outcomes contribute to this score, mirroring the point values stated directly to the planner in both experiments' system prompts (Appendices~\ref{app:system-prompt-blue} and~\ref{app:system-prompt-blue-vision}):

\begin{itemize}
    \item \textbf{Neutralize} (worth $1.0$ point): a single pursuer assigned \textsc{Neutralize} makes direct contact with its target. This resolves the threat immediately and requires only one agent, but that agent is permanently disabled afterward and takes no further part in the episode.
    \item \textbf{Capture} (worth $2.0$ points): at least $\kappa$ pursuers assigned \textsc{Capture} simultaneously hold within the capture radius of their target, without contact. This is worth twice as many points as \textsc{Neutralize} and leaves every participating pursuer fully operational, but requires coordinating multiple agents on the same target at once and resolves more slowly.
    \item \textbf{Escape} (worth $1.0$ point): a target that leaves the arena boundary unresolved -- neither neutralized nor captured -- is credited to the defending team as having exited without breaching the protected region.
\end{itemize}

A fourth outcome, \textbf{touchdown}, is not a point value but a terminal condition: if any attacking agent reaches the defended region before being resolved, the episode ends immediately in a loss for the defending team, regardless of the score accumulated up to that point.

This scoring asymmetry between \textsc{Neutralize} and \textsc{Capture} is deliberate: it is presented to the strategic planner explicitly as a tradeoff between speed and certainty on the one hand (\textsc{Neutralize}, one agent, lower score, permanent asset loss) and force preservation on the other (\textsc{Capture}, $\kappa$ agents, higher score, no asset loss), rather than as an incidental byproduct of the simulation mechanics. The planner is expected to weigh this tradeoff dynamically against the number of active pursuers remaining, the number of active attackers, and each attacker's proximity to the defended region.

\subsection{Experiment II}
\subsubsection{Defending Team (BLUE, Vision-Augmented)}
\label{app:system-prompt-blue-vision}

The RED team's system prompt for Experiment II is identical to Experiment~\ref{app:system-prompt-red} and is not repeated here.

\begin{quote}
\ttfamily\small
You are the tactical commander for the BLUE team defending a region against intrusions in a 2D pursuit-evasion simulation. Each planning cycle, you assign every active blue agent exactly one role:

- DEFEND: hold a guard station on the perimeter of the defended region.

\medskip

- NEUTRALIZE: a single blue closes to physical contact with its assigned target contact. No collision-avoidance safety margin is applied against that contact for this role -- contact is allowed and intended. Resolves fast, with just one agent. Consequence: that blue is permanently disabled afterward (it takes no further part in the engagement) and the kill is worth only 1.0 points.

\medskip

- CAPTURE: 2 or more blues converge on the same target contact simultaneously and hold within capture\_radius while maintaining a collision-avoidance safety margin -- captured by proximity without any contact. Slower and costlier up front (2 committed agents instead of 1), but every participating blue remains fully operational afterward, and the kill is worth 2.0 points -- more than NEUTRALIZE.

\medskip

- RECON: actively patrols toward a coverage region you choose (see the coverage list below) instead of holding a perimeter station or engaging -- it walks there while sweeping its sensor back and forth, then holds and keeps sweeping once it arrives. Use it to re-establish eyes on ground your team hasn't swept recently, not as a way to park an agent idle.

\medskip

This is a real tradeoff, not a preference: NEUTRALIZE trades a permanently lost asset and lower points for speed and certainty with a single agent. CAPTURE preserves your whole force and scores higher, but ties up 2 agents at once and only resolves once they arrive together, so a fast or already-close threat may breach or escape before CAPTURE converges. RECON commits an agent away from engagement or perimeter duty in exchange for reducing the chance an undetected threat is sitting in a region you haven't looked at recently. Weigh this against how many active (non-disabled) blues you have left and how urgent a given contact is. Disabled blues are listed in the observation and are unavailable -- do not assign them a role.

Critically: every object under "contacts" is UNIDENTIFIED. Most are real antagonistic drones -- genuine threats, the only kind that count toward your win condition. Some are harmless birds, and some are harmless civilian drones -- neither is a threat, neither is part of the win condition, and engaging one wastes that agent's role for the cycle while real threats go unaddressed. Position and velocity alone will often not tell the three apart -- a bird or a civilian drone can be moving just as fast, in just as plausible a direction, as a real antagonistic drone. Each contact has an image attached below the text observation; use it to judge. A bird is usually easy to rule out from its organic winged silhouette. A civilian drone is the hard case: it is a quadcopter too, and can look very similar to the antagonistic drone at a glance -- look closely at each image for anything that actually distinguishes them (color, markings, build, condition) rather than assuming shape alone settles it, and weigh how confident the image genuinely makes you against the tactical picture before committing NEUTRALIZE or CAPTURE.

Your team does not have full battlefield awareness: the contacts list below only includes contacts detected by your team's combined field of view (each blue senses a conic region ahead of its current heading) -- there may be additional contacts beyond what's listed that your team simply cannot currently see. A short or empty contacts list does not mean the defended region is safe, only that nothing is currently detected. The coverage list shows, for a fixed grid of regions spanning the arena, how many steps it's been since any blue's sensor last swept each one (-1 means never swept) -- higher numbers mean a threat could have been sitting there undetected for longer, and are the main signal for where to send RECON.

For every agent you assign NEUTRALIZE or CAPTURE, you must also set target\_contact\_id to the id of the specific contact (from the contacts list above) it should engage -- you decide which threat each agent goes after, not just what role it plays. target\_contact\_id must be one of the ids currently listed under contacts (you can only target a contact your team has actually detected); for DEFEND and RECON, set target\_contact\_id to -1. When assigning CAPTURE, coordinate: multiple agents assigned CAPTURE against the same target\_contact\_id converge on that one contact together, since capture requires several blues on the same target at once.

For every agent you assign RECON, you must also set target\_region\_id to the id of the specific coverage region (from the coverage list above) it should patrol toward -- prefer regions with a high (or -1, i.e. never-swept) steps\_since\_seen value. target\_region\_id must be one of the ids currently listed under coverage; for DEFEND, NEUTRALIZE, and CAPTURE, set target\_region\_id to -1. If multiple agents are assigned RECON, prefer spreading them across different regions rather than sending them all to the same one.

Assign roles based on the full tactical picture: how many contacts you can currently see, how close each is to blues and to the defended region, and how many blues you can commit to intercepting without leaving the perimeter undefended or over-spending your force. You do not need to use every role every cycle. Every active blue agent id provided must receive exactly one role, and for each one you must give a one-sentence reasoning explaining specifically why that agent got that role and target (referencing what you actually see in its image and the tactical picture -- distances, force levels, urgency, coverage staleness -- not a generic restatement of the role's definition).

Respond only with a valid JSON object matching the required schema.
\end{quote}

\begin{figure*}[t]
    \centering
    \begin{minipage}[t]{0.31\linewidth}
        \centering
        \includegraphics[width=\linewidth]{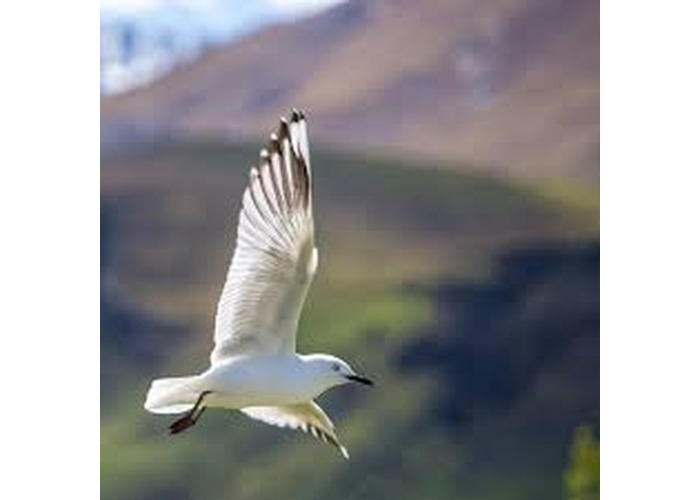}\\
        \textbf{(a) Bird}
        \vspace{0.4em}

        \small
        \textbf{Claude Haiku 4.5:} \textit{``Contact 0 is a bird (white winged silhouette clearly organic, not a drone), so agent 2 should conduct recon toward the never-swept southwestern region 0 to search for actual threats.''}
        \vspace{0.6em}

        \textbf{GPT-5 Nano:} \textit{``Agent 0 remains on the northeast perimeter; the detected contact is a bird (image shows a seagull), so no engagement and continued perimeter coverage is prudent.''}
        \vspace{0.6em}

        \textbf{Gemini 2.5 Flash:} \textit{``With only one detected contact that is clearly a bird, agent 0 should defend its perimeter station.''}
    \end{minipage}
    \hfill
    \begin{minipage}[t]{0.31\linewidth}
        \centering
        \includegraphics[width=\linewidth]{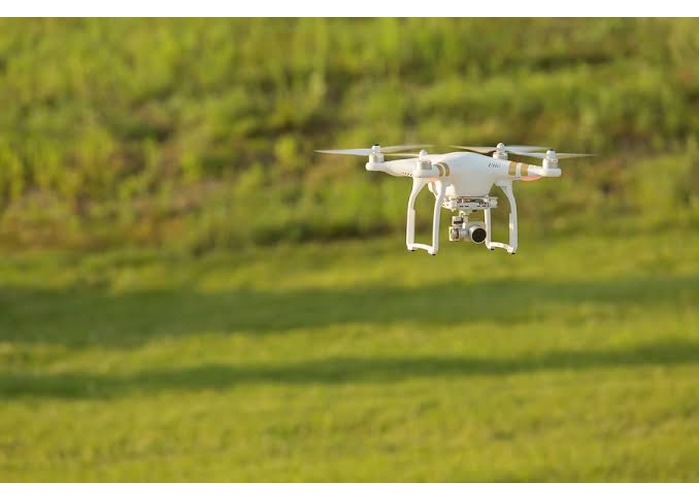}\\
        \textbf{(b) Civilian drone}
        \vspace{0.4em}

        \small
        \textbf{Claude Haiku 4.5:} \textit{``Contact 1 image is a civilian white quadcopter (not a threat); agent 3 at (75.41, 45.53) should patrol toward region 11 (center 62.50, 87.50, steps\_since\_seen=174) to re-establish coverage of the long-unseen northern perimeter area.''}
        \vspace{0.6em}

        \textbf{GPT-5 Nano:} \textit{``Agent 0 is near the defense\_center (82,82) at (70.54, 88.95) and should hold a perimeter guard; the detected contact (ID 1) sits roughly 42 units away and the attached image appears to be a civilian quadcopter, not an immediate threat, so no engagement this cycle is warranted.''}
        \vspace{0.6em}

        \textbf{Gemini 2.5 Flash:} \textit{``Contact 1 appears to be a harmless civilian drone, and with two agents committed to a high-priority threat, agent 2 should hold a defensive position near the perimeter.''}
    \end{minipage}
    \hfill
    \begin{minipage}[t]{0.31\linewidth}
        \centering
        \includegraphics[width=\linewidth]{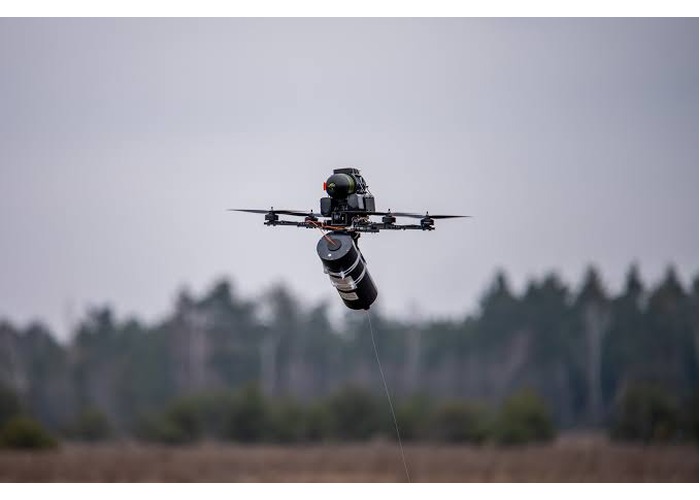}\\
        \textbf{(c) Antagonistic drone}
        \vspace{0.4em}

        \small
        \textbf{Claude Haiku 4.5:} \textit{``Agent 2 is closest to contact\_id=1 (the genuine antagonistic drone with visible payload and quadcopter arms)---rapid single-agent neutralization is appropriate given the contact's clear hostile signature and distance.''}
        \vspace{0.6em}

        \textbf{GPT-5 Nano:} \textit{``Agent 0 will join Agent 2 to CAPTURE Contact 1 because it is the only current real threat detected (drone with payload) and CAPTURE preserves both agents while yielding higher points.''}
        \vspace{0.6em}

        \textbf{Gemini 2.5 Flash:} \textit{``Agent 2 will initiate a capture of contact 1, which is clearly an antagonistic drone, to preserve assets and score higher points.''}
    \end{minipage}
    \caption{Each of the three vision-capable models' reasoning against the same contact image, for all three contact categories. All nine responses are confirmed correct against ground truth. Ground-truth confusability is asymmetric by design: the bird (a) is silhouette-distinguishable and never misidentified in this sample, while the civilian drone (b) shares the same quadcopter airframe as the genuine threat (c), making it the harder discrimination case the multi-modal experiment is built around.}
    \label{fig:reasoning-by-category}
\end{figure*}

\begin{figure*}[t]
    \centering
    \begin{minipage}[t]{0.31\linewidth}
        \centering
        \includegraphics[width=\linewidth]{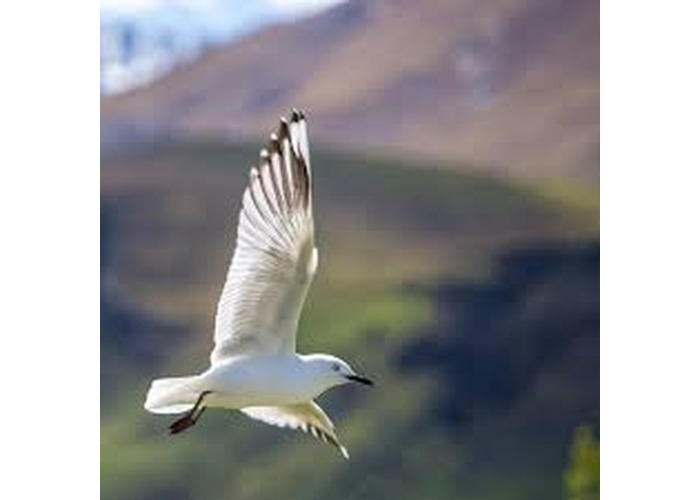}\\
        \textbf{(a) bird 1}
        \vspace{0.4em}

        \small
        \textbf{Claude Haiku 4.5:} \textit{``Agent 3 should recon region 6 (209 steps unseen) to check a stale sector; contact 0 is confirmed a bird and contact 2 appears to be an antagonistic drone but is still distant, so reconnaissance of blind spots takes priority over committing scarce forces prematurely.''}
        \vspace{0.6em}

        \textbf{GPT-5 Nano:} \textit{``Agent 0 is roughly 11 units from the defense center at [82,82] and should hold a guard station on the near-perimeter while the others sweep the high-staleness regions; a non-threat bird (contact 2) is present in view and does not require engagement.''}
        \vspace{0.6em}

        \textbf{Gemini 2.5 Flash:} \textit{``Agent 3 is assigned to defend the perimeter, as contacts 0 is being handled, contact 2 is a bird, and contact 3 is a civilian drone not requiring engagement.''}
    \end{minipage}
    \hfill
    \begin{minipage}[t]{0.31\linewidth}
        \centering
        \includegraphics[width=\linewidth]{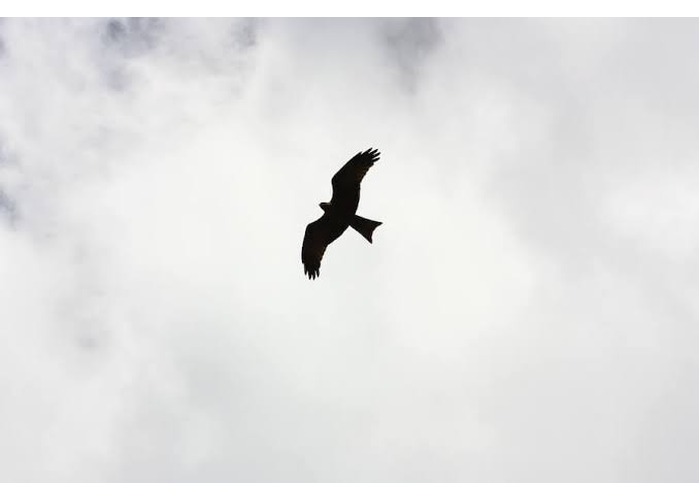}\\
        \textbf{(b) bird 2}
        \vspace{0.4em}

        \small
        \textbf{Claude Haiku 4.5:} \textit{``Agent 3 should patrol region 0 (bottom-left corner, never swept at -1 steps) to search for undetected threats; contacts 0 and 1 are either a bird or civilian drone (contact 0 has external payload, contact 1 has distinct bird silhouette), making RECON the priority to cover blind spots rather than commit to invalid engagement.''}
        \vspace{0.6em}

        \textbf{GPT-5 Nano:} \textit{``Agent 0 is assigned RECON to region 10, which has 70 steps since last sweep and is in a central east quadrant; this ensures a fresher sensor sweep while not engaging the detected contact, which is a non-threat bird (contact 1) visible in the image.''}
        \vspace{0.6em}

        \textbf{Gemini 2.5 Flash:} \textit{``Agent 2 will defend the perimeter, as contact 1 is a bird and not a threat, and agent 1 is engaging the other contact.''}
    \end{minipage}
    \hfill
    \begin{minipage}[t]{0.31\linewidth}
        \centering
        \includegraphics[width=\linewidth]{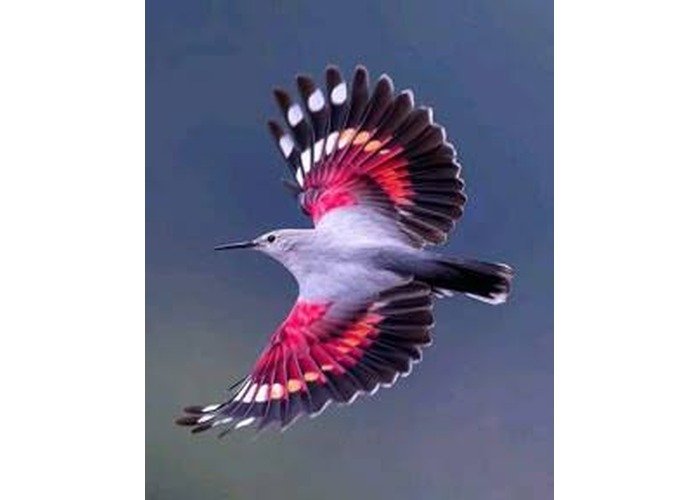}\\
        \textbf{(c) bird 3}
        \vspace{0.4em}

        \small
        \textit{This image appeared in only one game across the entire sweep and was never detected by any blue agent within that episode -- no model ever received or reasoned about it, for any of the three vision-capable models.}
    \end{minipage}
    \caption{Reasoning from each of the three vision-capable models against each of the three bird images. Correctness is not filtered for here -- some responses above misjudge distance or priority even when the bird itself is correctly ruled out as a non-threat.}
    \label{fig:reasoning-birds}
\end{figure*}

\begin{figure*}[t]
    \centering
    \begin{minipage}[t]{0.31\linewidth}
        \centering
        \includegraphics[width=\linewidth]{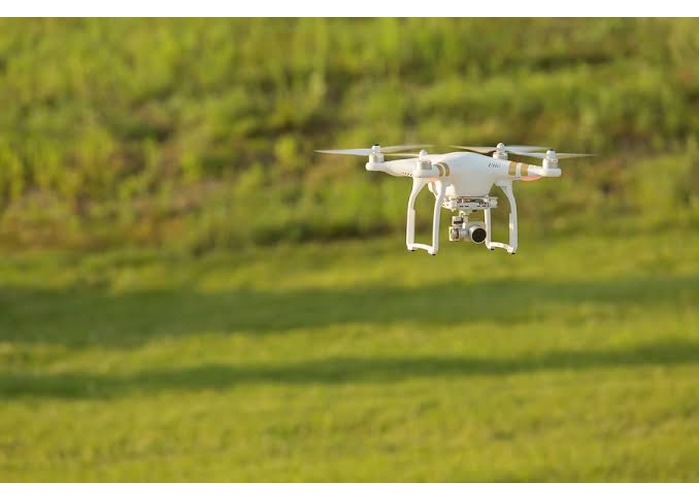}\\
        \textbf{(a) Civilian drone 1}
        \vspace{0.4em}

        \small
        \textbf{Claude Haiku 4.5:} \textit{``Contact 1 is a white quadcopter with visible camera and distinctive construction -- a real antagonistic drone -- currently at [35.36, 77.32] moving slowly; agent 2 at [55.34, 71.21] can intercept quickly before it threatens the defense center.''}
        \vspace{0.6em}

        \textbf{GPT-5 Nano:} \textit{``Agent 2 is closest to contact 1 (about 11 units away); neutralizing this contact swiftly removes a near-left threat before it can maneuver toward the defended zone.''}
        \vspace{0.6em}

        \textbf{Gemini 2.5 Flash:} \textit{``Contact 1 appears to be a civilian drone, so agent 1 will attempt to capture it along with agent 2 to preserve assets and score higher.''}
    \end{minipage}
    \hfill
    \begin{minipage}[t]{0.31\linewidth}
        \centering
        \includegraphics[width=\linewidth]{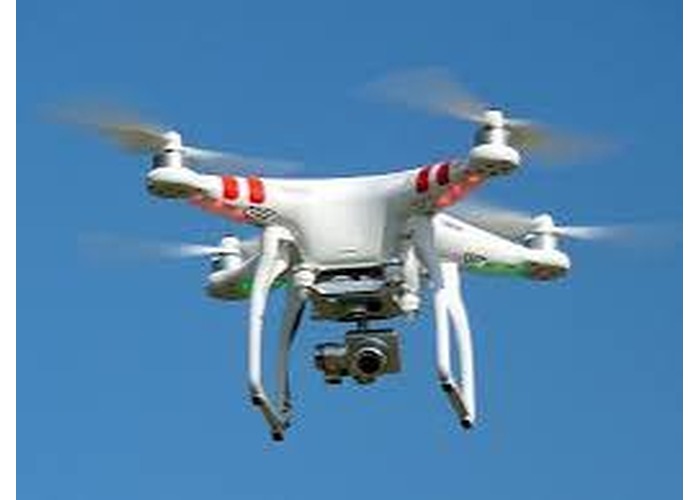}\\
        \textbf{(b) Civilian drone 1}
        \vspace{0.4em}

        \small
        \textbf{Claude Haiku 4.5:} \textit{``Agent 2 is closest to contact 3 (approximately 16 units away); the contact image shows a civilian-marked quadcopter with red/white color scheme inconsistent with antagonistic drones, but given the uncertain identification and positioning, a single neutralize action is most efficient to resolve the ambiguity while preserving perimeter coverage.''}
        \vspace{0.6em}

        \textbf{GPT-5 Nano:} \textit{``Agent 2 is the closest to contact 3 (roughly 8--9 units away), enabling a quick neutralization of the quadcopter-like threat shown in contact 3's image.''}
        \vspace{0.6em}

        \textbf{Gemini 2.5 Flash:} \textit{``Agent 1 is assigned to capture contact 3, which appears to be an antagonistic drone based on its appearance, working in concert with agent 2 to preserve forces.''}
    \end{minipage}
    \hfill
    \begin{minipage}[t]{0.31\linewidth}
        \centering
        \includegraphics[width=\linewidth]{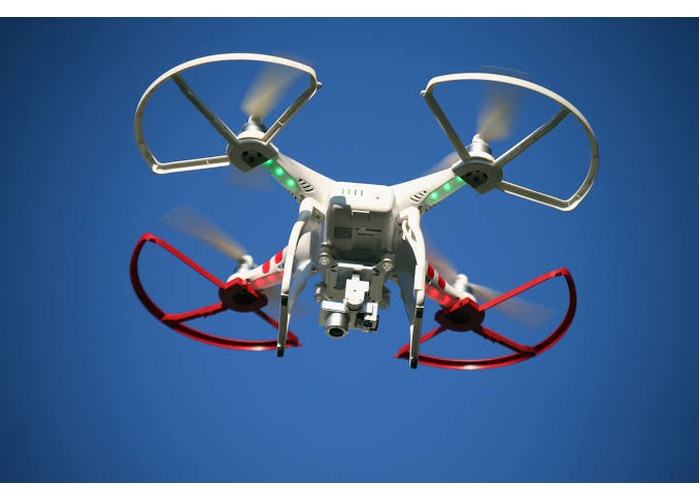}\\
        \textbf{(c) Civilian drone 3}
        \vspace{0.4em}

        \small
        \textbf{Claude Haiku 4.5:} \textit{``Contact 1 is a white civilian quadcopter (clear camera housing, non-threatening appearance) at [64.51, 75.21] -- assign 2 agents to CAPTURE it safely rather than risking permanent loss via NEUTRALIZE on what appears to be a harmless civilian drone.''}
        \vspace{0.6em}

        \textbf{GPT-5 Nano:} \textit{``Agent 2 is the closest to contact 1 (about 6.9 units away) and will close to neutralize it immediately.''}
        \vspace{0.6em}

        \textbf{Gemini 2.5 Flash:} \textit{``Contact 1 appears to be an antagonistic drone based on its clear quadcopter design with propeller guards, and Agent 2 is well-positioned to initiate a capture with another agent.''}
    \end{minipage}
    \caption{Reasoning from each of the three vision-capable models against each of the three civilian drone images. Correctness is not filtered for: Claude misidentifies (a) as antagonistic, Gemini misidentifies (b) and (c) as antagonistic, illustrating that this decoy category is confused in both directions and by every model at least once.}
    \label{fig:reasoning-civilian}
\end{figure*}

\begin{figure*}[t]
    \centering
    \begin{minipage}[t]{0.31\linewidth}
        \centering
        \includegraphics[width=\linewidth]{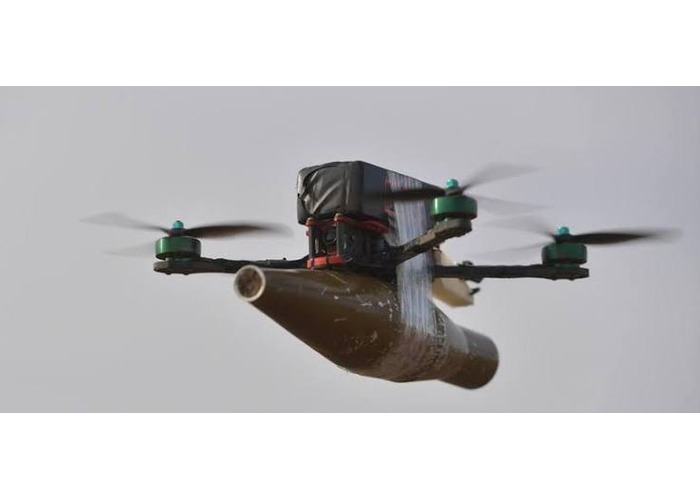}\\
        \textbf{(a) Red drone 1}
        \vspace{0.4em}

        \small
        \textbf{Claude Haiku 4.5:} \textit{``Contact 1 is a quadcopter with distinctive industrial/payload design confirming it as an antagonistic drone moving inbound at high velocity toward the defended region, warranting immediate capture engagement to preserve force and score higher points.''}
        \vspace{0.6em}

        \textbf{GPT-5 Nano:} \textit{``Agent 1 is roughly 10 units from contact 1 and can neutralize quickly for a decisive kill, reducing immediate danger.''}
        \vspace{0.6em}

        \textbf{Gemini 2.5 Flash:} \textit{``Agent 2 is assigned to CAPTURE contact 1, another antagonistic drone with an explosive payload, to address a second immediate threat with a coordinated approach.''}
    \end{minipage}
    \hfill
    \begin{minipage}[t]{0.31\linewidth}
        \centering
        \includegraphics[width=\linewidth]{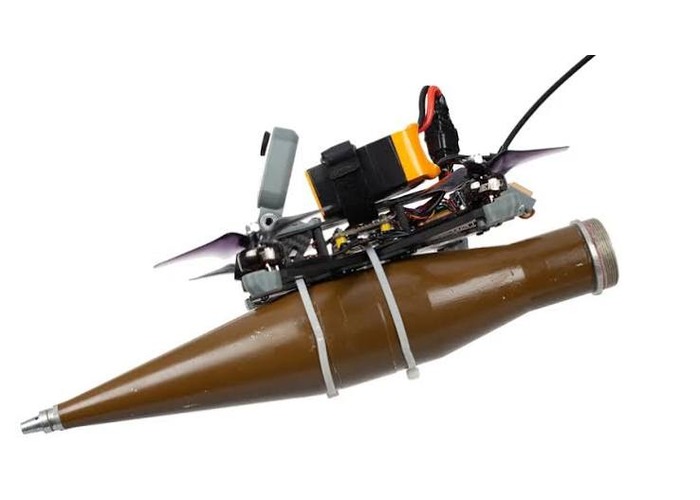}\\
        \textbf{(b) Red drone 2}
        \vspace{0.4em}

        \small
        \textbf{Claude Haiku 4.5:} \textit{``Agent 2 is positioned southwest and can intercept contact 0; the image clearly shows an antagonistic armed drone with a payload attachment, warranting immediate engagement via capture with agent 3 to preserve operational capability.''}
        \vspace{0.6em}

        \textbf{GPT-5 Nano:} \textit{``Agent 1 is closest to contact 0 (about 10--11 units away) and can attempt a fast neutralization; contact 0's image shows a drone-like threat.''}
        \vspace{0.6em}

        \textbf{Gemini 2.5 Flash:} \textit{``Contact 0 is clearly an antagonistic drone with an explosive payload attached, and is moving towards the defended area, so Agent 0 will help capture it with Agent 2 to ensure it is stopped without losing any blues.''}
    \end{minipage}
    \hfill
    \begin{minipage}[t]{0.31\linewidth}
        \centering
        \includegraphics[width=\linewidth]{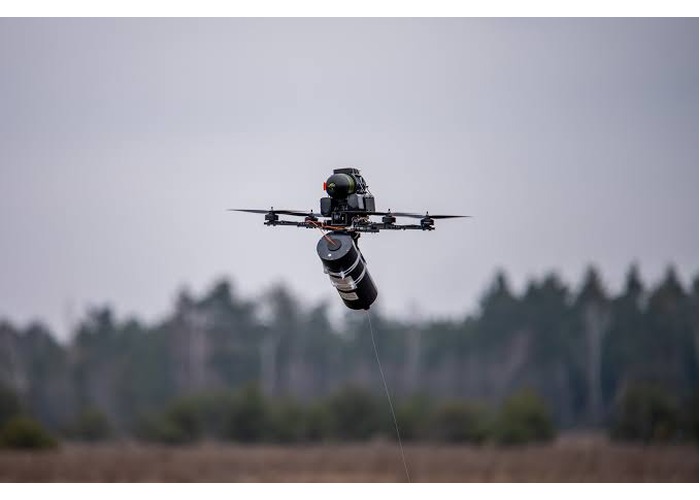}\\
        \textbf{(c) Red drone 3}
        \vspace{0.4em}

        \small
        \textbf{Claude Haiku 4.5:} \textit{``Agent 2 is already moving eastward and can intercept contact 2 (identified as an antagonistic drone with suspended payload in the image) along with agent 3 for a high-value 2-point capture.''}
        \vspace{0.6em}

        \textbf{GPT-5 Nano:} \textit{``Agent 0 is $\sim$33--34 units from contact 2 and will converge with Agent 3 to CAPTURE this threat quickly while the rest of the team holds the perimeter.''}
        \vspace{0.6em}

        \textbf{Gemini 2.5 Flash:} \textit{``Contact 2 appears to be another antagonistic drone carrying a payload, and neutralizing it quickly is critical to eliminate the threat, even at the cost of one blue agent.''}
    \end{minipage}
    \caption{Reasoning from each of the three vision-capable models against each of the three antagonistic drone images. All nine responses correctly identify the contact as a genuine threat.}
    \label{fig:reasoning-drones}
\end{figure*}

\begin{figure*}[t]
    \centering
    \begin{subfigure}[b]{\textwidth}
        \centering
        \includegraphics[width=0.32\textwidth]{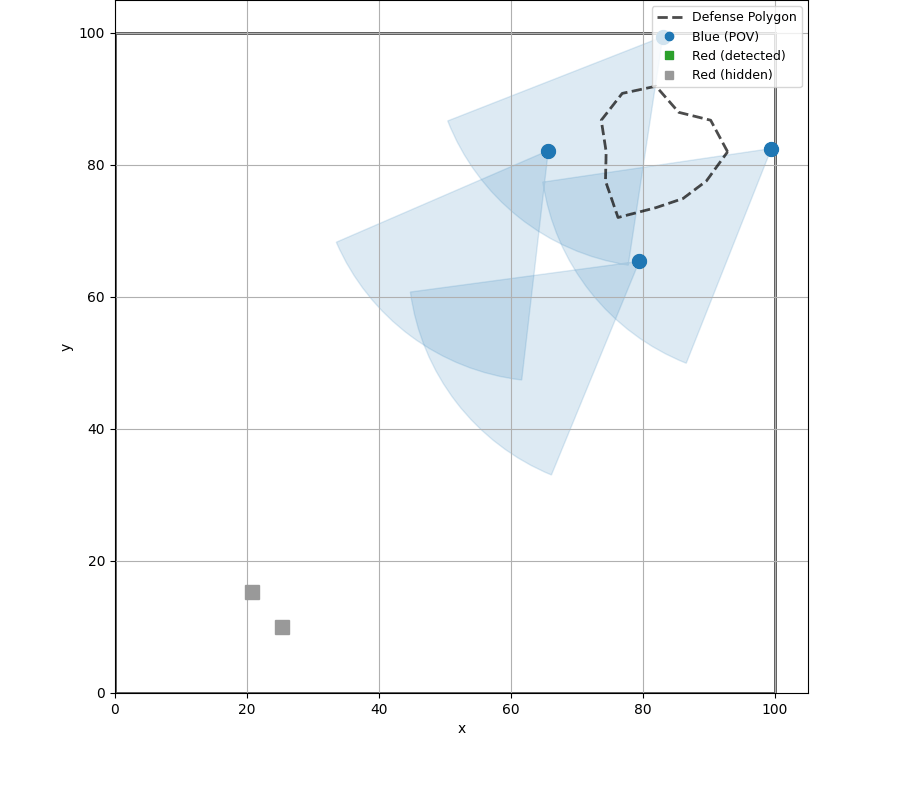}\hfill
        \includegraphics[width=0.32\textwidth]{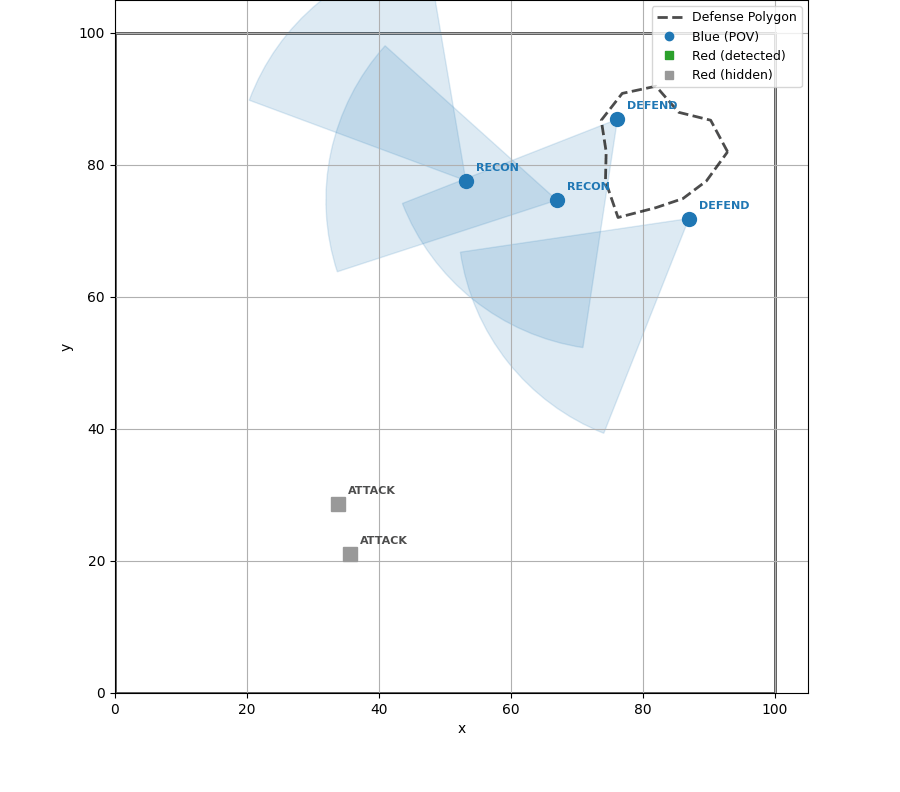}\hfill
        \includegraphics[width=0.32\textwidth]{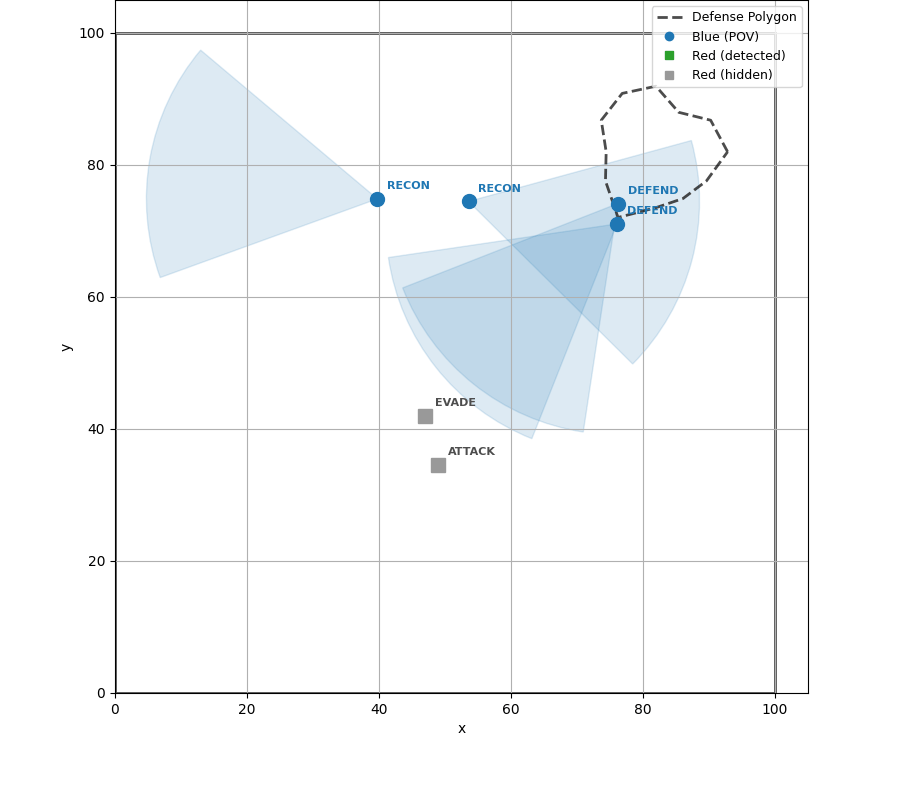}\\[0.4em]
        \includegraphics[width=0.32\textwidth]{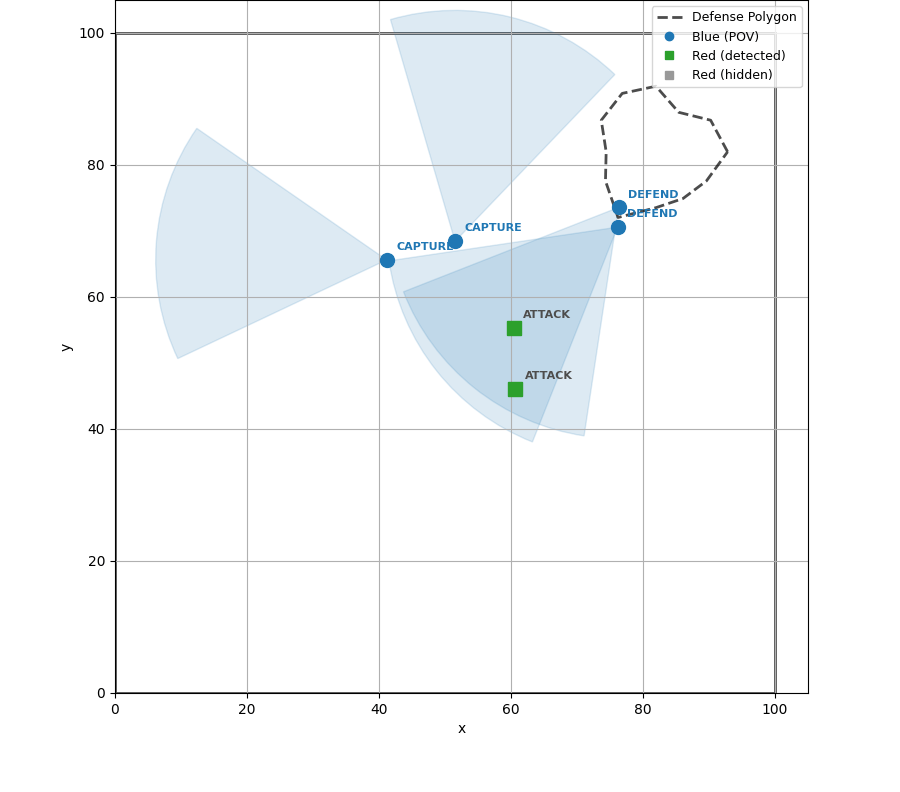}\hfill
        \includegraphics[width=0.32\textwidth]{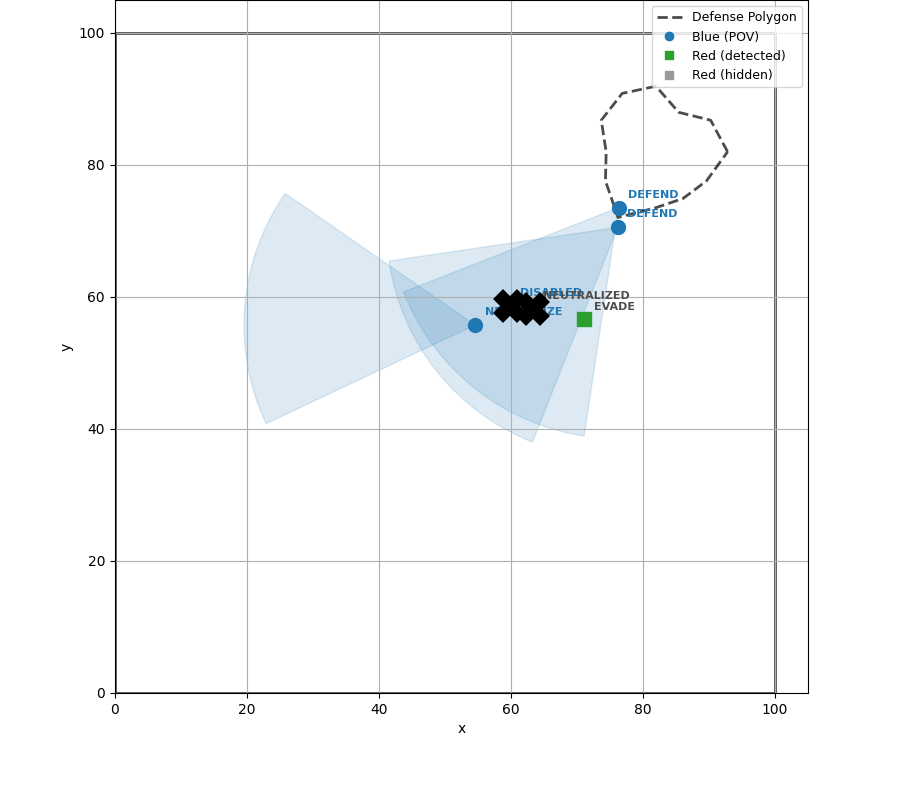}\hfill
        \includegraphics[width=0.32\textwidth]{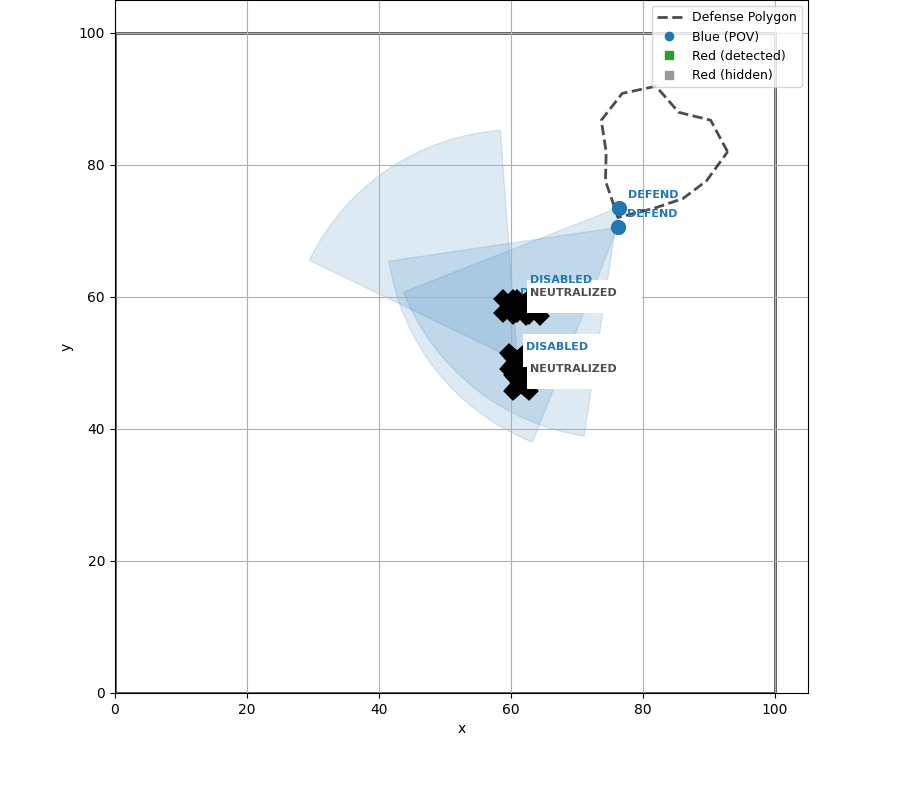}
        \caption{Blue POV -- each defender's field-of-view wedge and its live read of the two antagonistic contacts (green: detected, gray: hidden).}
        \label{fig:exp2-4v2-blue-pov}
    \end{subfigure}

    \vspace{1em}

    \begin{subfigure}[b]{\textwidth}
        \centering
        \includegraphics[width=0.32\textwidth]{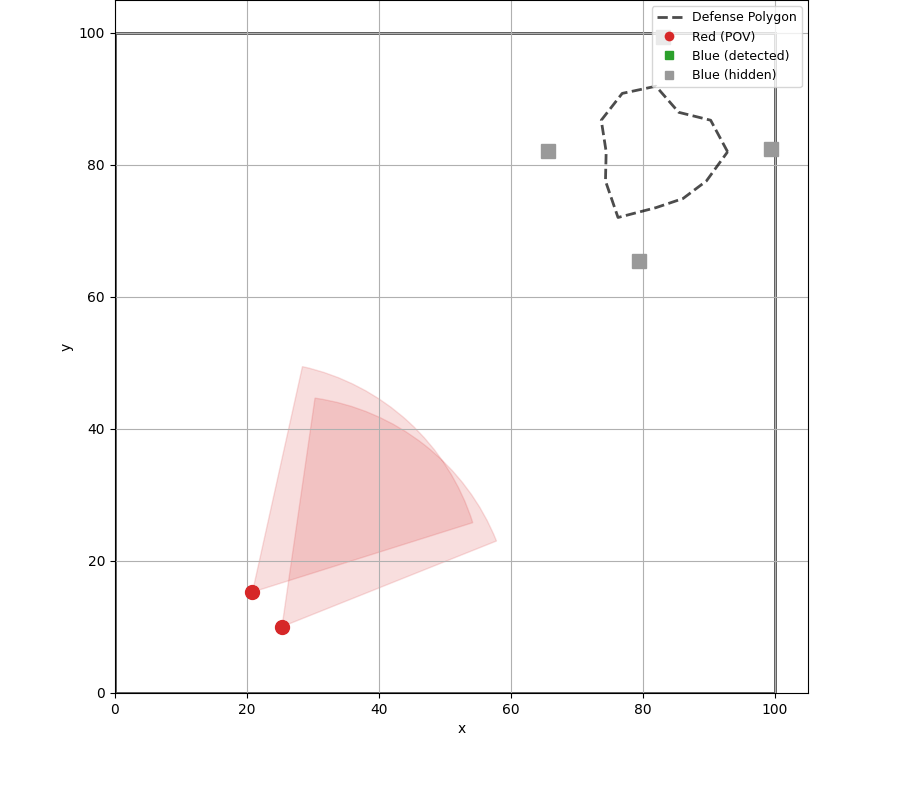}\hfill
        \includegraphics[width=0.32\textwidth]{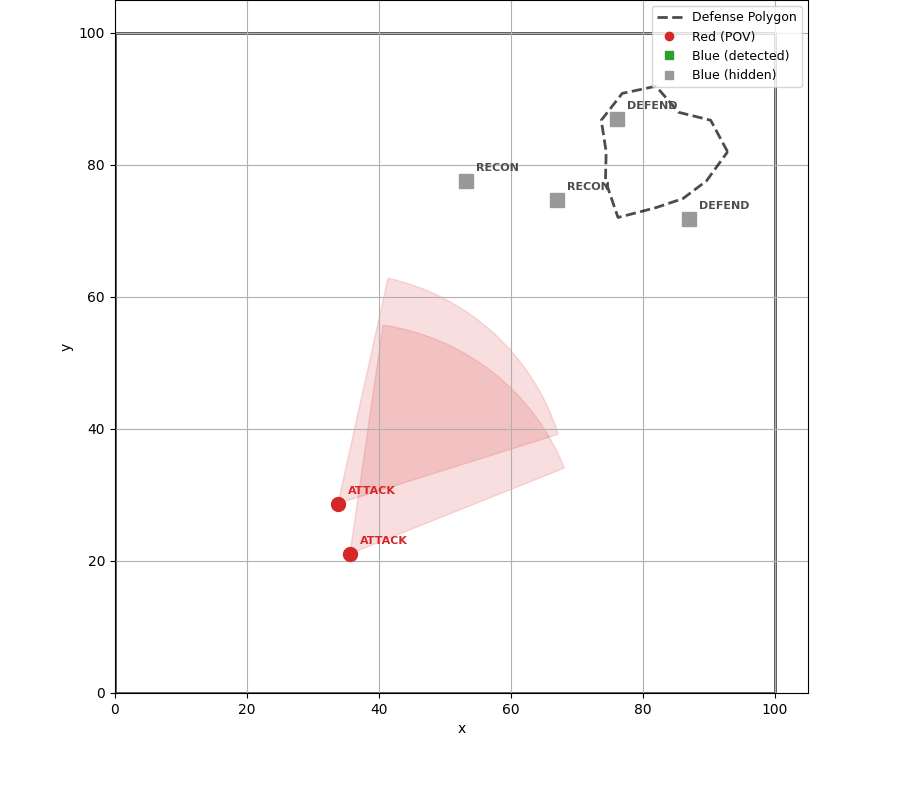}\hfill
        \includegraphics[width=0.32\textwidth]{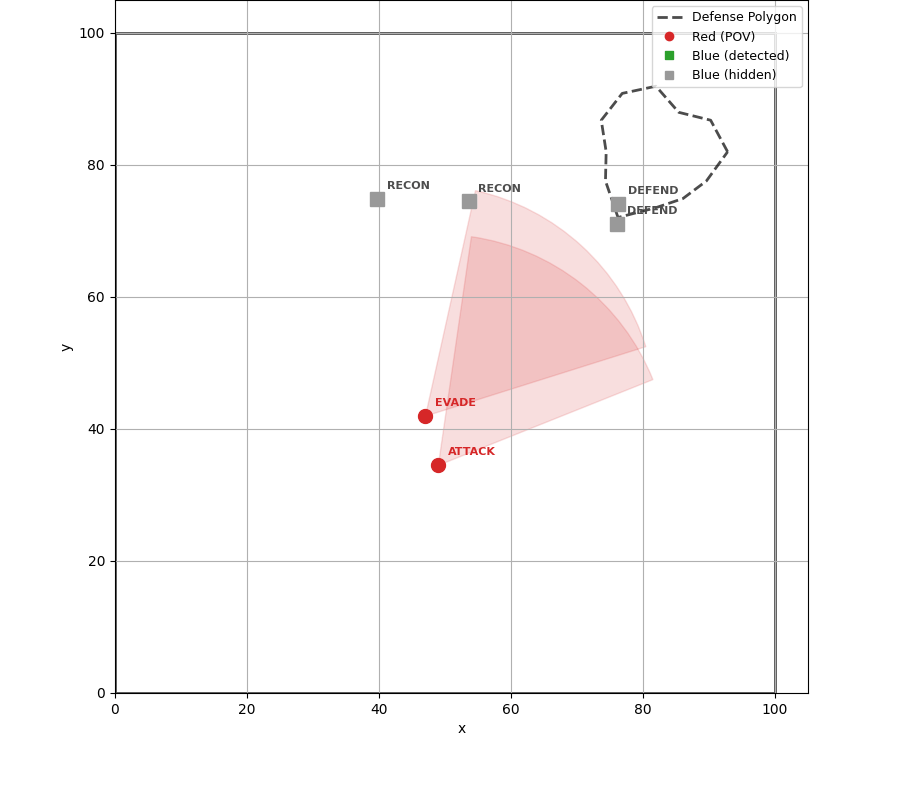}\\[0.4em]
        \includegraphics[width=0.32\textwidth]{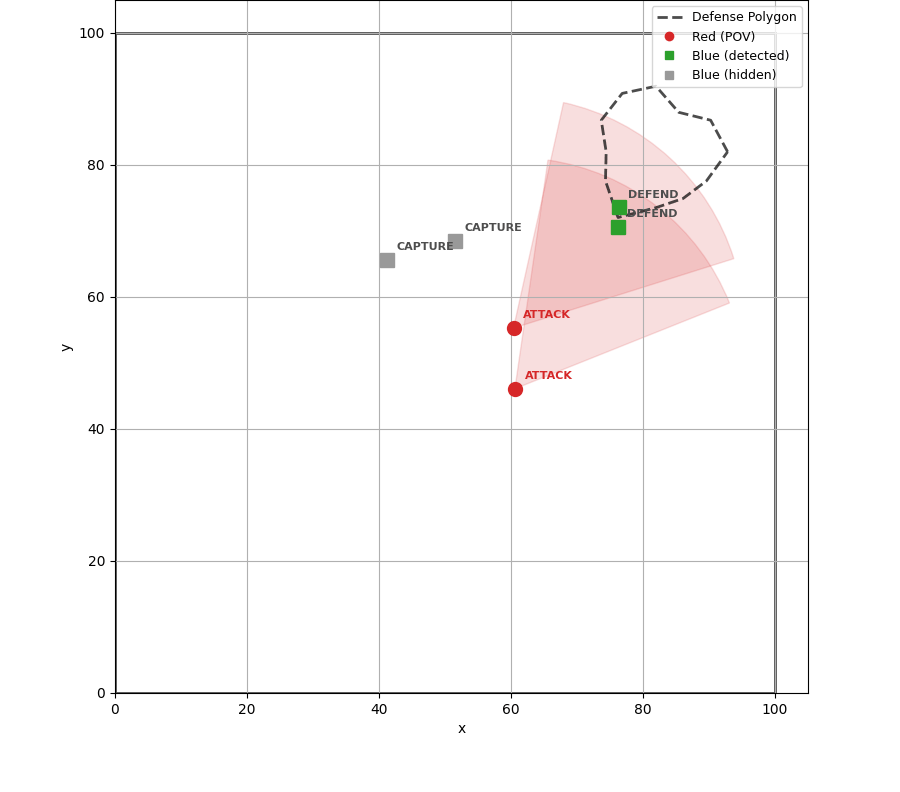}\hfill
        \includegraphics[width=0.32\textwidth]{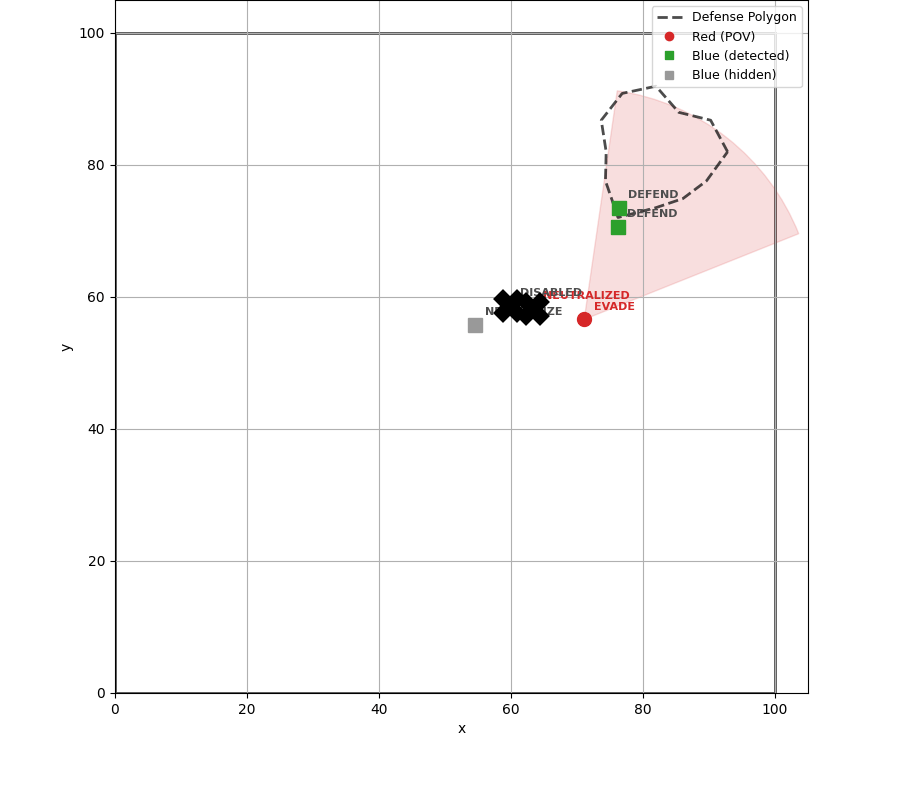}\hfill
        \includegraphics[width=0.32\textwidth]{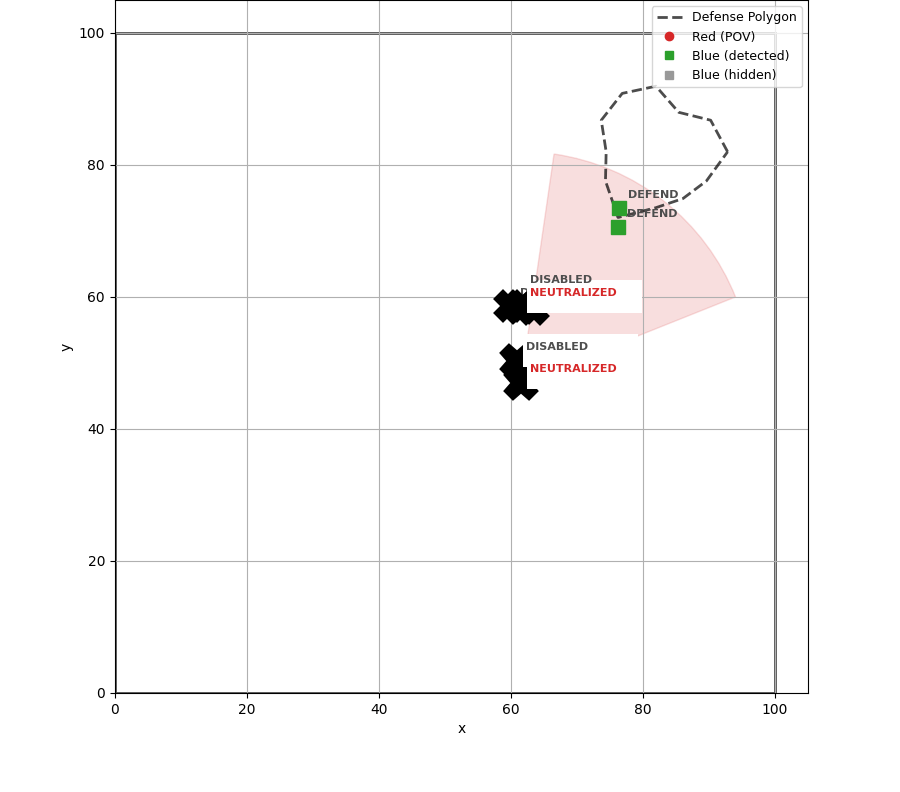}
        \caption{Red POV -- the same episode from the two antagonistic drones' own sensors.}
        \label{fig:exp2-4v2-red-pov}
    \end{subfigure}

    \caption{Frame-by-frame progression of a representative Experiment II engagement (4 blue defenders vs.\ 2 real antagonistic drones, no decoy contacts present), shown from each team's own point of view. Frames are sampled evenly from the first step (leftmost) to the final resolved step (rightmost), in which both drones are neutralized via direct contact and the blue team wins.}
    \label{fig:exp2-4v2-povs}
\end{figure*}

% Requires: \usepackage{graphicx} and \usepackage{subcaption} in the preamble.
\begin{figure*}[t]
    \centering
    \begin{subfigure}[b]{\textwidth}
        \centering
        \includegraphics[width=0.32\textwidth]{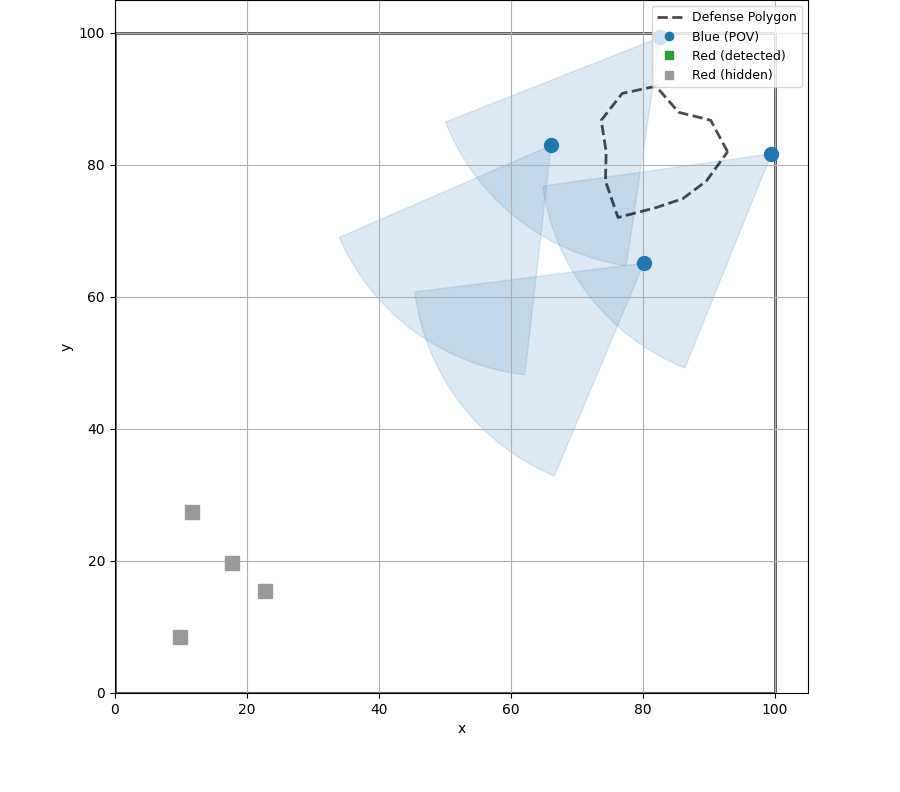}\hfill
        \includegraphics[width=0.32\textwidth]{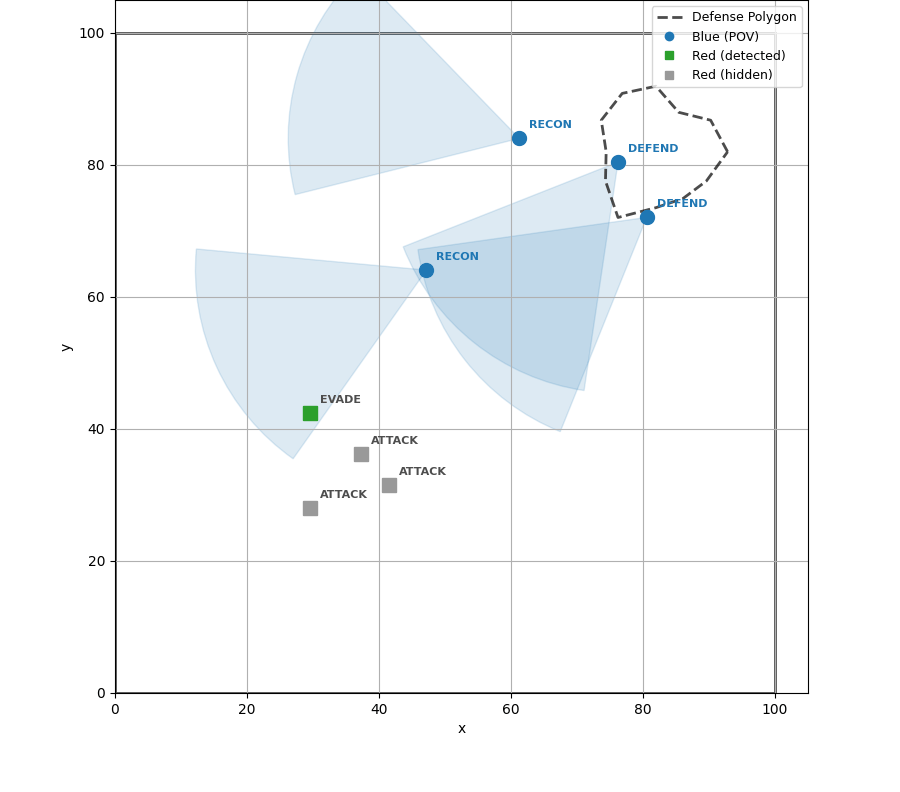}\hfill
        \includegraphics[width=0.32\textwidth]{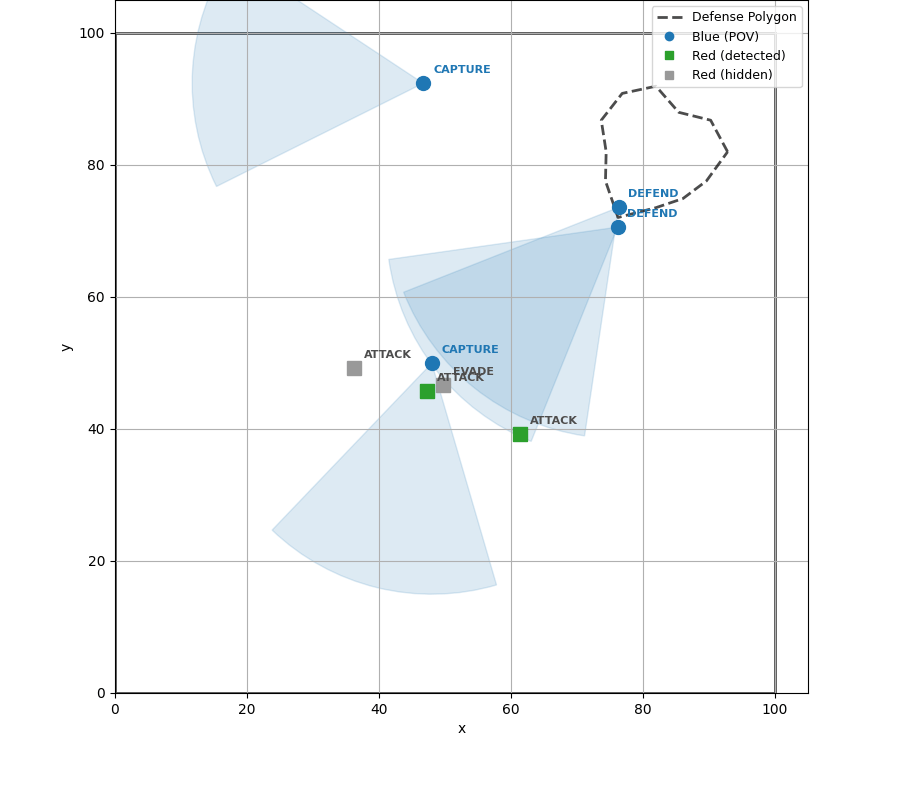}\\[0.4em]
        \includegraphics[width=0.32\textwidth]{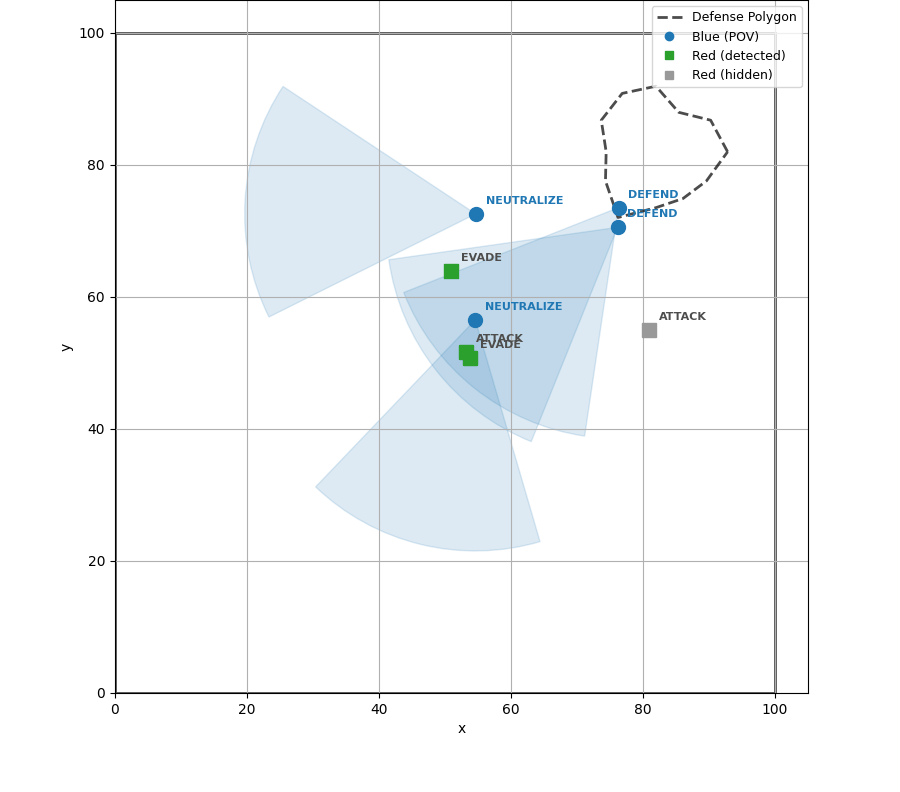}\hfill
        \includegraphics[width=0.32\textwidth]{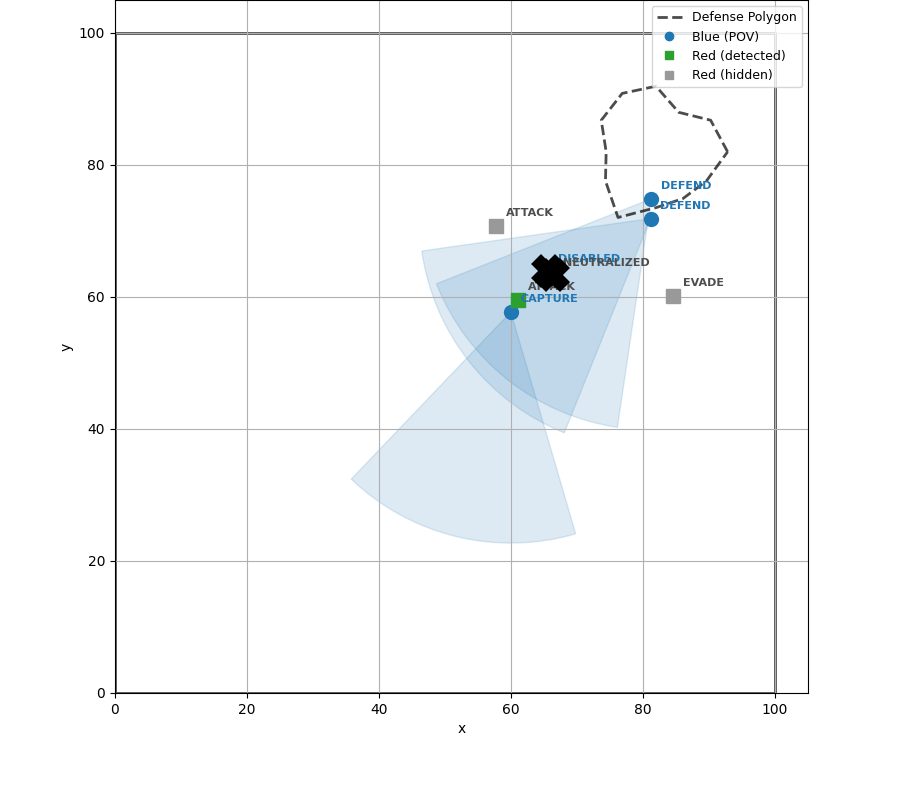}\hfill
        \includegraphics[width=0.32\textwidth]{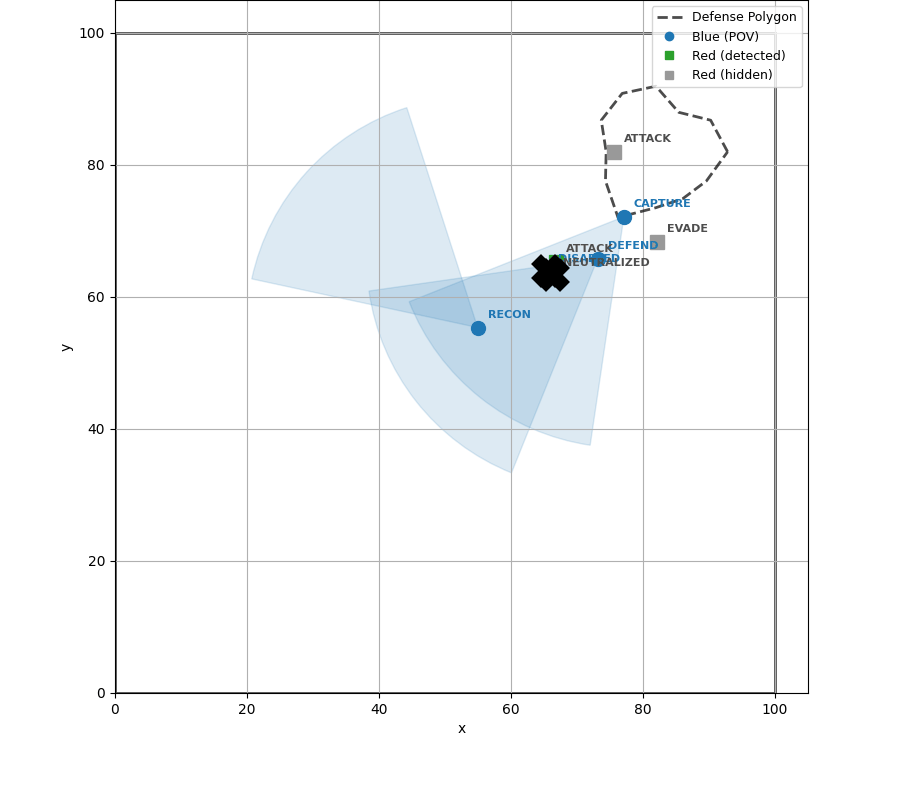}
        \caption{Blue POV -- each defender's field-of-view wedge and its live read of the four antagonistic contacts (green: detected, gray: hidden).}
        \label{fig:exp2-4v4-blue-pov}
    \end{subfigure}

    \vspace{1em}

    \begin{subfigure}[b]{\textwidth}
        \centering
        \includegraphics[width=0.32\textwidth]{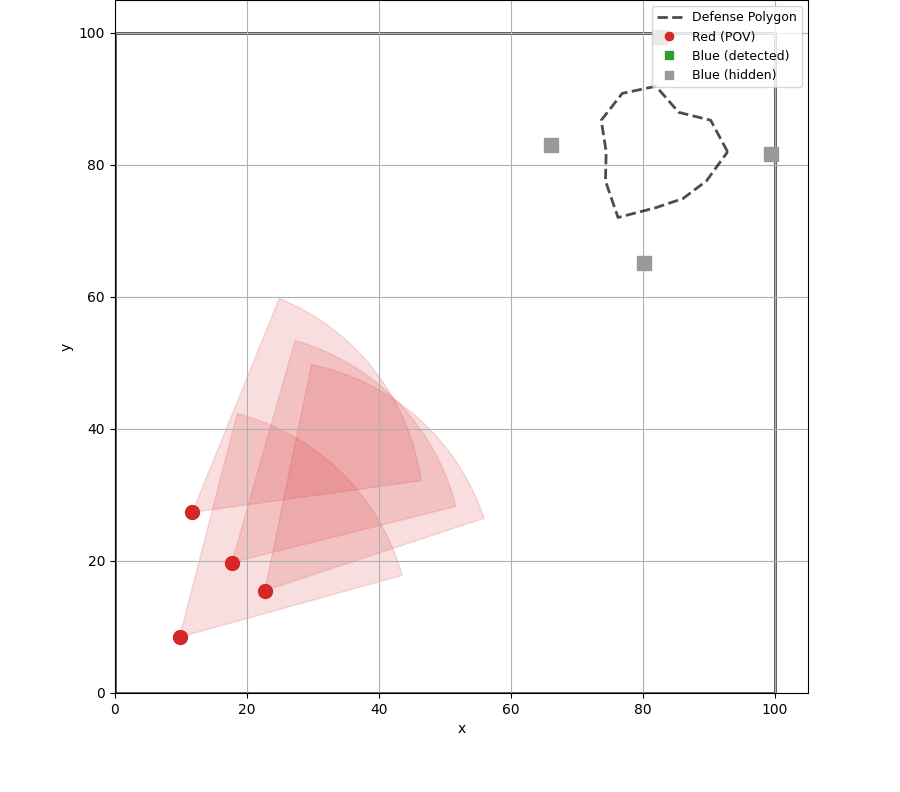}\hfill
        \includegraphics[width=0.32\textwidth]{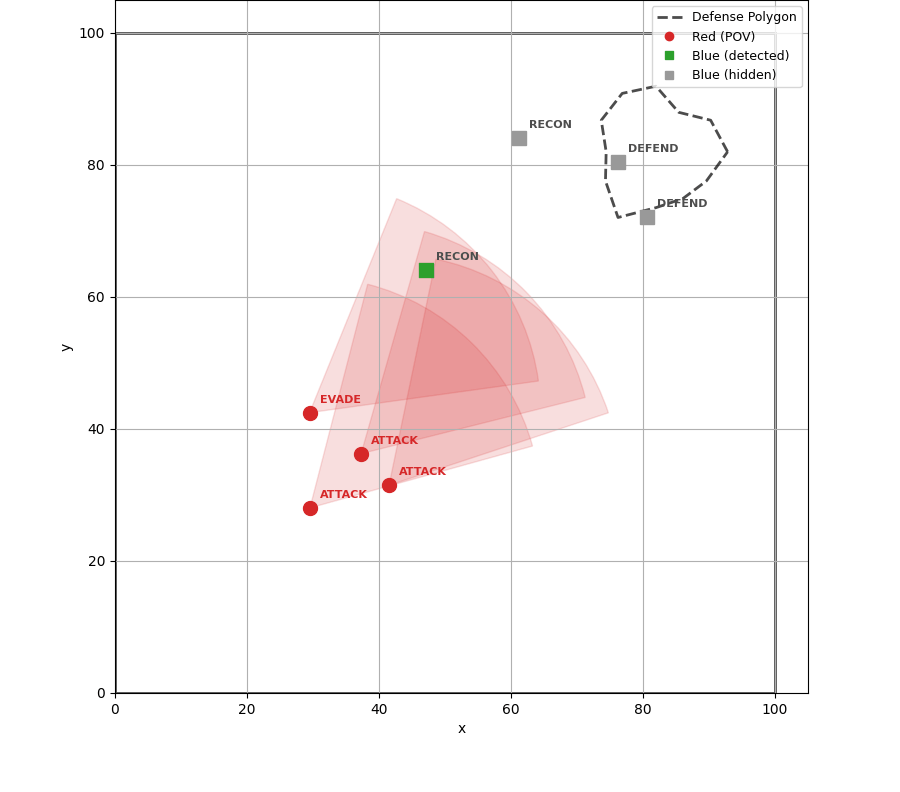}\hfill
        \includegraphics[width=0.32\textwidth]{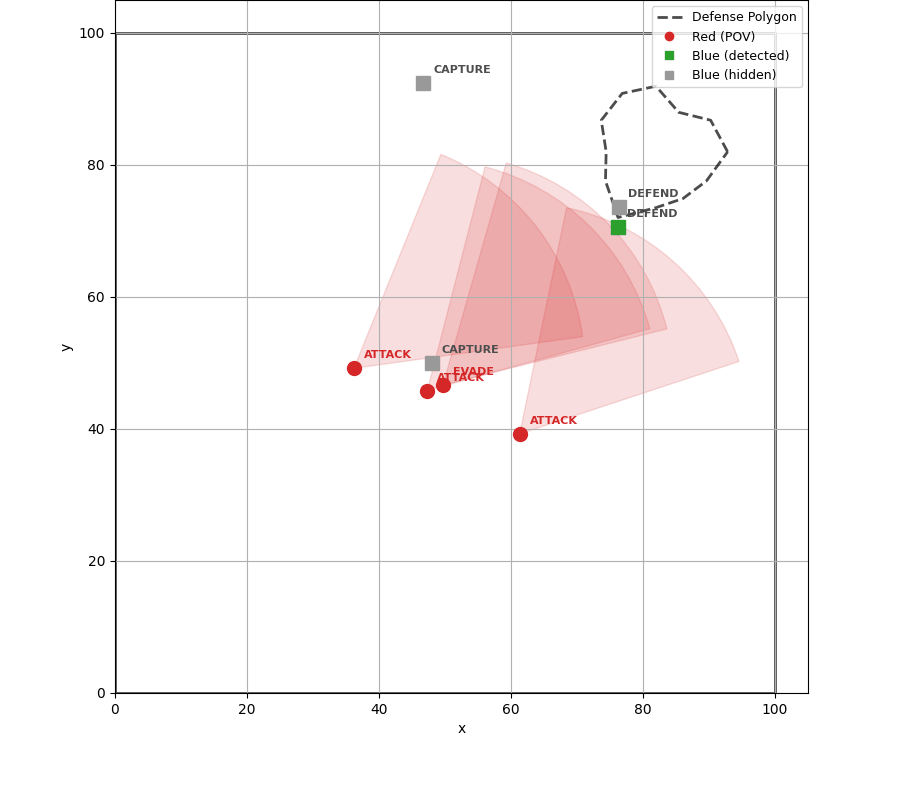}\\[0.4em]
        \includegraphics[width=0.32\textwidth]{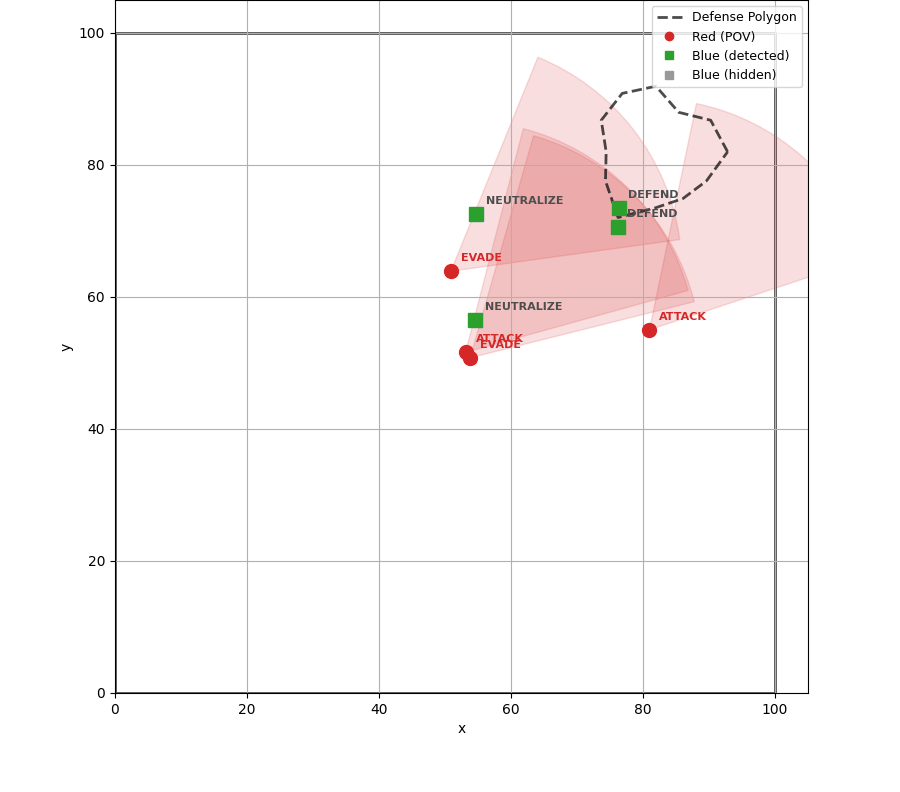}\hfill
        \includegraphics[width=0.32\textwidth]{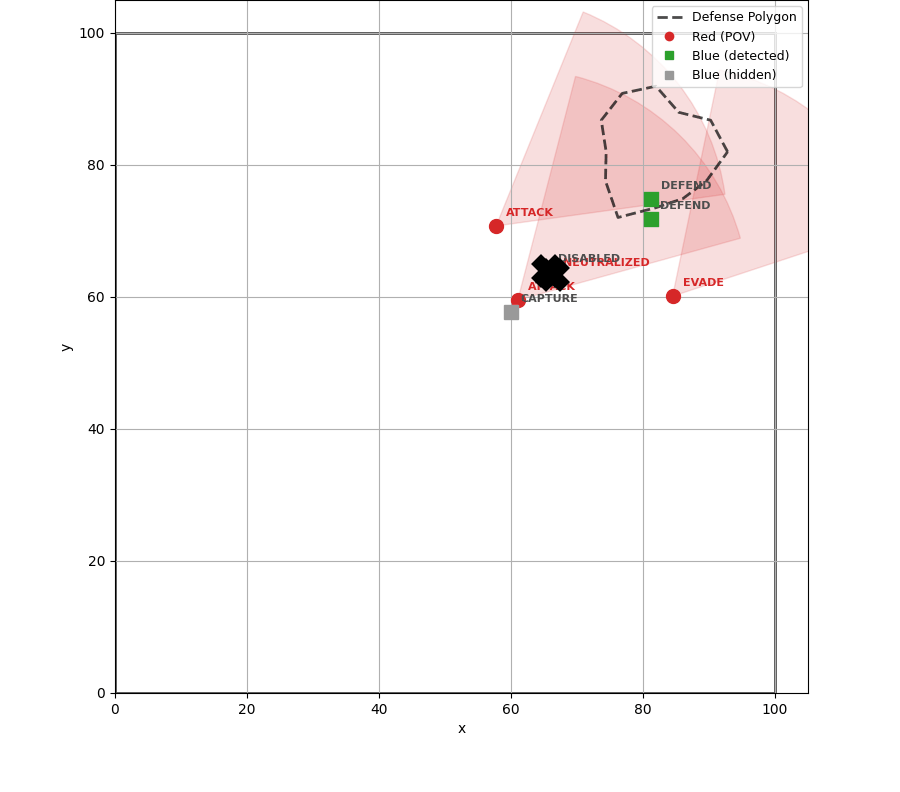}\hfill
        \includegraphics[width=0.32\textwidth]{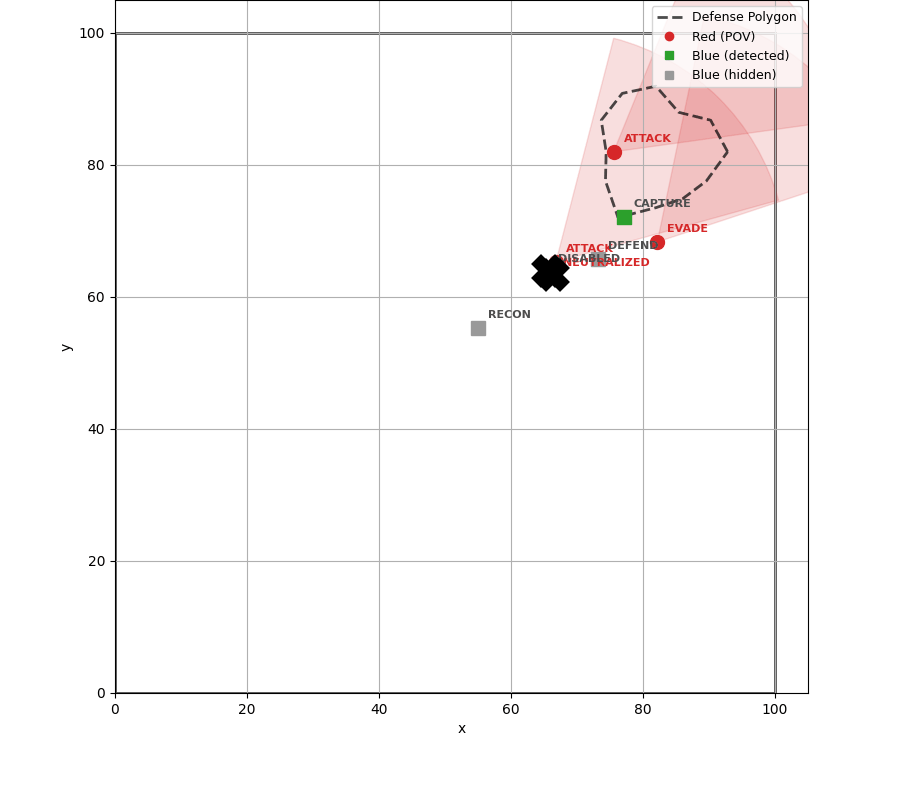}
        \caption{Red POV -- the same episode from the four antagonistic drones' own sensors.}
        \label{fig:exp2-4v4-red-pov}
    \end{subfigure}
`
    \caption{Frame-by-frame progression of a representative Experiment II engagement (4 blue defenders vs.\ 4 real antagonistic drones, no decoy contacts present), shown from each team's own point of view. Frames are sampled evenly from the first step (leftmost) to the final resolved step (rightmost), in which one drone breaches the defended region before the remaining defenders can resolve the engagement, ending in a red victory.}
    \label{fig:exp2-4v4-povs}
\end{figure*}

\end{document}